\documentclass{jfm}

\usepackage{natbib}
\usepackage{amsmath}
\usepackage{color}

\ifx\pdfoutput\undefined
\usepackage{graphicx}
\DeclareGraphicsExtensions{.ps,.eps,.cps}
\newcommand\DOT{.}
\else
\usepackage[pdftex]{graphicx}
\DeclareGraphicsExtensions{.png,.jpg,.pdf}
\newcommand\DOT{.}
\fi

\ifCUPmtlplainloaded \else
  \checkfont{eurm10}
  \iffontfound
    \IfFileExists{upmath.sty}
      {\typeout{^^JFound AMS Euler Roman fonts on the system,
                   using the 'upmath' package.^^J}%
       \usepackage{upmath}}
      {\typeout{^^JFound AMS Euler Roman fonts on the system, but you
                   dont seem to have the}%
       \typeout{'upmath' package installed. JFM.cls can take advantage
                 of these fonts,^^Jif you use 'upmath' package.^^J}%
       \providecommand\upi{\pi}%
      }
  \else
    \providecommand\upi{\pi}%
  \fi
\fi

\ifCUPmtlplainloaded \else
  \checkfont{msam10}
  \iffontfound
    \IfFileExists{amssymb.sty}
      {\typeout{^^JFound AMS Symbol fonts on the system, using the
                'amssymb' package.^^J}%
       \usepackage{amssymb}%

      }{}
  \fi
\fi

\ifCUPmtlplainloaded \else
  \IfFileExists{amsbsy.sty}
    {\typeout{^^JFound the 'amsbsy' package on the system, using it.^^J}%
     \usepackage{amsbsy}}
    {\providecommand\boldsymbol[1]{\mbox{\boldmath $##1$}}}
\fi

\providecommand\bnabla{\boldsymbol{\nabla}}
\providecommand\bcdot{\boldsymbol{\cdot}}

\newcommand\Real{\mbox{Re}} 
\newcommand\Imag{\mbox{Im}} 
\newcommand\Rey{\mbox{\textit{Re}}}  

\newcommand\slsR{\mathsfbi{R}} 
\newcommand\slsT{\mathsfbi{T}} 

\newcommand\bita{\mathit{\boldsymbol{a}}}   
\newcommand\bitu{\mathit{\boldsymbol{u}}}   
\newcommand\bitv{\mathit{\boldsymbol{v}}}   
\newcommand\bitx{\mathit{\boldsymbol{x}}}   
\newcommand\bitr{\mathit{\boldsymbol{r}}}   
\newcommand\bitt{\boldsymbol{\theta}}   
\newcommand\bitz{\mathit{\boldsymbol{z}}}   

\newcommand\sLambda{{\it \Lambda}}   

\newsavebox{\astrutbox}
\sbox{\astrutbox}{\rule[-5pt]{0pt}{20pt}}

\newcommand\eg{e.g.\ }
\newcommand\ie{i.e.\ }

\graphicspath{{./}{./Figs/}}

\title[Supercritical Bifurcation Cascade in Pipe Flow]{From travelling
  waves to mild chaos: a supercritical bifurcation cascade in pipe flow}

\author[F. Mellibovsky and B. Eckhardt]%
{F.\ns M\ls E\ls L\ls L\ls I\ls B\ls O\ls V\ls S\ls K\ls Y$^1$
\and B.\ns E\ls C\ls K\ls H\ls A\ls R\ls D\ls T$^{2,3}$}

\affiliation{$^1$Departament de F{\'{i}}sica Aplicada, Universitat
  Polit{\`{e}}cnica de Catalunya, 08034, Barcelona, Spain
  \\[\affilskip] $^2$Fachbereich Physik, Philipps-Universit\"{a}t
  Marburg, D-35032 Marburg, Germany
  \\[\affilskip] $^3$J.M. Burgerscenter, Delft University of Technology, 2638 CD Delft,
  The Netherlands}

\pubyear{????}
\volume{???}
\pagerange{???--???}
\date{2011}  

\begin{document}

\maketitle

\begin{abstract}
We study numerically a succession of transitions in pipe Poiseuille
flow that leads from simple travelling waves to waves with
chaotic time-dependence.
The waves at the origin of the bifurcation cascade possess a
shift-reflect symmetry and are both axially and azimuthally periodic
with wave numbers $\kappa=1.63$ and $n=2$, respectively.
As the Reynolds number is increased, successive transitions result in
a wide range of time-dependent solutions that includes spiralling,
modulated-travelling, modulated-spiralling,
doubly-modulated-spiralling and mildly chaotic waves.  We show that
the latter spring from heteroclinic tangles of the stable and unstable
invariant manifolds of two shift-reflect-symmetric
modulated-travelling waves. The chaotic set thus produced is confined
to a limited range of Reynolds numbers, bounded by the occurrence of
manifold tangencies.  The states studied here belong to a subspace of
discrete symmetry which makes many of the bifurcation and
path-following investigations presented technically feasible. However,
we expect that most of the phenomenology carries over to the full
state-space, thus suggesting a mechanism for the formation and
break-up of invariant states that can sustain turbulent dynamics.

\end{abstract}

\section{Introduction}\label{sec:intro}

The problem of the transition to turbulence 
in shear flows is one of great
theoretical complexity and paramount practical relevance
\cite[][]{Gro00,ESHW07,Eckhardt09}.  Since Osbourne \cite{Rey1883}
published his fundamental work on the onset of turbulent motion in a
straight pipe of circular cross-section, this problem has become a
paradigm for this class of flows. 
Many
theoretical \cite[][]{BoBr88,BrGr99}, numerical
\cite[][]{ScHe_JFM_94,Zik_PoF_96,SMZN_JFM_99} and experimental
\cite[][]{WyCh_JFM_73,WySoFr_JFM_75,DaMu_JFM_95,HSWE_NATURE_06}
studies have focused on elucidating how turbulence can appear
for finite amplitude perturbations despite 
the linear stability of the laminar 
Hagen-Poiseuille flow \cite[][]{Pfenniger_B_61,MeTr_JCP_03}. 

The computation of finite amplitude solutions in the form of
travelling waves \cite[][]{FaEc_PRL_03,WeKe_JFM_04,PrKe_PRL_07} has
opened up a window for further studies from a nonlinear dynamics
perspective. This parallels similar developments in other flows such
as plane Couette \cite*[][]{Nagata_PRE_97,WaGiWa_PRL_07} or plane
Poiseuille \cite*[][]{PuSa_JFM_88,SoMe_JFM_91,EhKo_JFM_91}.
These waves typically appear in saddle-node bifurcations and come in a
variety of symmetries \cite*[][]{PrKe_PRL_07,PrDuKe_PTRSA_09}. The
upper-branch solutions, with wall friction values closer to turbulent
ones, seem to play a role in developed turbulence
\cite*[][]{ESHW07,KeTu_JFM_07,ScEcVo_PRE_07} and have been observed in
experiments \cite[][]{HVWNFEWKW_SCI_04}. Their lower-branch
counterparts appear to be connected to the critical threshold
\cite*[][]{DuWiKe_JFM_08,MeMe_PTRSA_09} separating the basins of
attraction of laminar and turbulent flows, and are relevant for the
transition process \cite[][]{SkYoEc_PRL_06,ScEcYo_PRL_07}.
Some of the latest developments
for the case of pipe flow are compiled in \cite{Eckhardt09}.

The presence of finite amplitude persistent structures is a prerequisite
for the appearence of turbulence, but further structural information 
is needed to explain or understand the dynamics. Specifically, 
further bifurcations are needed in order to increase the temporal
complexity and to change the nature of the dynamics from 
persistent to transient or vice versa. Accordingly, we here focus on
the transitions of these waves in order to explore the origins of the
increasing complexity observed in pipe flow.

Since tracking of states with several unstable directions in
high-dimensional spaces is technially difficult and computationally
demanding, we focus on the subspace with $2$-fold azimuthal
periodicity that was also explored in \citet*[][]{MeEc_JFM_10}. The
symmetry eliminates some unstable directions so that a direct
identification of the relevant structures without the need for
advanced tracking or stabilisation methods is possible. Specifically,
we have identified a family of states in this subspace for which the
lower-branch solutions have a single unstable direction within the
subspace so that a symmetry-restricted time evolution converges when
combined with edge tracking techniques
\cite*[][]{SkYoEc_PRL_06,ScEcYo_PRL_07,DuWiKe_JFM_08,ScEc_09}. A
thorough study of the Takens-Bodganov bifurcation in which this family
of waves is created showed that, for wavenumbers beyond a certain
value, the upper-branch waves are stable within the subspace and,
therefore, within grasp of direct time-evolution
\cite*[][]{MeEc_JFM_10}. We here will study the evolution of the upper
branch solutions and their secondary instabilities that lead to
time-dependent behaviour and, eventually, to chaos and turbulence.

Due to the broken fore-aft symmetry of pipe flow, the simplest
possible solutions already exhibit a time-dependence in the form of an
axial downstream drift with a constant phase speed. Some spiralling
(rotating-travelling) waves, which rotate due to the breach of all
azimuthal reflection ($Z_2$) symmetries, have also been identified
\cite[][]{PrKe_PRL_07,MeMe_PTRSA_09}. \cite*{DuPrKe_PoF_08} took
time-dependency in pipe flow one step further by computing a branch of
modulated travelling waves (or, as they called them, relative periodic
orbits), although their role in transition or sustained turbulence was
assessed as inconclusive. A second family arose naturally from the
study of the Takens-Bogdanov bifurcation in \cite{MeEc_JFM_10}, but
again its existence seemed at best incidental. 
No further solutions of higher complexity that bridge the gap
between the steady laminar flow and turbulence have been observed or 
computed in pipe flow.

\begin{figure}
  \begin{center}
    \includegraphics[width=0.9\linewidth,clip]{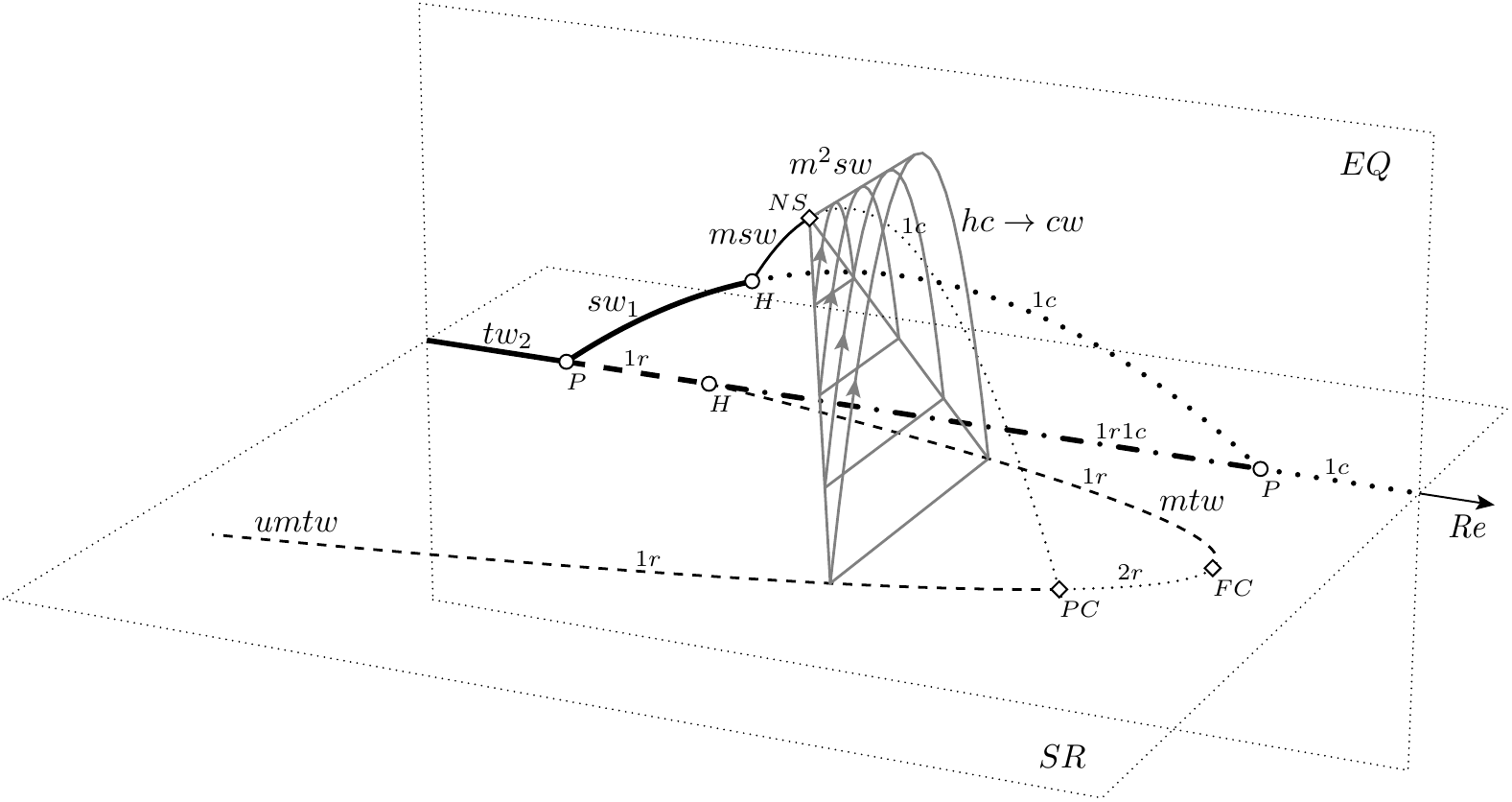}
  \end{center}
  \caption{Sketch summarising the bifurcation structure identified
    and discussed in the present manuscript.  The three axes indicate
    the Reynolds number ($\Rey$, the control parameter), the departure
    from the space of relative equilibria ($EQ$ vertical plane) and
    the departure from the shift-reflect subspace (horizontal $SR$
    plane). 
    %
    %
    The various states found are travelling ($tw$),
    spiralling ($sw$), modulated-travelling ($mtw$), unstable
    modulated-travelling ($umtw$), modulated-spiralling ($msw$),
    doubly-modulated-spiralling ($m^2sw$) and chaotic ($cw$)
    waves. Pitchfork ($P$), Hopf ($H$), Neimark-Sacker ($NS$), fold-of
    -cycles ($FC$), pitchfork-of-cycles ($PC$), and a global
    heteroclinic-cyclic ($hc$) bifurcations are indicated. Solid,
    dashed, dotted and dash-dotted lines represent solutions with $0$
    (stable), $1$, $2$ and $3$-dimensional unstable manifolds. Real
    and complex pairs are indicated with $r$ and $c$.}
  \label{fig:BifDia}
\end{figure}

The aim of this study is precisely to take advantage of the stability
of upper-branch waves in the family of shift-reflect symmetric states
to explore a bifurcating
cascade and to produce, along the way, an array of solutions
exhibiting ever increasing time-dependent
complexity. Figure~\ref{fig:BifDia} summarises in a sketch the bifurcation
cascade that the travelling waves undergo as $\Rey$ is increased.
All relative equilibria ($EQ$ vertical plane) as well as all stable
states (solid lines) and part of the shift-reflect ($SR$ horizontal
plane) unstable states are readily accessible and give evidence of the
bifurcation scenario proposed. We will argue that, at the final stage,
the clean heteroclinic-cyclic bifurcation shown here is replaced by a
wedge of manifold tangencies giving rise to mildly chaotic waves (as
defined and explained in \S\ref{sec:discu}).

The states discussed here may be stable in their symmetry-restricted
subspace, but they are typically unstable when considered within the
full state space. Since there is nothing to suggest that the
bifurcation sequence observed and discussed here requires the symmetry
reduction (except for reducing the effort that has to go into their
identification), it can be expected that it also appears for
travelling waves in the full state space. In this sense the present
study also suggests paths for the formation and break up of other
invariant structures in pipe flow.


The outline of the paper is as follows. In \S\ref{sec:formet}, we
present the pipe Poiseuille flow problem and sketch a numerical scheme
for the integration of the resulting equations. The symmetries of the
problem are discussed and numerical methods for the computation and
stability analysis of relative equilibria are briefly summarised. Some
aspects of bifurcation theory in the case of relative equilibria are
also reviewed in \S\ref{sec:formet}. A detailed exploration of
parameter space delimiting the region of existence of relevant
travelling and spiralling waves and an analysis of their stability is
presented in \S\ref{sec:releq}. In \S\ref{sec:timedep} we set out to
uncover successive transitions from these waves into increasingly
complex types of solutions and describe their main features. The
detailed aspects of the global transition leading to the creation of a
chaotic set are then discussed in \S\ref{sec:discu}. Conclusions and
ensuing remarks are summarised in \S\ref{sec:conclu}.

\section{Formulation and methods}\label{sec:formet}


In this section we briefly summarise the methods section from a
previous contribution so that the symbols and key elements are
introduced.  Further details can be found in \S2 of
\cite{MeEc_JFM_10}.

We consider the flow of an incompressible Newtonian fluid through a pipe
of circular cross section at constant mass flow. The Reynolds number
$\Rey = U D / \nu$ is based on the mean axial flow speed $U$, the pipe 
diameter $D$ and the kinematic viscosity $\nu$. The base profile,
in nondimensional cylindrical coordinates ${\bitx}=(r,\theta,z)$, 
is ${\bitu}_{\rm b}=u_r^{\rm b}\;\hat{\bitr}+u_{\theta}^{\rm b}\;\hat{\bitt}+
u_z^{\rm b}\;\hat{\bitz}=(1-r^2)\;\hat{\bitz}$. 
The equations for the perturbations in velocity ${\bitu}=(u_r,u_{\theta},u_z)$ and pressure $p$ 
are the Navier-Stokes equations
\begin{equation}
  \partial_t{\bitu} = -\bnabla p + \frac{1}{\Rey}\Delta {\bitu} -
  ({\bitu} \bcdot \bnabla) ({\bitu}_{\rm b} +{\bitu}) - ({\bitu}_{\rm b}
  \bcdot \bnabla) {\bitu} + f \; \hat{\bitz},\\
  \label{eq:NS}
\end{equation}
together with the incompressibility constraint,
\begin{equation}
  \bnabla \bcdot  {\bitu} = 0,
  \label{eq:DV}
\end{equation}
the condition for constant mass flux,
\begin{equation}
  Q(\bitu) = \int_0^{2 \pi} \int_0^1 (\bitu \bcdot \hat{\bitz}) \; r
  {\rm d}r {\rm d}\theta = 0,
  \label{eq:CMc}
\end{equation}
and the boundary conditions,
\begin{equation}
  {\bitu}(1,\theta,z;t)={\boldsymbol 0}, \;\; {\bitu}(r,\theta+2\pi/{n_s},z,t)={\bitu}(r,\theta,z+2\upi/\kappa,t)={\bitu}(r,\theta,z,t)
\label{eq:BC}.
\end{equation}
The adjustable axial forcing $f=f(t)$ in (\ref{eq:NS}) ensures the
constant mass-flux constraint (\ref{eq:CMc}). In addition to the
non-slip boundary condition at the wall we have periodicity in the
azimuthal and axial directions. In the azimuthal direction we take
$n_s=2$ throughout, confining the analysis to $2$-fold
azimuthally-periodic fields. All solutions in this subspace are also
solutions to the full Navier-Stokes equations, but the imposed
discrete symmetry supresses some undesired unstable directions. The
axial wave number $\kappa$ is an additional parameter that was
important in unfolding the Takens-Bodganov bifurcation in
\cite{MeEc_JFM_10}. We here focus on the values $\kappa=1.63$
($\sLambda \simeq 1.927 D$), for which the bifurcation cascade presents
its maximum complexity and richness of solutions, and on the interval
$\kappa \in [1.5,1.7]$. The Reynolds number is varied in the range
$\Rey \in [1800,2800]$, with most of the bifurcations concentrated in
a small range around $\Rey \simeq 2200$.

For the spatial discretisation of (\ref{eq:NS}-\ref{eq:BC}) we use a
solenoidal spectral Petrov-Galerkin scheme thoroughly described and
tested by \cite*{MeMe_ANM_07}.
The velocity field is expanded in the form
\begin{equation}
  \bitu(r,\theta,z;t)=\sum_{l=-L}^{L} \sum_{n=-N}^{N}
  {e}^{-{\rm i} (\kappa l z + n_s n \theta)} \;\bitu_{ln}(r;t)
\label{eq:uexp1},
\end{equation}
\begin{equation}
  \bitu_{ln}(r;t)=
  \sum_{m=0}^{M} a_{lnm}^{(1,2)}(t) \; \bitv_{lnm}^{(1,2)}(r)
\label{eq:uexp2},
\end{equation}
with $a_{lnm}^{(1,2)}$ the complex expansion coefficients which are
collected in a state vector $\bita$.  The spectral resolution, checked
as adequate for the computations performed in this study, has been set
to $L=16$, $N=12$ and $M=36$, corresponding to $\pm 16$ axial and $\pm
12$ azimuthal Fourier modes, and to $37$ Chebyshev collocation points
for the radial coordinate. For the time-evolution, we take a 4th order
linearly implicit method with stepsize $\Delta t = 1 \times 10^{-2}\,
D/(4 U)$.

Fluid flow solutions lie in infinite dimensional space, of which we
compute finite, yet high-dimensional representations. Elegant
approaches to project these solutions onto low-dimensional spaces to
aid visualisation have been devised \cite*[][]{GiHaCv_JFM_08} and
successfully applied to other flows such as plane Couette
\cite*[][]{GiHaCv_JFM_09}. We take the simpler approach of using some
random expansion coefficients and global derived quantities to
represent the flow. To free the representation from the drifting
degeneracy associated to the travelling and/or rotating component of
the waves, it is more convenient to use the absolute value rather than
the coefficients themselves. In this study we will be making extensive
use of the moduli of a streamwise-independent ($\lvert a_{010}^{(2)}
\rvert$), an axisymmetric ($\lvert a_{100}^{(2)} \rvert$) and a mixed
($\lvert a_{110}^{(2)} \rvert$) coefficient. In what follows, the
absolute value symbols will be omitted for simplicity.

The basic solutions come in the form of travelling waves,
which are best described as relative equilibria and possess the continuous
space-time symmetry
\begin{equation}
(u,v,w)_{\rm tw}(r,\theta,z;t)=(u,v,w)_{\rm tw}(r,\theta,z - c_z t;0)
\label{eq:SpaceTimeSym},
\end{equation}
where $c_z$ is the axial drift speed. In a comoving reference frame
travelling downstream with speed $c_z$ travelling waves appear as
stationary solutions. Near a relative equilibrium the drift dynamics
is trivial and decouples from the dynamics orthogonal to the drift. As
a result, bifurcations of relative equilibria can be analysed in two
steps, first describing the bifurcations associated to the orthogonal
dynamics, then adding the corresponding drift along the travelling
direction \cite*[][]{Krupa_SJMA_90}.

In order to avoid the transformation to the co-moving frame of 
reference, it is convenient to have observables that are translationally
invariant in the axial direction: then the phase motion of the travelling
wave drops out and they appear as fixed points in this observable without
further action. One such quantity is the normalised energy,
\begin{equation}
    \varepsilon({\bitu}) = \frac{1}{2\epsilon_{\rm b}} \int_0^{2
    \upi/\kappa} {\rm d}z \int_0^{2 \pi} {\rm d}\theta \int_0^1 r{\rm
    d}r \; {\bitu}^* \bcdot {\bitu}
\label{eq:energy1},
\end{equation}
with $\epsilon_{\rm b}=\upi^2/(3\kappa)$ the energy of the basic flow
and $^*$ symbolising complex conjugation.  This energy decouples
exactly into the sum of its axial-azimuthal Fourier components
$\varepsilon_{ln}=\varepsilon(\bitu_{ln})$. As a measure of
three-dimensional structure, we define the non-axisymmetric
streamwise-dependent component as $\varepsilon_{\rm 3D}=
\sum_{l,n\ne0}\varepsilon(\bitu_{ln})$.

Another such quantity is the mean axial pressure gradient needed to
drive the flow at constant mass-flux, normalised by the pressure
gradient for the corresponding laminar flow,
\begin{equation}
  (\bnabla p)_z = \left. 
  	\left\langle \iint_{r,\theta} \frac{\partial p}{\partial z}
  \right\rangle_z 
  \right/ 
  \left( \frac{dp}{dz} 
  \right)_{lam}=1+\frac{Re f}{4}
\label{eq:gradp}.
\end{equation}
It is closely related to the wall friction factor
\cite*[][]{SchGer_B_68}:
\begin{equation}
  C_f = \frac{2 D}{\rho U^2} \left\langle \iint_{r,\theta} \frac{\partial
    p}{\partial z}\right\rangle_z = 64 \frac{(\bnabla p)_z}{Re}
\label{eq:gradp},
\end{equation}
which constitutes a good indicator of whether the flow is laminar or
turbulent.

In the azimuthal direction, all the solutions studied here are
invariant under the cyclic group $C_{2}$ (rotations by integer
multiples of $\pi$). On top of this symmetry, the travelling wave
family that constitutes the departing point for the present study,
possesses an additional discrete symmetry: a combined shift-reflect
symmetry. Solutions invariant under this symmetry operation,
\begin{equation}
S{\bitu}({\bitx})=S(u,v,w)(r,\theta_i+\theta,z;t)=(u,-v,w)(r,\theta_i-\theta,z+\upi/\kappa;t)
\label{eq:ShiftRefSym},
\end{equation}
are left unaltered when shifted half a wavelength downstream and then
reflected with respect to any of two diametral planes tilted with
$\theta_{0}$ and $\theta_0+\upi/2$, where $\theta_0$ parametrises the
azimuthal degeneracy of solutions. It can be shown that the expansion
coefficient $a_{100}^{(2)}$ vanishes exactly when this symmetry is
present, its norm giving a fairly good notion of how far apart from
the symmetry space any given flow field is. The $S$ symmetry is a
remnant version of the $Z_2$ reflections group implied by the broken
$O(2)=SO(2) \times Z_2$ azimuthal symmetry. Waves that break all
left-right symmetries incorporate an azimuthal precessing motion on
top of the axial drift, so that they spiral and modify
(\ref{eq:SpaceTimeSym}) into
\begin{equation}
(u,v,w)_{\rm tw}(r,\theta,z;t)=(u,v,w)_{\rm tw}(r,\theta - c_{\theta}
  t,z - c_z t;0)
\label{eq:SpaceTimeSym2},
\end{equation}
where $c_{\theta}$ is the azimuthal drift speed. Evidently, the
decoupling from the drift dynamics of the drift-orthogonal dynamics
discussed for travelling waves holds true for waves that also rotate.

Modulated waves arise from Hopf bifurcations of travelling and
spiralling waves. No general result is available when it comes to their
stability. The effects on the degenerate frequency
(or frequencies) associated to the travelling (or spiralling)
component of the wave when a modulational frequency comes into play
cannot be discarded at once. Nonetheless, as our computations will
show, the approach of choosing an appropriate comoving frame of
reference remains accurate, and modulated travelling
(spiralling) waves seem to behave as relative periodic orbits rather
than as generic quasiperiodic orbits on a $2$-torus ($3$-torus).

In order to choose this appropriate frame of reference we define the
instantaneous axial and azimuthal phase speeds \cite*[][]{MeEc_JFM_10} as
the values $c_z$ and $c_\theta$, respectively, that minimize at any
given instant of time, the $2$-norm
\begin{equation}
  \min_{c_z,c_{\theta}} \left\Vert \bita(t+\Delta t) - \slsT(c_z \Delta t)
  \slsR(c_{\theta} \Delta t) \; \bita(t) \right\Vert
  \label{eq:PhSpeeds},
\end{equation}
with $\slsT(\Delta z)$ and $\slsR(\Delta \theta)$ the diagonal
operators for translation and rotation, whose action on the components
of the state vector is defined as:
\begin{equation}
  \left.
  \begin{array}{rcl}
    \left( \slsT(\Delta z) \; \bita \right)_{lnm}^{(1,2)} & = &
    a_{lnm}^{(1,2)} \; \mathrm{e}^{-{\rm i} \kappa l \Delta z}, \\[0.2cm]
    \left( \slsR(\Delta \theta) \; \bita \right)_{lnm}^{(1,2)} & = &
    a_{lnm}^{(1,2)} \; \mathrm{e}^{-{\rm i} n_s n \Delta \theta}.
  \end{array}
  \right\}
  \label{eq:TransRotOp}
\end{equation}

Therefore, we will systematically work on a comoving reference frame
in which waves become equilibria and we will apply continuous
dynamical systems theory. Moreover, modulated and doubly-modulated
waves become periodic orbits and $2$-tori, so that they can be seen as
equilibria and periodic orbits, respectively, on a purposefully
designed comoving Poincar\'e section. This is accomplished by using
drift-independent quantities upon definition of the Poincar\'e
section. The stability of modulated waves can then be studied through
their associated Poincar\'e map, making use of discrete-time dynamical
systems theory.

Besides the complications introduced by the bifurcation of relative
equilibria, \ie group orbits invariant under the flow of equivariant
vector fields, we will be dealing here with additional symmetries that
modify or replace well known bifurcating scenarios that are typical of
low dimensional dynamical systems. To overcome this, we will take
advantage of the substantial developments that have been achieved in
the last two decades regarding bifurcation in dynamical systems with
symmetries
\cite*[][]{GoStSc_B_88,CrKn_ARFM_91,ChossatLauterbach_B_00}.

\section{Relative equilibria}\label{sec:releq}
The relative equilibria considered here are travelling waves or
spiralling waves, i.e. states that become stationary in appropriately
translating and/or rotating frames of reference.  In addition to the
discrete $2$-fold azimuthal periodicity ($C_{2}$, \ie $n_s=2$), the
waves at the origin of this study also show a combined shift-reflect
symmetry ($S$).  They were first computed by \cite{FaEc_PRL_03} and
\cite{WeKe_JFM_04} using volume forcing homotopy.

When parametrised in terms of axial wavenumber ($\kappa$) and Reynolds
number ($\Rey$), they appear in a Takens-Bogdanov bifurcation,
thoroughly studied in \cite{MeEc_JFM_10}. Their existence extends in
Reynolds number to as low as $\Rey \simeq 1358.5$ for the optimal wave
number $\kappa \simeq 1.55$
\cite[][]{FaEc_PRL_03,WeKe_JFM_04}. Ensuing from this bifurcation
scenario there are parameter regions in which upper-branch solutions
are stable to $2$-fold azimuthally-periodic perturbations so that
time-dependent solutions which bifurcate at higher $\Rey$ can be
computed via straightforward symmetry-restricted time-evolution.

In what follows, we continue both upper- and lower-branch members of
this family of travelling waves to higher $\Rey$ and analyse their
stability in order to explore the framework in which increasingly
complex dynamics occur.

We do not study the shift-reflect mirror-symmetric family of
travelling waves \cite[][]{DuWiKe_JFM_08} that coexists in state space
with the solutions discussed here. They reside in a region of phase
space that is noticeably far from the present region of interest and
there are no indications that they interact with the solutions found
in this work, at least not in the parameter ranges studied here.

\subsection{Travelling waves}

\begin{figure}
  \begin{center}
    \begin{tabular}{cccccc}
      \raisebox{0.16\linewidth}{(\textit{a})}\hspace{-0.6cm} &
      \includegraphics[height=0.15\linewidth,clip]{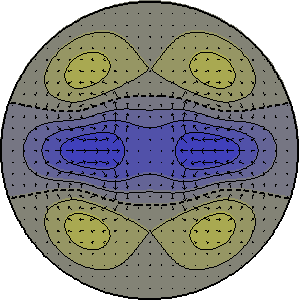} &
      \includegraphics[height=0.15\linewidth,clip]{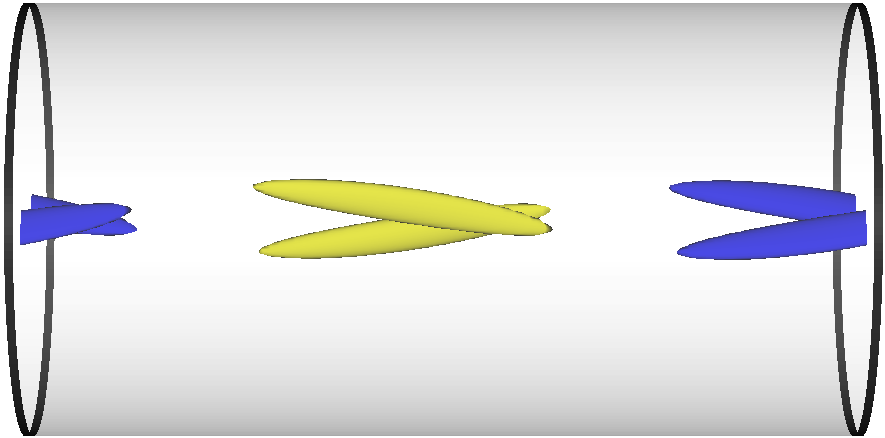} &
      \raisebox{0.16\linewidth}{(\textit{b})}\hspace{-0.6cm} &
      \includegraphics[height=0.15\linewidth,clip]{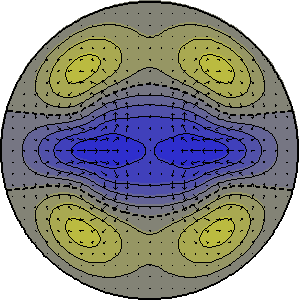} &
      \includegraphics[height=0.15\linewidth,clip]{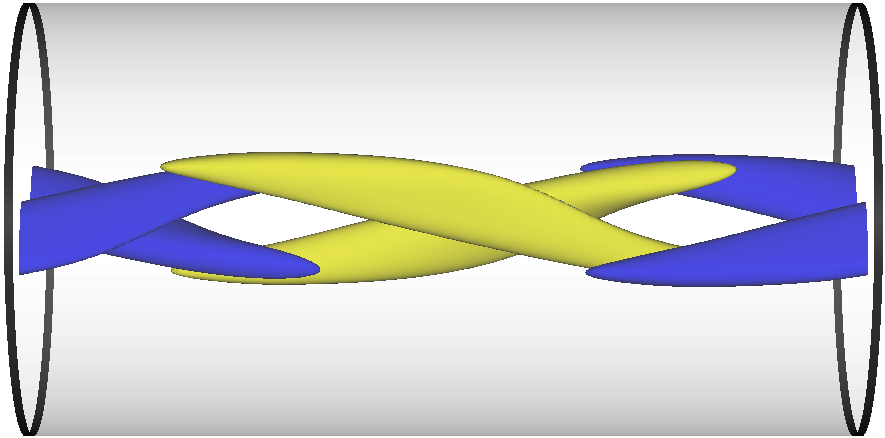}\\
      \raisebox{0.16\linewidth}{(\textit{c})}\hspace{-0.6cm}
      &
      \includegraphics[height=0.15\linewidth,clip]{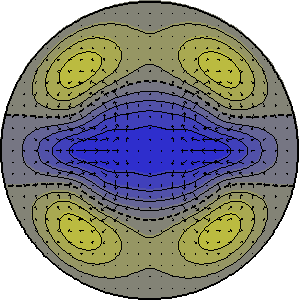} &
      \includegraphics[height=0.15\linewidth,clip]{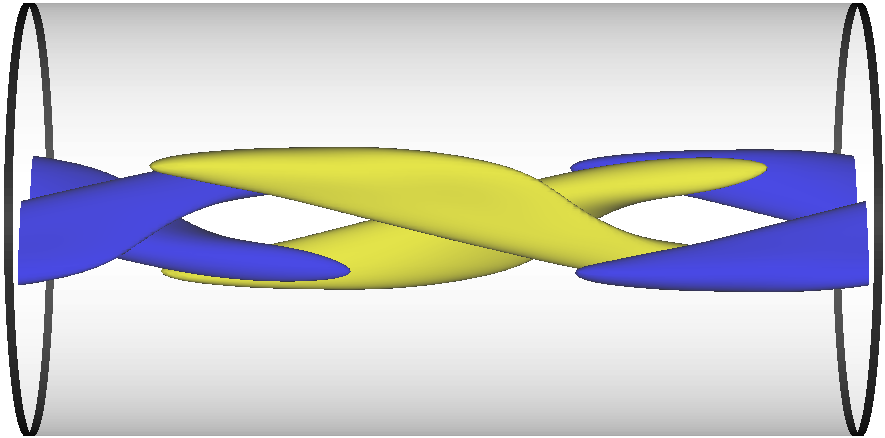} &
      \raisebox{0.16\linewidth}{(\textit{d})}\hspace{-0.6cm} &
      \includegraphics[height=0.15\linewidth,clip]{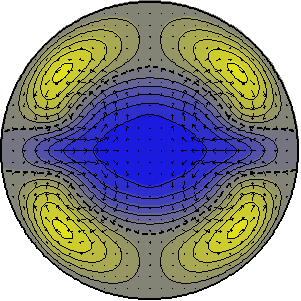} &
      \includegraphics[height=0.15\linewidth,clip]{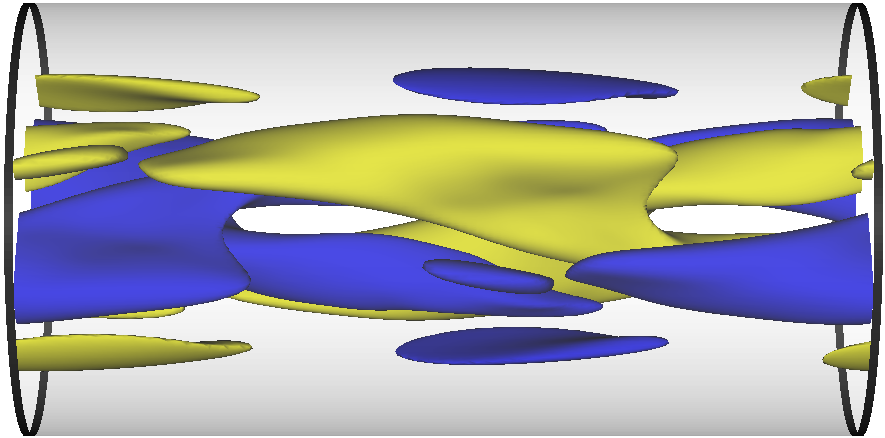}\\
    \end{tabular}
  \end{center}
  \caption{Travelling waves at $(\kappa,\Rey)=(1.63,2215)$. Left:
    $z$-averaged cross-sectional axial velocity contours spaced at
    intervals of $\Delta \langle u_z\rangle_z = 0.1 U$. In-plane
    velocity vectors are also displayed. Right: axial vorticity
    isosurfaces at $\omega_z=\pm 1 U/D$. Fluid flows
    rightwards. Blue (dark gray) for negative, Yellow (light) for
    positive. (\textit{a}) Lower-branch ($tw_1$), (\textit{b})
    lower-middle-branch ($tw_2$), (\textit{c}) upper-middle-branch
    ($tw_3$) and (\textit{d}) upper-branch ($tw_4$) travelling waves.}
  \label{fig:tws}
\end{figure}

Lower-branch travelling waves of this family can be continued to
extremely high $\Rey$ without much noticeable change to their
stability properties, regardless of $\kappa$.
As many other lower-branch travelling waves, they seem to develop a
critical layer as $\Rey$ is increased \cite*[][]{Vis_PTRSA_09} and
exhibit a single unstable eigenmode when considered in the azimuthal
subspace they inhabit. They are edge states \cite[][]{ScEcYo_PRL_07}
of the $2$-fold azimuthally-periodic pipe within this subspace.  In
the full state space they may be part of the edge and may provide the
symmetric states needed to connect the stable manifolds of
non-symmetric but symmetry-related edge states
\cite*[][]{VoScEc_NJP_09}.
%
Figure~\ref{fig:tws}(\textit{a}) shows
$z$-averaged cross-sectional axial velocity contours ($\langle
u_z\rangle_z$, left) and a couple of axial vorticity isosurfaces
($\omega_z$, right) of a lower-branch travelling wave at
$(\kappa,\Rey)=(1.63,2215)$.
The shift-reflect symmetry of the solution is evidenced by the two
orthogonal diametral reflection planes (at $\theta_i=0,\upi/2$) of the
$z$-averaged cross-sectional contours. High- and low-speed streaks are
clearly visible. In-plane velocity vectors show the location, on
average, of the two pairs of counter-rotating vortices that are
displayed in the three-dimensional view.

\begin{figure}
  \begin{center}
    \begin{tabular}{cc}
      \raisebox{0.5\linewidth}{(\textit{a})}\hspace{-0.6cm} &
      \includegraphics[width=0.9\linewidth,clip]{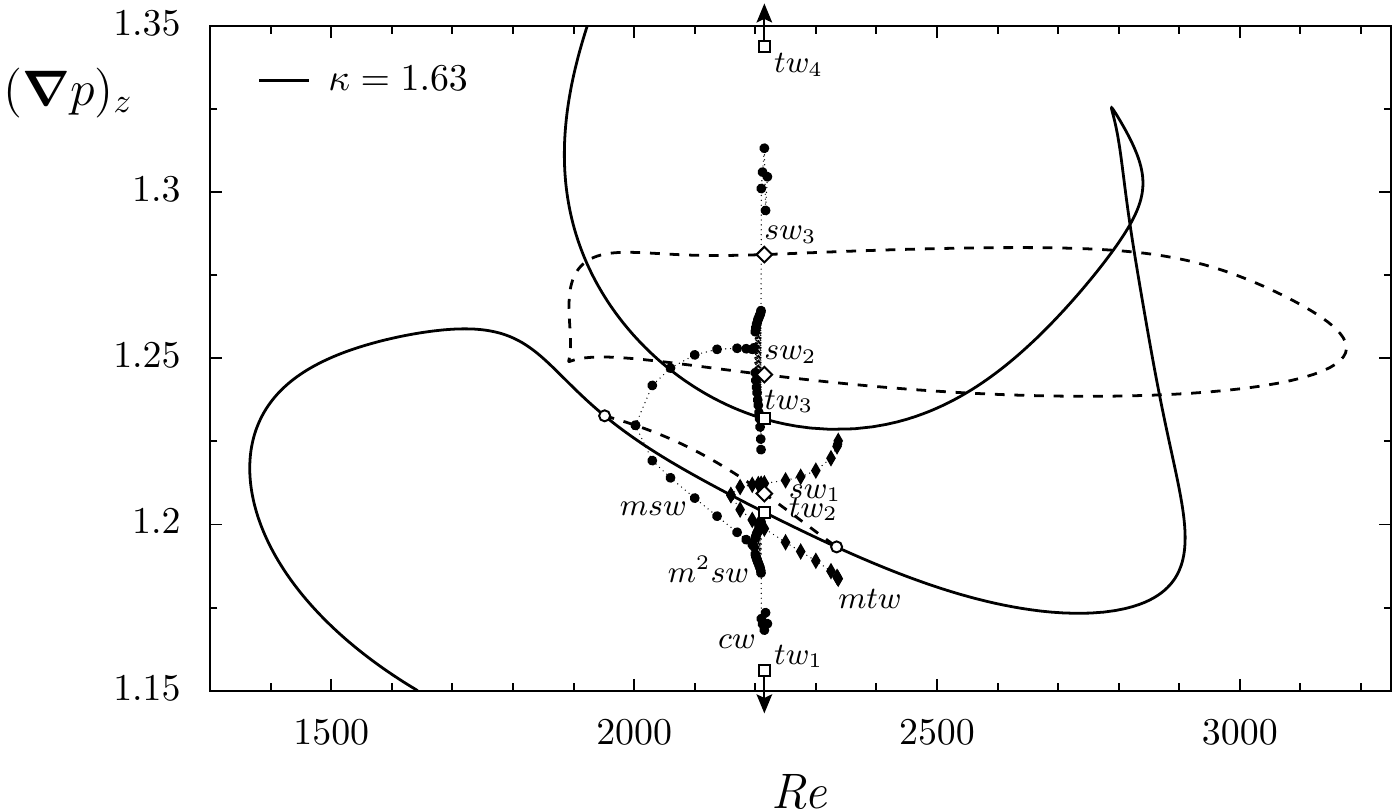}\\
      \raisebox{0.5\linewidth}{(\textit{b})}\hspace{-0.6cm} &
      \includegraphics[width=0.9\linewidth,clip]{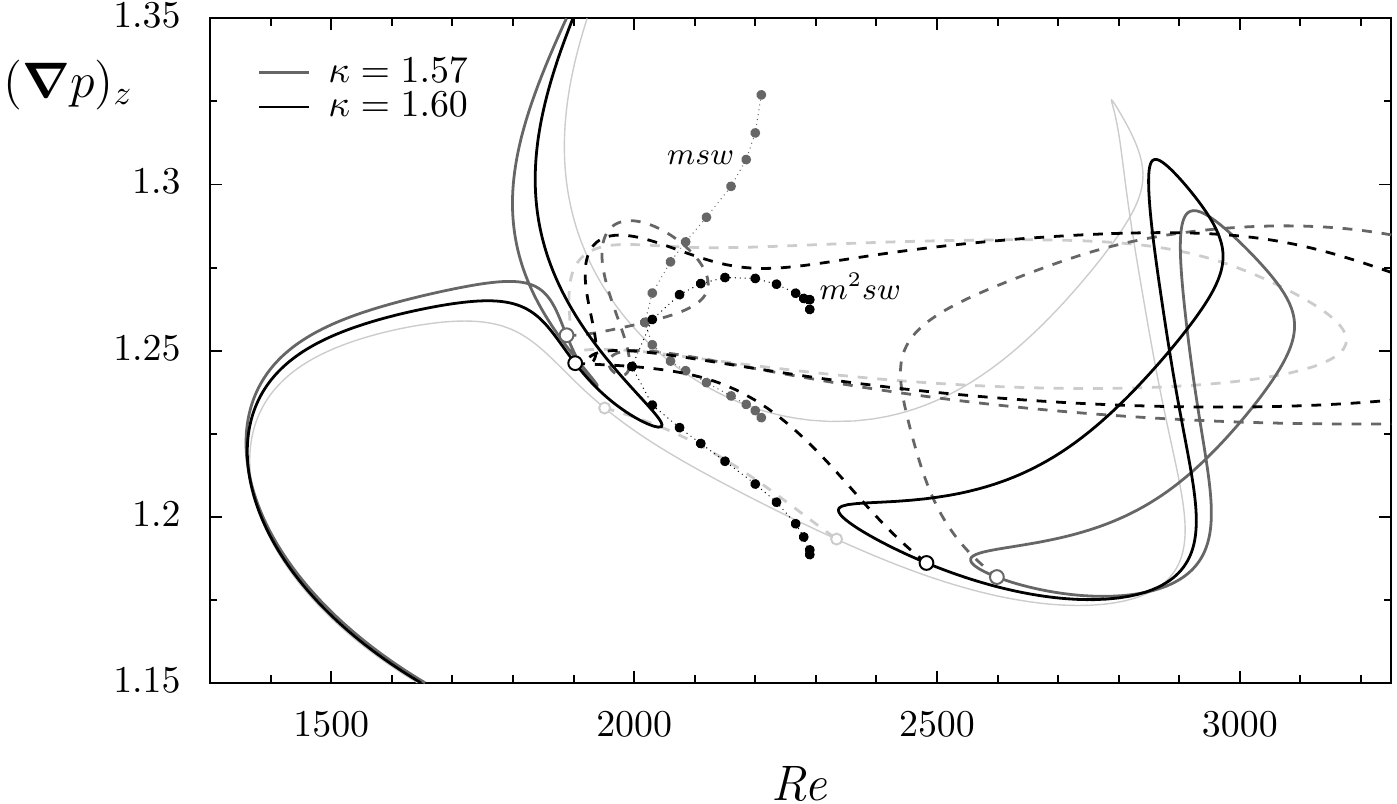}\\
    \end{tabular}
  \end{center}
  \caption{Mean axial pressure gradient ($(\bnabla p)_z$) as a
    function of $\Rey$ for (\textit{a}) $\kappa=1.63$ and (\textit{b})
    $\kappa=1.57,1.60$ (shading as explained in the legend). 
    Solid lines represent shift-reflect travelling waves, dashed for
    spiralling waves. Extrema of time-dependent travelling/spiralling
    solutions are marked as filled diamonds/circles, which denote Hopf
    bifurcations when placed on equilibria continuation curves. Open
    circles denote pitchfork bifurcations. Relative equilibria at
    $(\kappa,\Rey)=(1.63,2215)$ have been marked with open squares
    ($tw_s$, travelling waves from figure~\ref{fig:tws}) and diamonds
    ($sw_s$, spiralling waves from figure~\ref{fig:sws}). Modulated
    travelling, modulated spiralling, doubly-modulated spiralling and
    chaotic waves are labelled $mtw$, $msw$, $m^2sw$ and $cw$,
    respectively.}
  \label{fig:grpvsRe}
\end{figure}

The behaviour of upper-branch waves depends much more strongly on
$\kappa$. This is apparent from figure~\ref{fig:grpvsRe}, where
$(\bnabla p)_z$ has been plotted against $\Rey$ for three different
values of $\kappa$.
At high $\kappa$ (figure~\ref{fig:grpvsRe}\textit{a}), upper-branch
waves extend to moderate $\Rey\sim 2900$, then turn back in a
contorted fashion toward lower values of the parameter, undergo a
saddle-node bifurcation at about $\Rey\simeq 1885$ and finally
progress towards much higher $\Rey$.  As a result, four instances of
the same solution coexist over a fairly wide $\Rey$-range. They are
shown in figure~\ref{fig:tws}
for $(\kappa,\Rey)=(1.63,2215)$ and marked with open squares as
$tw_1$, $tw_2$, $tw_3$, and $tw_4$ in
figure~\ref{fig:grpvsRe}(\textit{a}).  As already mentioned, the wave
requiring the lowest driving axial pressure gradient ($tw_1$, on the
lower branch, out of scale in figure~\ref{fig:grpvsRe}\textit{a} for
clarity) lives on the laminar flow basin boundary. At the other end of
the continuation curve, the waves exhibit driving pressure gradients
in the region of turbulent flow values ($tw_4$,
figure~\ref{fig:tws}\textit{d}, on the upper branch, also out of range
in figure~\ref{fig:grpvsRe}\textit{a})).  Structures resembling these
and other upper-branch travelling waves have been observed
experimentally in developed turbulence \cite[][]{HVWNFEWKW_SCI_04} as
well as in numerical simulation
\cite*[][]{KeTu_JFM_07,ScEcVo_PRE_07,WiKe_PRL_08}.  The shift-reflect
symmetry is clearly preserved but vorticity and axial velocity
gradients in the vicinity of the wall are much more pronounced than
for the lower-branch waves, as expected for turbulent solutions.

Due to the contorted shape of the curve, two additional waves exist in a
region that extends up to $\Rey \simeq 2800$. They are
$tw_2$ and $tw_3$ of figures~\ref{fig:tws}\textit{b} and
\ref{fig:tws}\textit{c}, which we dub lower- and upper-middle-branch
waves, respectively, because of the relative values of their
pressure gradients. Both waves are also shift-reflect symmetric.
The main properties of all members of this travelling-waves family at
$(\kappa,\Rey)=(1.63,2215)$ are summarised in table~\ref{tab:tsws}.
\begin{table}
  \begin{center}

    \begin{tabular}{@{}cccccccccccc@{}}
      & $(\bnabla p)_z$ & $\nu$ & $c_z$ & $10^{3} \times c_{\theta}$ &
      $10^{4} \times \varepsilon_{\rm 2\theta}$ & $10^{2} \times
      \varepsilon_{\rm 2z}$ & $10^{4} \times \varepsilon_{\rm 3D}$ &
      $ev \in S $ & $ev \in \bar{S}$\\

      $tw_1$ & $1.1171$ & $0.03228$ & $1.5481$ & $0.0$ & $0.3940$ &
      $1.0508$ & $4.0964$ & $1r$ & \\

      $tw_2$ & $1.2037$ & $0.03478$ & $1.4600$ & $0.0$ & $1.0010$ &
      $1.3027$ & $7.5781$ & $1c$ & $1r$\\

      $tw_3$ & $1.2318$ & $0.03559$ & $1.4349$ & $0.0$ & $1.3086$ &
      $1.3051$ & $8.8704$ & $1r$ & $1r$\\

      $tw_4$ & $1.4738$ & $0.04258$ & $1.2891$ & $0.0$ & $4.3161$ &
      $1.4333$ & $19.280$ & $1c$ & $1r+2c$\\

      $sw_1$ & $1.2093$ & $0.03494$ & $1.4583$ & $\pm 0.9192$ &
      $1.0331$ & $1.3632$ & $7.6261$ & & $1c$\\

      $sw_2$ & $1.2813$ & $0.03702$ & $1.4413$ & $\pm 1.5964$ &
      $1.6748$ & $1.9689$ & $8.4999$ & & $1r+1c$\\

      $sw_3$ & $1.2451$ & $0.03597$ & $1.5015$ & $\pm 1.9248$ &
      $1.7667$ & $2.4578$ & $5.3584$ & & $1c$\\

    \end{tabular}
  \end{center}
  \caption{Relative equilibria of the $2$-fold azimuthally periodic
    subspace at $(\kappa,\Rey)=(1.63,2215)$. The last two columns,
    $ev \in S $ and $ev \in \bar{S}$, count real ($r$) and complex
    pairs ($c$) of eigenmodes corresponding to the shift-reflect and
    shift-reflect-orthogonal subspaces, respectively. Tabulated
    travelling/spiralling waves are marked with open squares/diamonds
    in figure~\ref{fig:grpvsRe}(\textit{a}).}
  \label{tab:tsws}
\end{table}

\begin{figure}
  \begin{center}
    \begin{tabular}{cccc}
      \raisebox{0.31\linewidth}{(\textit{a})}\hspace{-0.6cm} &
      \includegraphics[height=0.3\linewidth,clip]{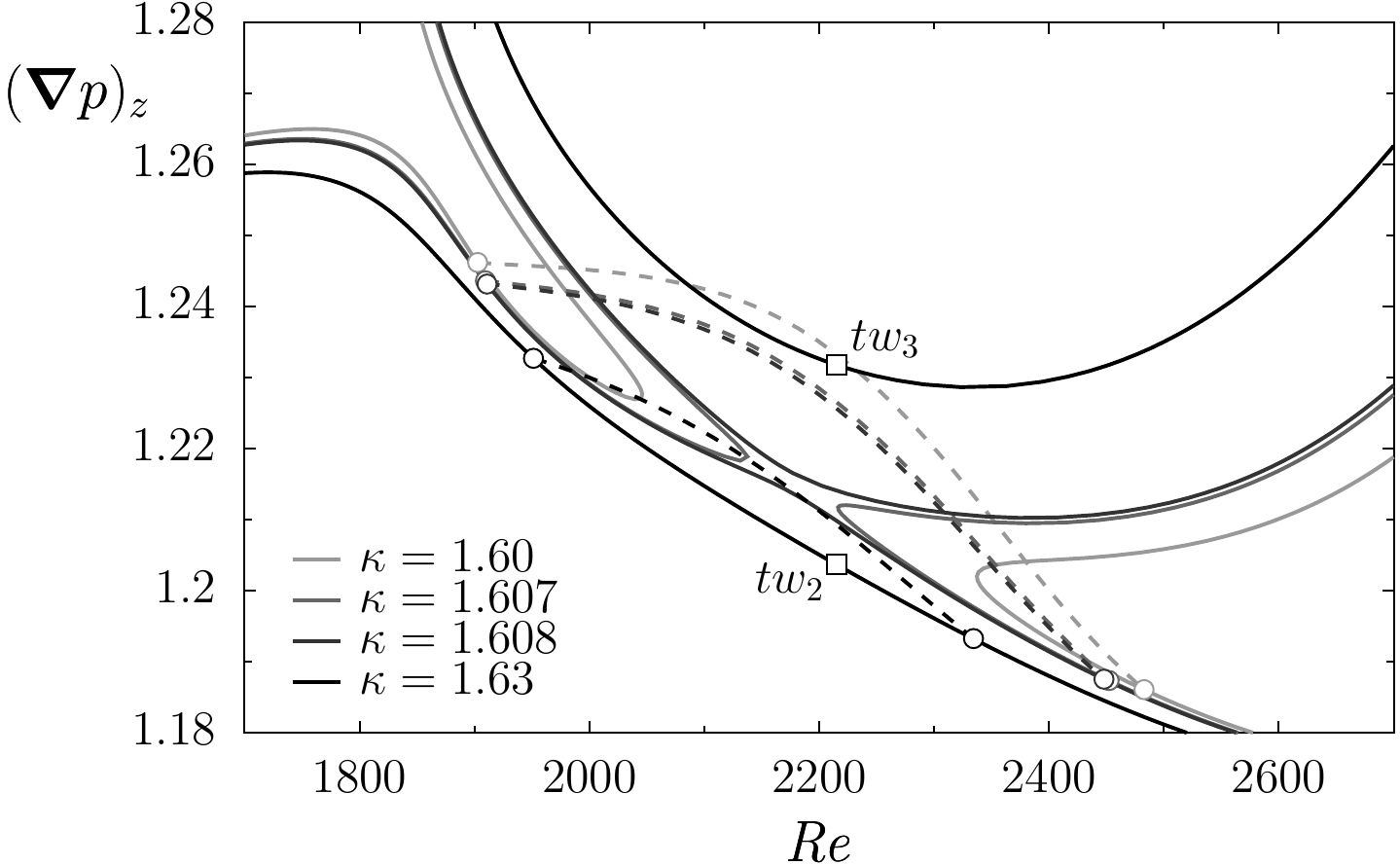} &
      \raisebox{0.31\linewidth}{(\textit{b})}\hspace{-0.6cm} &
      \includegraphics[height=0.3\linewidth,clip]{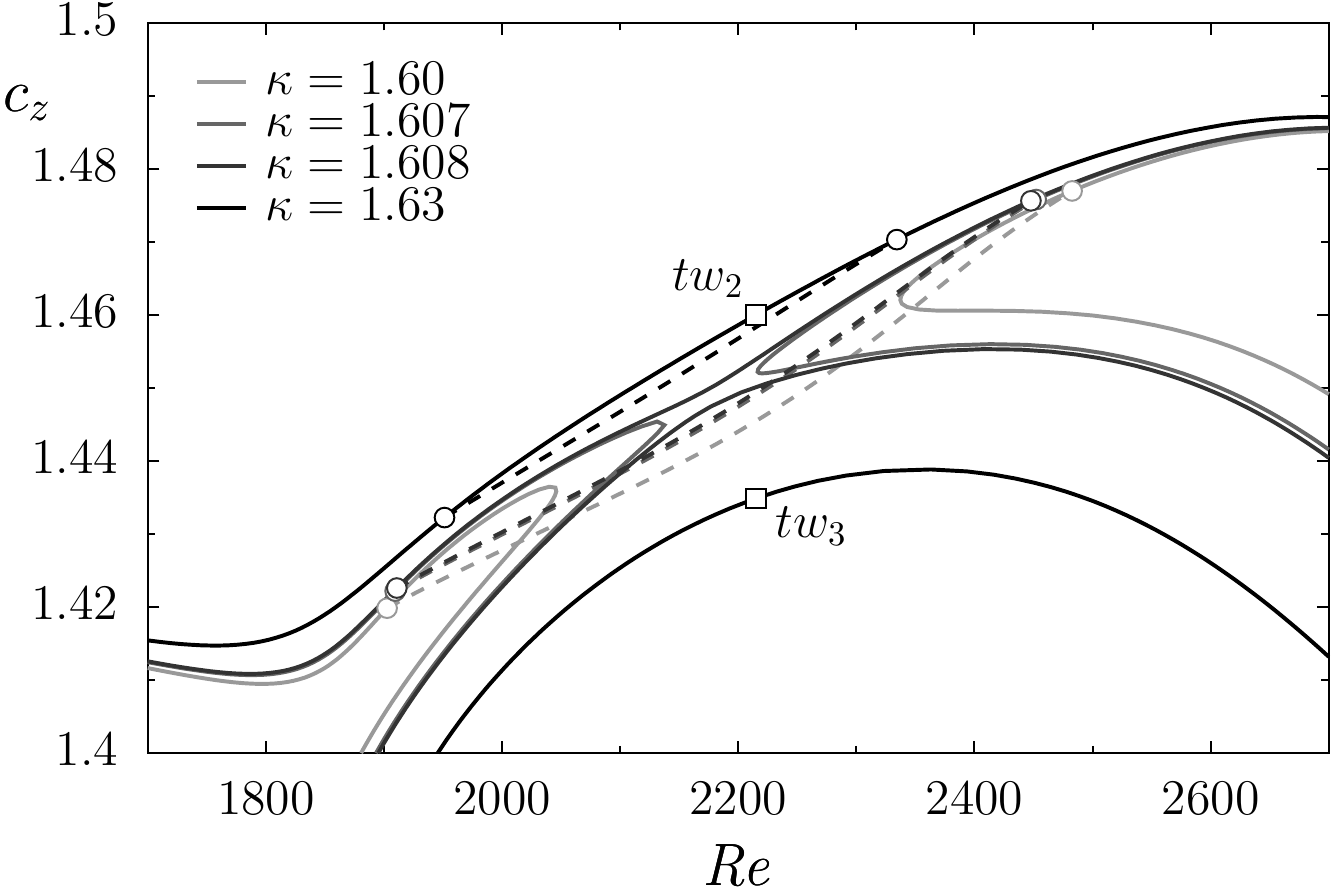}
    \end{tabular}
  \end{center}
  \caption{Detail near $(\kappa,\Rey)\sim(1.61,2200)$ highlighting the
    change in behaviour of the continuation curves. Solid lines
    represent travelling waves (darker for increasing $\kappa$)
    whereas dashed lines correspond to spiralling waves. Lower-middle
    branch ($tw_2$) and upper-middle branch ($tw_3$) waves are
    indicated with squares. Open circles denote pitchfork
    bifurcations. (\textit{a}) Mean axial pressure gradient $(\bnabla
    p)_z$ and (\textit{b}) axial advection speed ($c_z$), as a
    function of $\Rey$.}
  \label{fig:grpczvsRe_po}
\end{figure}

For lower $\kappa$ (figure~\ref{fig:grpvsRe}\textit{b}), the
continuation curve splits into two, adding two extra saddle-node points
and leaving an island of solutions at moderate
$\Rey$. Figure~\ref{fig:grpczvsRe_po} shows a close-up of the
parameter region where this topological change takes place.
It becomes clear in figure~\ref{fig:grpczvsRe_po}(\textit{a}) that for
$\kappa \lesssim 1.607$, the segments of the continuation curve
corresponding to $tw_2$ and $tw_3$ collide and produce a gap in $\Rey$
where the multiplicity of solutions is reduced to just two. Waves
$tw_2$ and $tw_3$ at $(\kappa,\Rey)=(1.63,2215)$
(figures~\ref{fig:tws}\textit{b},\textit{c}) already look much alike
and, indeed, collide in a saddle-node point and disappear as $\kappa$ is
reduced, leaving $tw_1$ and $tw_4$
(figures~\ref{fig:tws}\textit{a},\textit{d}) as the only surviving
members of the family.

\subsection{Spiralling waves}

A pair of symmetry-conjugate branches of spiralling waves branches 
off the lower-middle-branch of shift-reflect waves in a
symmetry-breaking pitchfork bifurcation (dashed lines in
figures~\ref{fig:grpvsRe} and
\ref{fig:grpczvsRe_po}\textit{a},\textit{b}). As a result of the
loss of symmetry they start precessing in the azimuthal direction, but
remain stationary in an appropriately spiralling frame of reference.

\begin{figure}
  \begin{center}
    \begin{tabular}{cccc}
      \raisebox{0.33\linewidth}{(\textit{a})}\hspace{-0.6cm} &
      \includegraphics[height=0.32\linewidth,clip]{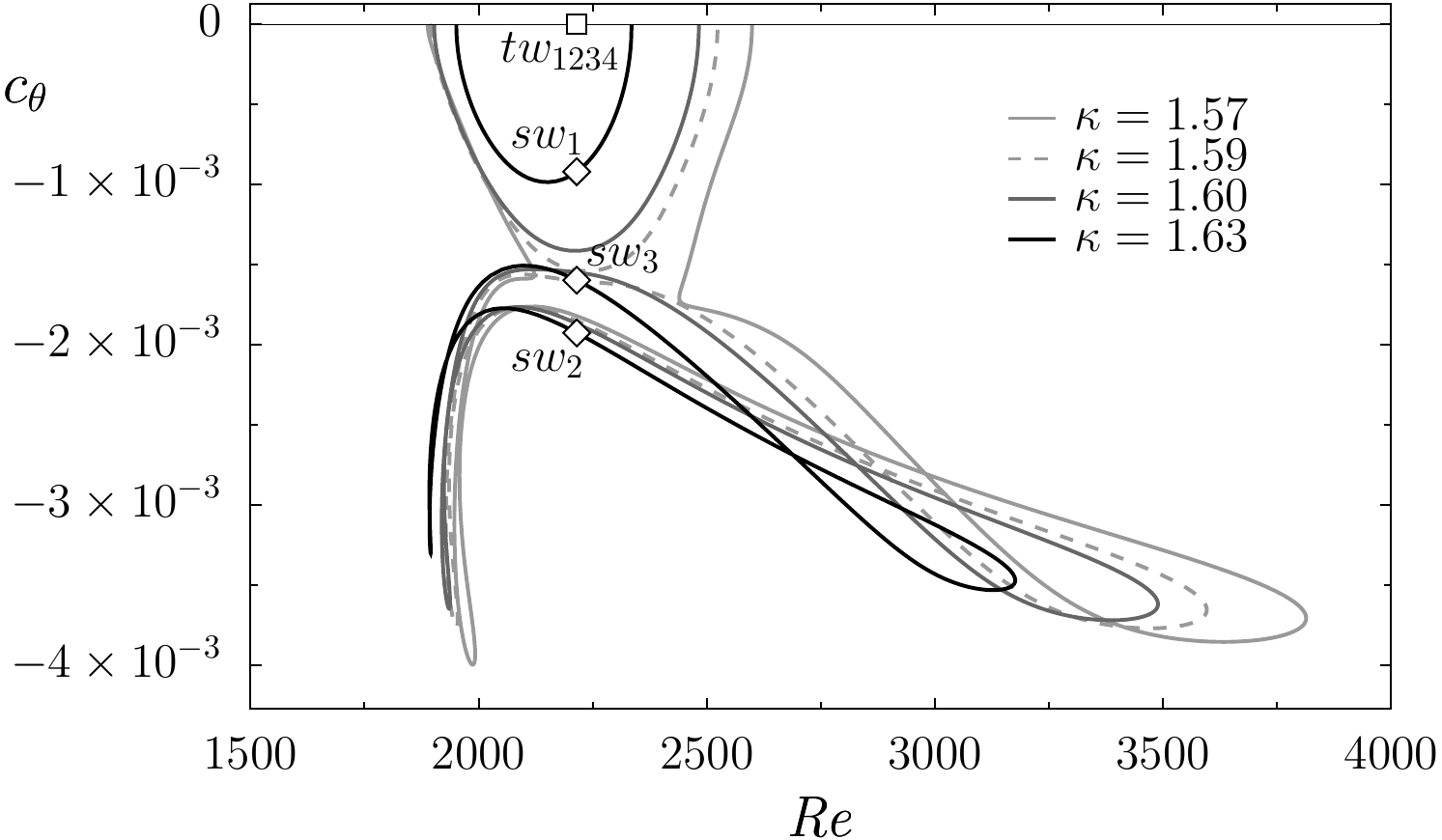} &
      \raisebox{0.33\linewidth}{(\textit{b})}\hspace{-0.6cm} &
      \includegraphics[height=0.32\linewidth,clip]{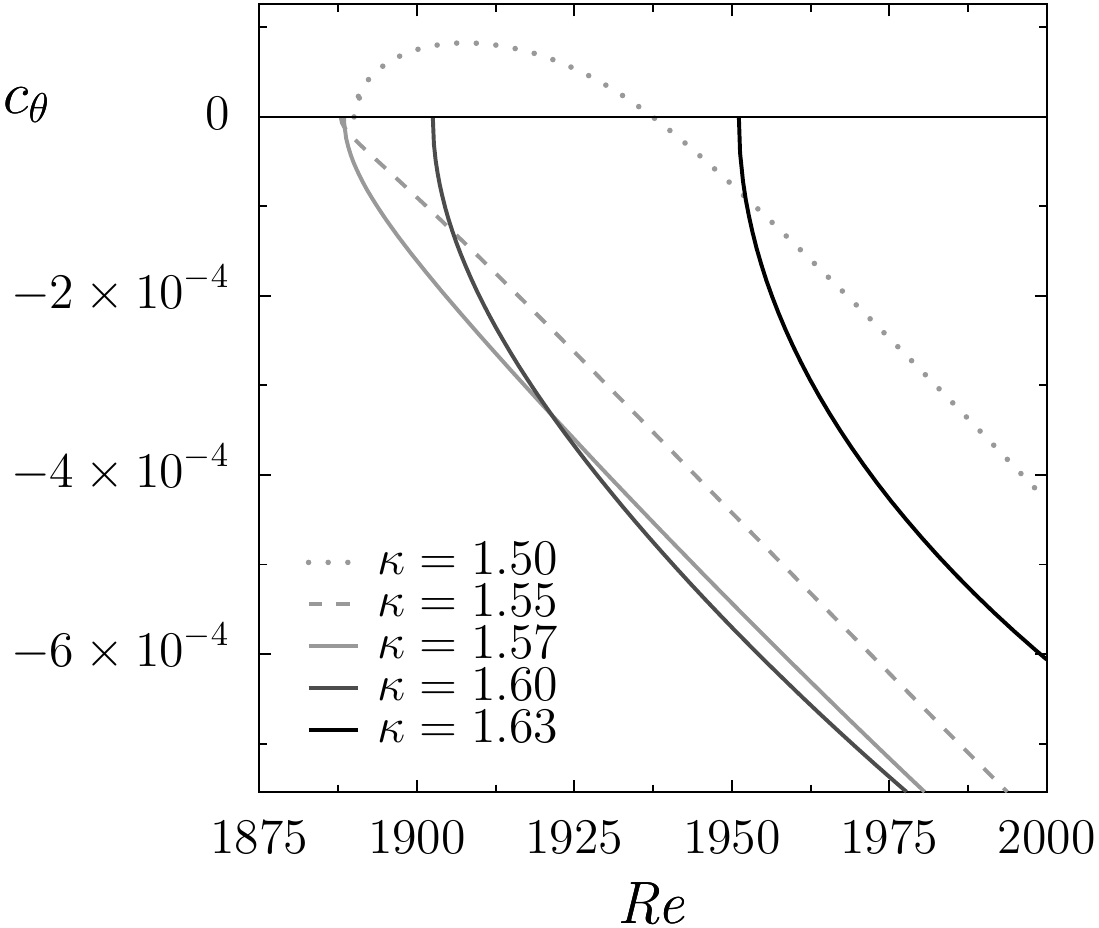}
    \end{tabular}
  \end{center}
  \caption{(\textit{a}) Rotation phase speed ($c_{\theta}$) as a
    function of $\Rey$ and $\kappa$ (shading) for spiralling
    waves. Shift-reflect travelling waves collapse on the
    $x$-axis. Relative equilibria at $(\kappa,\Rey)=(1.63,2215)$ have
    been marked with open squares ($tw_s$, travelling waves from
    figure~\ref{fig:tws}) and diamonds ($sw_s$, spiralling waves from
    figure~\ref{fig:sws}). (\textit{b}) Detail of the bifurcating
    point including lower $\kappa$-curves.}
  \label{fig:ctvsRe}
\end{figure}

For the lower $\kappa$ shown (light gray), the $\Rey$-continuation
curve starts at a pitchfork point, exhibits a counterclockwise loop to
higher values of the pressure gradient and then a clockwise loop to
lower pressure gradients before continuing to higher $\Rey$. The
branch extends beyond the region of existence of middle-branch
travelling waves, where it exhibits an additional fold (outside the
range of the figure) that takes them back in $\Rey$. At
this point the curve bends back to end up branching back with
middle-branch travelling waves in a second pitchfork. In this way, the
spiralling waves bridge the $\Rey$-gap left by middle-branch
travelling waves described above. At the two pitchfork ends of the
continuation curve, spiralling waves bifurcate with no rotation
speed. As the asymmetry increases, rotation speed builds up reaching
significant precessing rates, as shown in figure~\ref{fig:ctvsRe}.
The waves continued have negative rotation speed, but, obviously,
their shift-reflected counterparts exhibit opposite azimuthal
rotation. The rotation drift of the waves is much slower than their
axial drift. As a matter of fact, the spiralling waves reported here 
travel a minimum of $700$ wavelengths or $1350$ diameters
in the time they complete a full rotation.

\begin{figure}
  \begin{center}
    \begin{tabular}{cccccc}
      \raisebox{0.16\linewidth}{(\textit{a})}\hspace{-0.6cm} &
      \includegraphics[height=0.15\linewidth,clip]{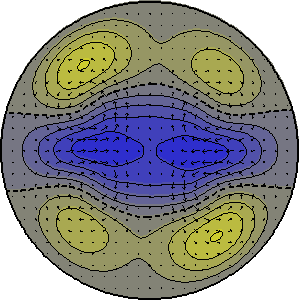} &
      \includegraphics[height=0.15\linewidth,clip]{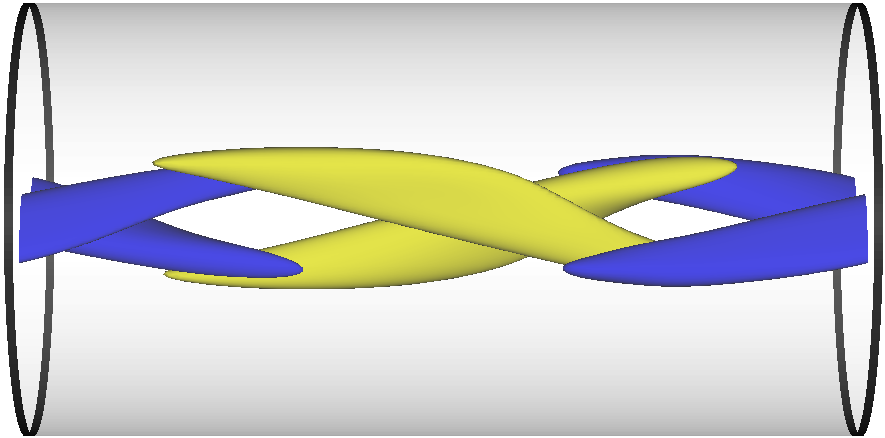} & & &
      \\ \raisebox{0.16\linewidth}{(\textit{b})}\hspace{-0.6cm} &
      \includegraphics[height=0.15\linewidth,clip]{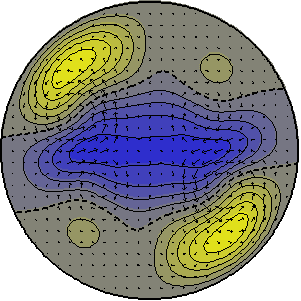} &
      \includegraphics[height=0.15\linewidth,clip]{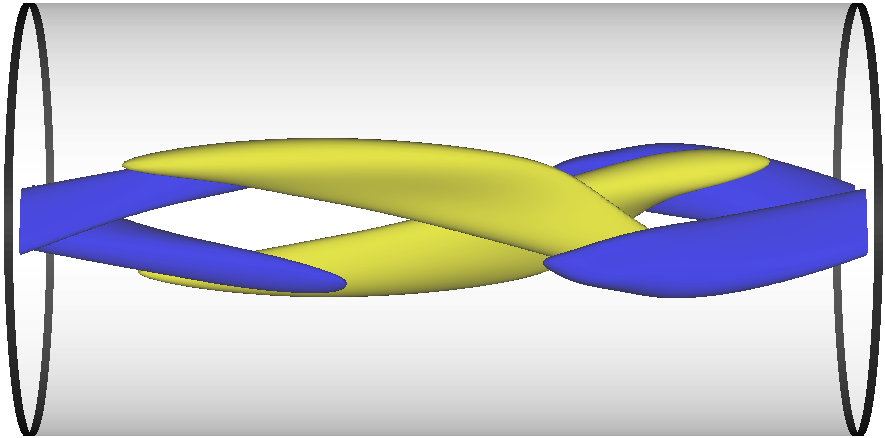} &
      \raisebox{0.16\linewidth}{(\textit{c})}\hspace{-0.6cm} &
      \includegraphics[height=0.15\linewidth,clip]{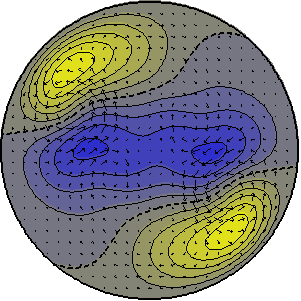} &
      \includegraphics[height=0.15\linewidth,clip]{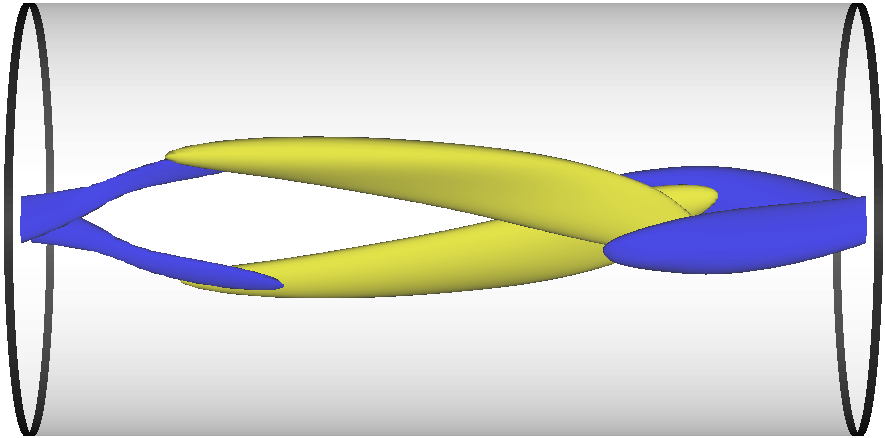}\\
    \end{tabular}
  \end{center}
  \caption{Spiralling waves at $(\kappa,\Rey)=(1.63,2215)$. Left:
    $z$-averaged cross-sectional axial velocity contours spaced at
    intervals of $\Delta \langle u_z\rangle_z = 0.1 U$. In-plane
    velocity vectors are also displayed. Right: axial vorticity
    isosurfaces at $\omega_z=\pm 1 U/D$. Fluid flows
    rightwards. Blue (dark gray) for negative, yellow (light) for
    positive. (\textit{a}) Lower-branch ($sw_1$), (\textit{b})
    middle-branch ($sw_2$) and (\textit{c}) upper-branch ($sw_3$)
    spiralling waves.}
  \label{fig:sws}
\end{figure}

For $\kappa \gtrsim 1.59$ (dashed curve in
figure~\ref{fig:ctvsRe}\textit{a}) the continuation curve splits in
two, leaving a disconnected island of strongly spiralling waves and a
monotonous curve connecting at both ends with the travelling-wave
family. At $(\kappa,\Rey)=(1.63,2215)$, three different spiralling
waves coexist: lower-, middle- and upper-branch waves, indicated by
open diamonds and named $sw_1$, $sw_2$ and $sw_3$. They are shown in
figure~\ref{fig:sws} and they all clearly break the shift-reflect
symmetry.
The spiralling waves most strongly related to $tw_2$ is $sw_1$. They
are close in parameter space (their continuation curves are connected)
and they also look very much alike. Cross-sectional contours are
strikingly similar, except that $sw_1$ has slightly broken the
shift-reflect symmetry. The other two spiralling waves, $sw_2$ and
$sw_3$, have completely broken the symmetry and their rotation rates
are much larger than for $sw_1$. The main properties of these
spiralling waves at $(\kappa,\Rey)=(1.63,2215)$ are summarised
alongside those of travelling waves in table~\ref{tab:tsws}.

It has already been pointed out that spiralling waves that break a
left-right symmetry, here in the form of a shift-reflect symmetry,
appear in symmetry-related pairs that rotate either clockwise or
counterclockwise with opposite rotation speeds. This is a general
remark, but the loss of symmetry does not preclude the existence of
non-robust asymmetric waves that balance the azimuthal driving force
and, consequently, do not rotate. As is the case for some helical
waves in pipe flow \cite[][]{PrKe_PRL_07}, non-shift-reflect waves
exist whose rotation rate cancels exactly. This is exemplified for
$\kappa=1.50$ (dotted line) in figure~\ref{fig:ctvsRe}(\textit{b}),
where a branch of spiralling waves first appears with counterclockwise
rotation that only for higher Reynolds numbers switches to the
clockwise rotation that characterises all other spiralling waves
continued in the plot. As the rotation speed crosses zero, two
mutually-symmetric spiralling waves switch the sense of
rotation. Since this crossing happens for a specific value of the
system parameters it is a non-robust, non-generic phenomenon.

\subsection{Stability of travelling and spiralling waves}

The Takens-Bogdanov bifurcation in which the travelling waves 
appear at the lower $\Rey$-range was studied in \cite{MeEc_JFM_10}. 
Lower-branch waves
were shown to always exhibit a single unstable direction, while
sufficiently far away from the bifurcation point, upper-branch waves
(here renamed lower-middle-branch for convenience) were shown to be
stable to all $2$-fold azimuthally-periodic perturbations.

The stability properties of travelling waves along the continuation
surface is quite involved, with many eigenvalues crossing back and
forth, and many states related by the multiple folds present.  A
detailed study of these bifurcations is beyond the scope of the
present analysis. Instead, we will concentrate on parameter regions where
interesting dynamics occur and on stable waves that
transitions to more complex flows.  Thus, we will initially keep
the wavenumber at $\kappa=1.63$ and search for bifurcations of the
lower-middle-branch travelling wave ($tw_2$) and the lower-branch
spiralling wave ($sw_1$) with Reynolds numbers in the range $\Rey \in
[1800,2600]$. Stability of other waves will be reported for
completeness, but their transitions cannot be analysed in detail since
the states resulting are all unstable at onset and not accessible
through time evolution.  The stability analysis will then be broadened
by varying $\kappa$ and $\Rey$ together in order to extend the
bifurcation points into bifurcation curves in
$(\Rey,\kappa)$-parameter space.

\begin{figure}
  \begin{center}
    \begin{tabular}{cccc}
      \raisebox{0.3\linewidth}{(\textit{a})}\hspace{-0.6cm} &
      \includegraphics[height=0.29\linewidth,clip]{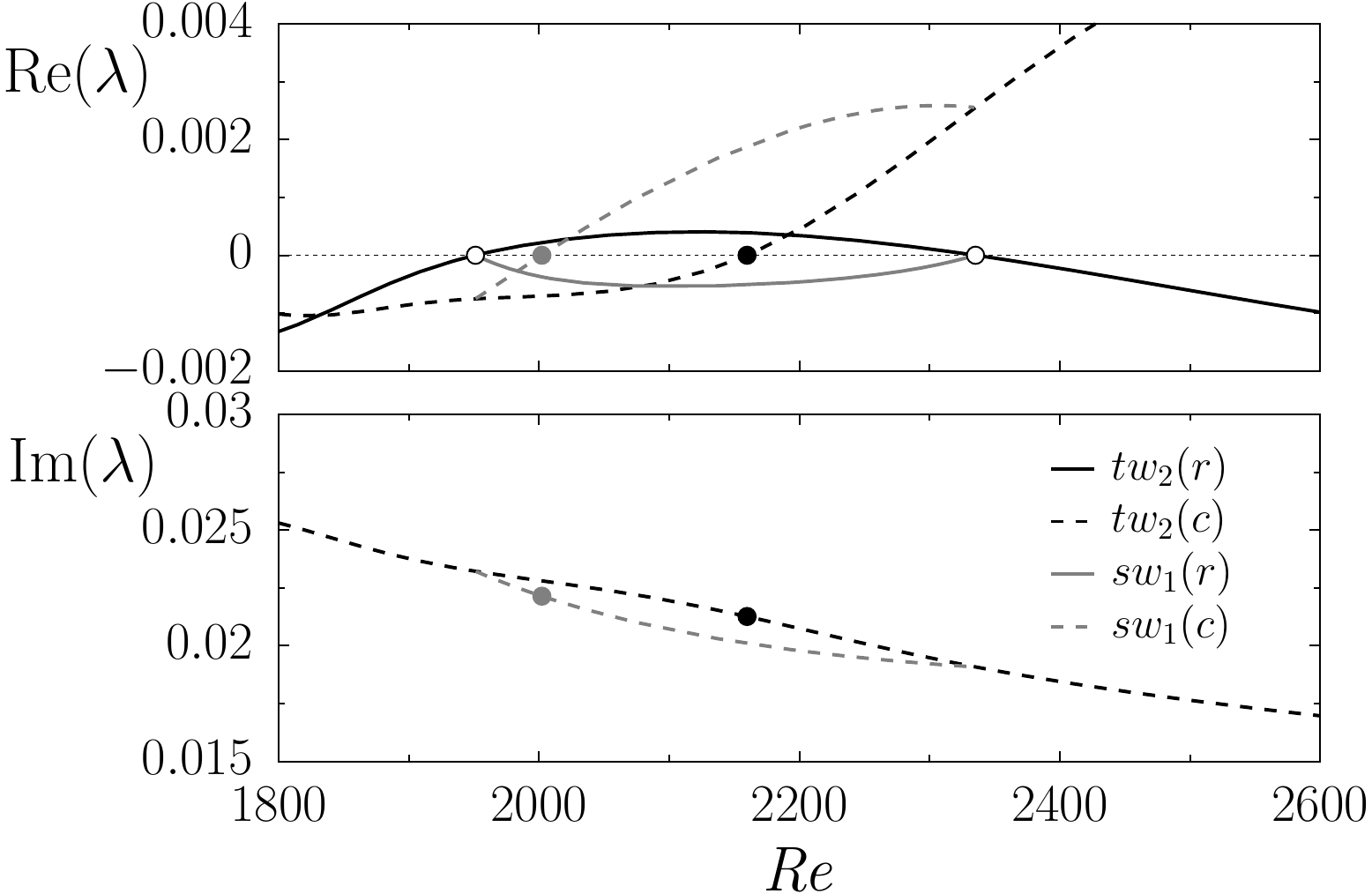} &
      \raisebox{0.3\linewidth}{(\textit{b})}\hspace{-0.6cm} &
      \includegraphics[height=0.29\linewidth,clip]{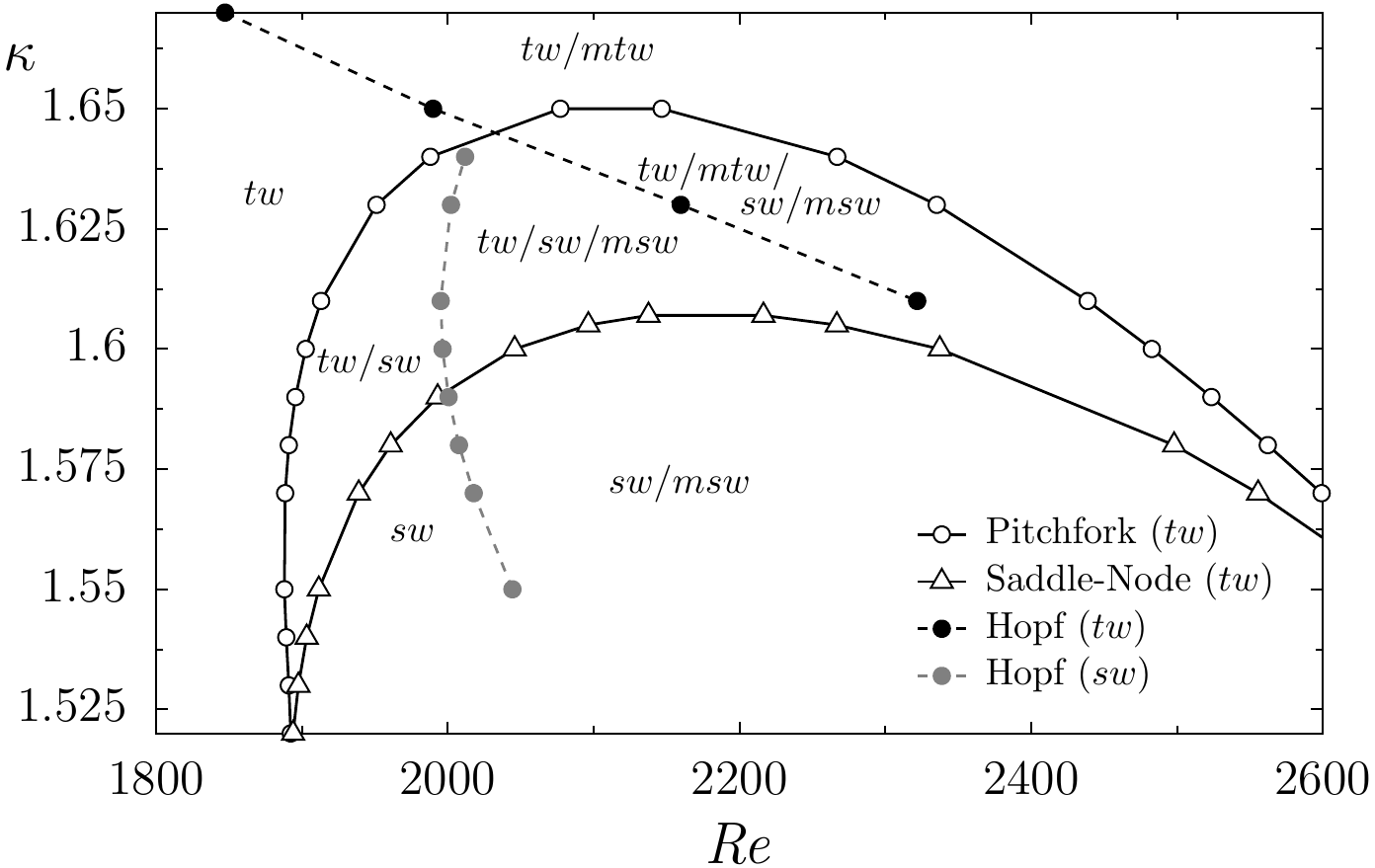}
    \end{tabular}
  \end{center}
  \caption{(\textit{a}) Real (solid) and complex (dashed) relevant
    eigenvalues of $tw_2$ (black) and $sw_1$ (gray) along their
    $\kappa=1.63$-continuation curves. Open and filled circles
    correspond to pitchfork and Hopf bifurcation points,
    respectively. (\textit{b}) Bifurcation curves on
    $(\Rey,\kappa)$-parameter space as explained in the
    legend. Regions of existence of travelling, spiralling,
    modulated-travelling and modulated-spiralling waves are labelled
    $tw$, $sw$, $mtw$ and $msw$, respectively.}
  \label{fig:stan}
\end{figure}

Figure~\ref{fig:stan}(\textit{a}) tracks the relevant (bifurcating)
eigenvalues of $tw_2$ and $sw_1$ for $\kappa=1.63$ and varying $\Rey$.
A real eigenvalue of $tw_2$ (black solid line) undergoes two
zero-crossings (open circles) at $\Rey=1951.1$ and $\Rey=2335.2$,
corresponding to the creation and destruction of a spiralling-wave
($sw_1$) branch. 
Both pitchfork bifurcations are supercritical and the spiralling
waves created have one less unstable direction than the travelling
waves they originate from. Inbetween the two pitchfork points, at
about $\Rey=2159.8$, a complex pair (black dashed line) crosses the
imaginary axis at a Hopf point (black circle), and a branch of 
modulated travelling waves ($mtw$) appears. The
initial modulational frequency of these waves is given by the
imaginary part of the eigenvalue at the crossing. Here,
$\omega=0.02127\;  (4U/D)$, corresponding to a period
$T=2\upi/\omega=295.5 D/(4U)$. Index theory dictates that, due to the
travelling waves stability properties just described, spiralling waves
must be stable at the low-$\Rey$ end and Hopf unstable at the
high-$\Rey$ end.


The complex pair duplicated on the
spiralling branch (gray dashed line) takes the lead from its
travelling counterpart and undergoes a Hopf bifurcation (gray circle)
somewhat earlier in $\Rey$, 
generating a branch of modulated
spiralling waves ($msw$). These waves bifurcate at $\Rey=2002.1$ 
with $\omega=0.02214 \;  (4U/D)$, corresponding to a period $T=283.8
D/(4U)$, slightly shorter than for $mtw$.
Both Hopf bifurcations are supercritical and the emerging 
modulated waves inherit the stability of the original waves.

Identical analyses have been carried out for several other
values of $\kappa$ in order to explore the topology of parameter space. 
Tracks of the bifurcation points from figure~\ref{fig:stan}(\textit{a}) 
in the $\kappa$-$\Rey$ parameter space are shown 
in figure~\ref{fig:stan}(\textit{b}). A saddle-node
bifurcation curve (solid line with triangles), only present for
$\kappa \lesssim 1.607$, delimits a region where no middle-branch
travelling waves exist, as described above and illustrated in
figure~\ref{fig:grpczvsRe_po}. A pitchfork bifurcation curve (solid
line with open circles) 
delimits the region of existence of lower-branch spiralling waves
($sw_1$). Middle- and upper-branch spiralling waves can extend beyond
this region and actually do so as shown in figure~\ref{fig:ctvsRe}. In
the region where travelling waves exist, modulated travelling waves
that preserve the shift-reflect symmetry appear along a Hopf line
(black dashed line with filled circles).  As the travelling-waves
Hopf-line crosses into the region of existence of spiralling waves
another Hopf line (gray dashed line with filled circles), duplicated
from the original, but related to the destabilisation of
spiralling waves, starts running on their continuation surface.
Along this line, modulated spiralling waves are created.  All these
modulated waves acquire the stability properties of the original
waves so that both Hopf lines are supercritical. Since the waves go
unstable when $\Rey$ is increased, the region of existence of
modulated waves always lies to the right of their respective
bifurcation curves.

\begin{figure}
  \begin{center}
    \begin{tabular}{cccc}
      \raisebox{0.33\linewidth}{(\textit{a})}\hspace{-0.6cm} &
      \includegraphics[height=0.32\linewidth,clip]{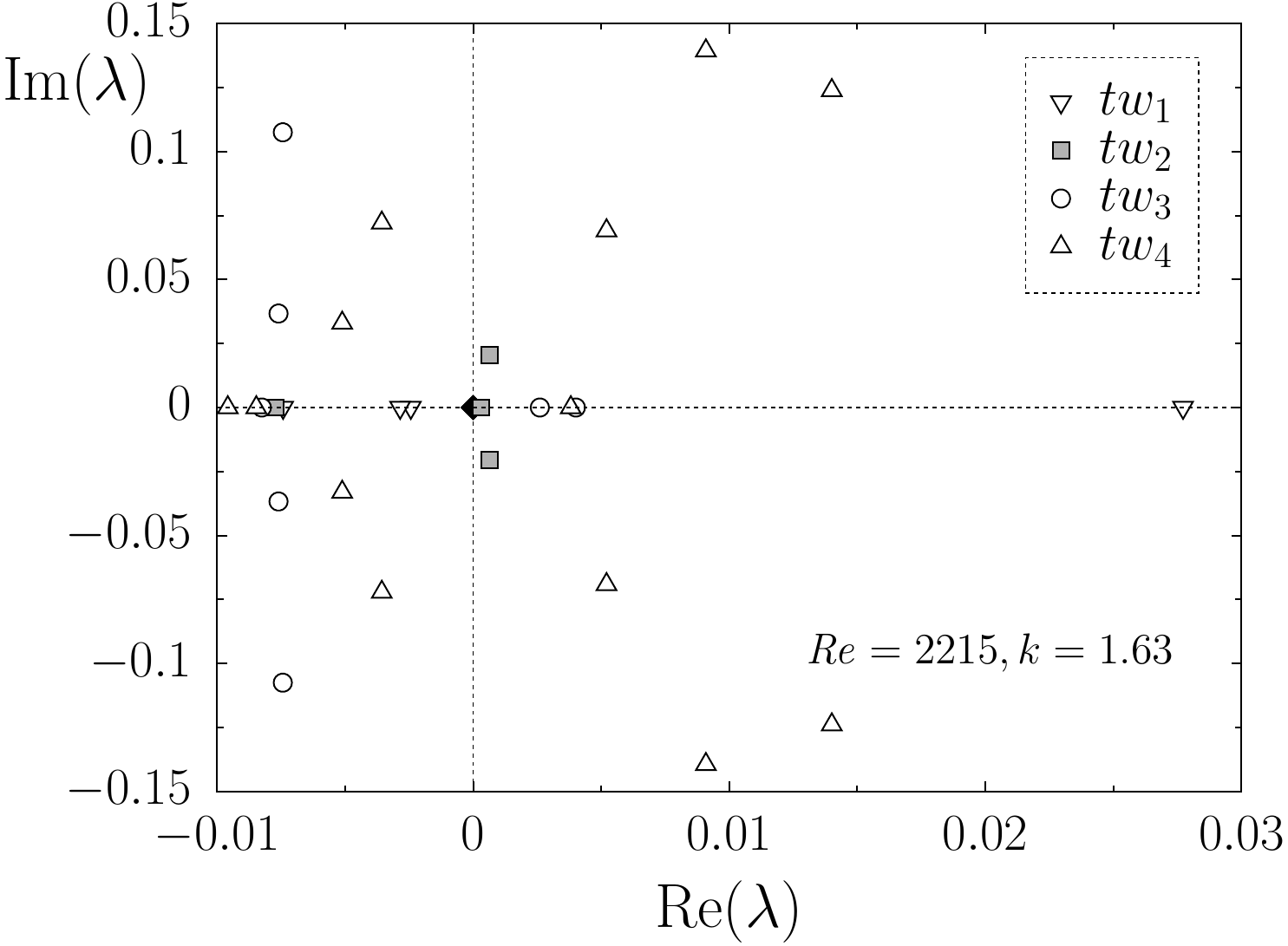} &
      \raisebox{0.33\linewidth}{(\textit{b})}\hspace{-0.6cm} &
      \includegraphics[height=0.32\linewidth,clip]{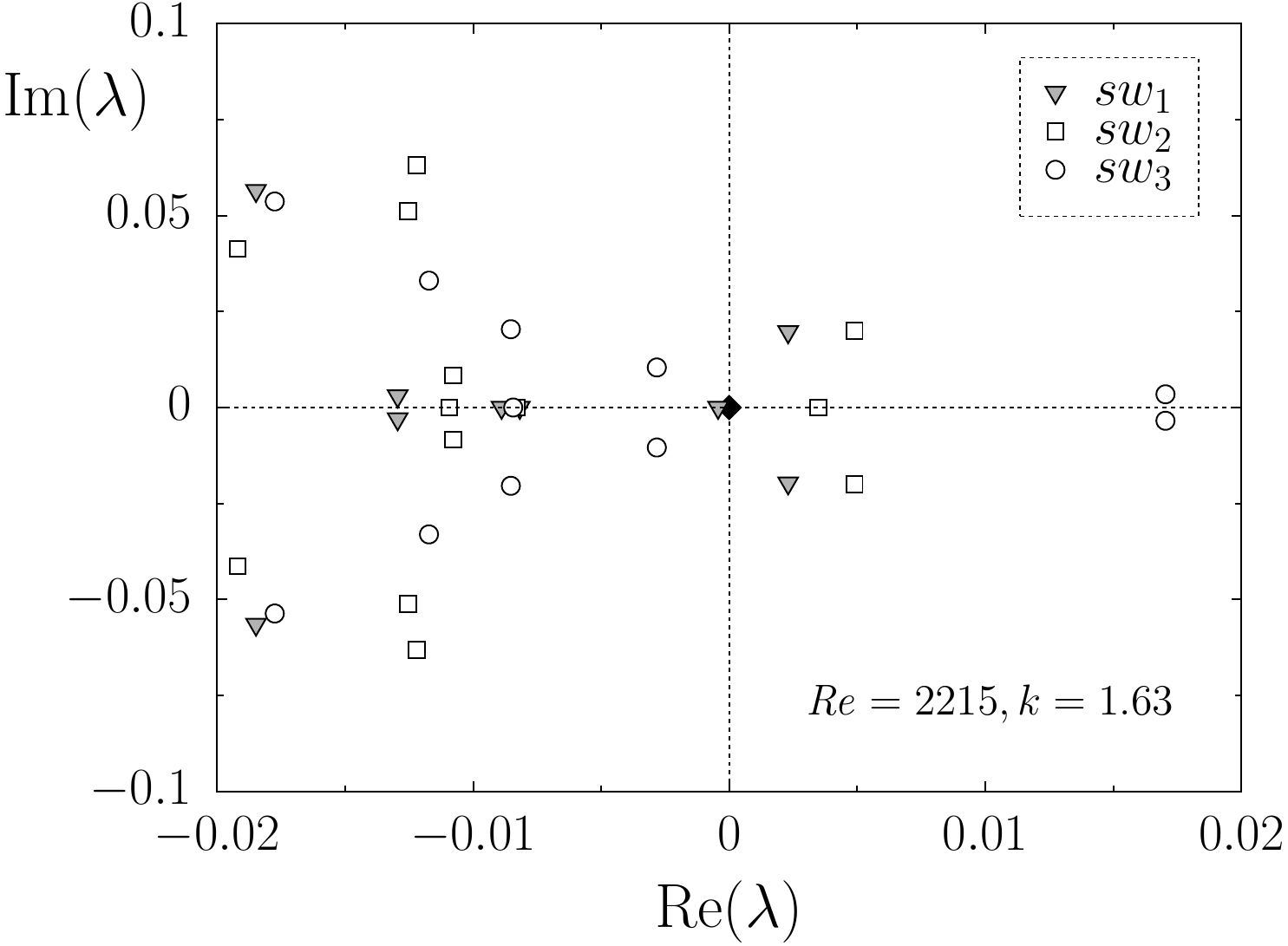}
    \end{tabular}
  \end{center}
  \caption{Spectra of (\textit{a}) travelling and (\textit{b})
    spiralling waves at $(\Rey,\kappa)=(2215,1.63)$ as indicated in
    figure~\ref{fig:grpvsRe}(\textit{a}). Eigenvalues are coded as
    explained in the legend. The black diamond at the origin represents
    degenerate eigenmodes corresponding to infinitesimal translation
    and rotation of the waves.}
  \label{fig:SAk1.63Re2215}
\end{figure}
\begin{figure}
  \begin{center}
    \begin{tabular}{cccccc}
      \raisebox{0.16\linewidth}{(\textit{a})}\hspace{-0.6cm} &
      \includegraphics[height=0.15\linewidth,clip]{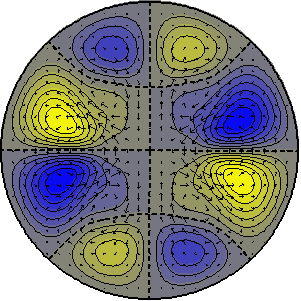} &
      \includegraphics[height=0.15\linewidth,clip]{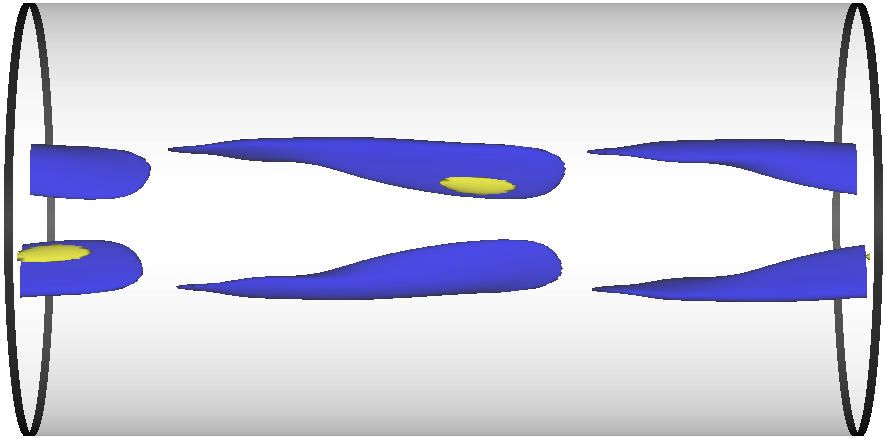}
      & \raisebox{0.16\linewidth}{(\textit{b})}\hspace{-0.6cm} &
      \includegraphics[height=0.15\linewidth,clip]{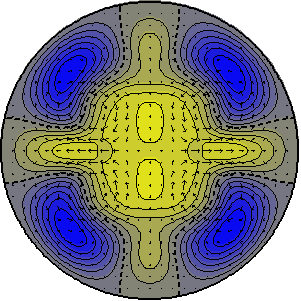} &
      \includegraphics[height=0.15\linewidth,clip]{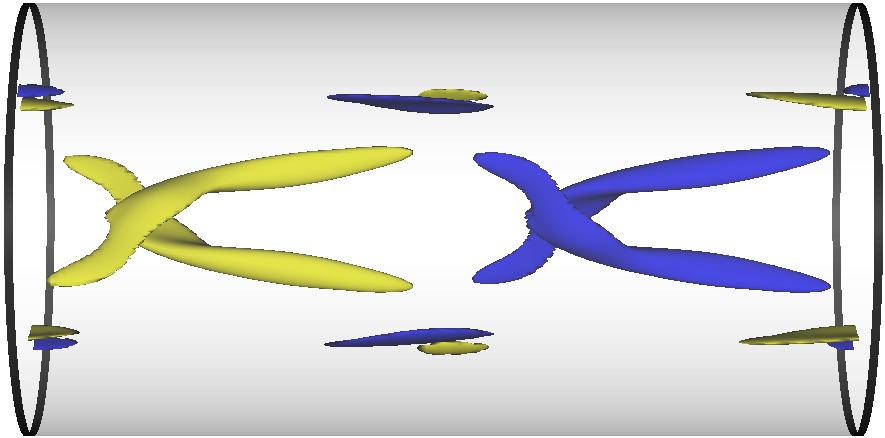}\\
      &
      &
      &
      &
      \includegraphics[height=0.15\linewidth,clip]{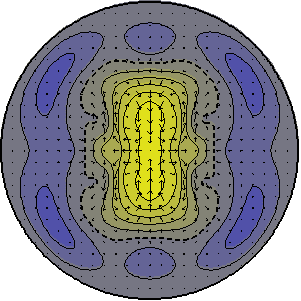} &
      \includegraphics[height=0.15\linewidth,clip]{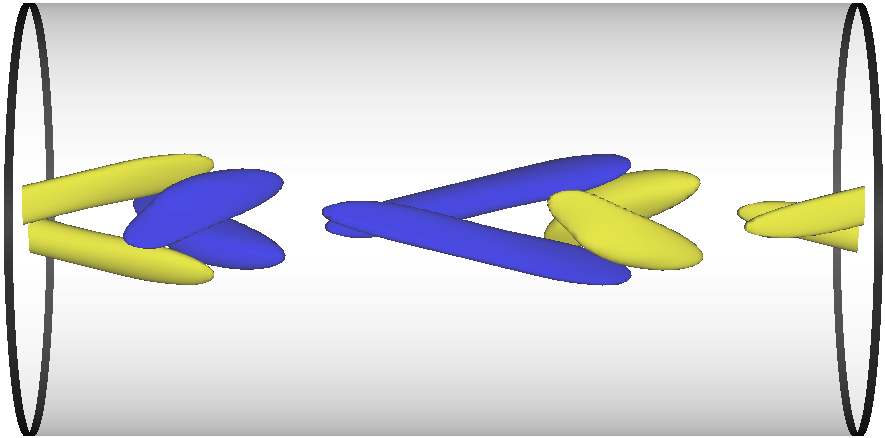}\\
    \end{tabular}
  \end{center}
  \caption{Unstable eigenmodes of $tw_2$ (see
    figure~\ref{fig:SAk1.63Re2215}\textit{a}) at
    $(\kappa,\Rey)=(1.63,2215)$. Shown are averaged cross-sectional
    axial velocity contours with in-plane velocity vectors and
    three-dimensional isosurfaces of axial vorticity (blue -dark- for
    negative, yellow -light- for positive). (\textit{a})
    Symmetry-breaking real eigenmode. (\textit{b}) Real (top) and
    imaginary (bottom) parts of the symmetry-preserving unstable
    complex pair. Axial vorticity has been scaled ($\omega_z(\Imag)=5
    \omega_z(\Real)$) to aid visualisation.}
  \label{fig:twev}
\end{figure}
\begin{figure}
  \begin{center}
    \begin{tabular}{cccccc}
      \raisebox{0.16\linewidth}{(\textit{a})}\hspace{-0.6cm} &
      \includegraphics[height=0.15\linewidth,clip]{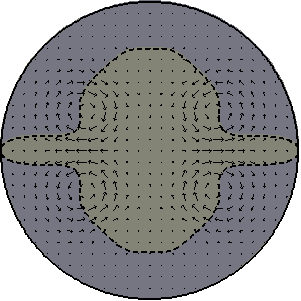} &
      \includegraphics[height=0.15\linewidth,clip]{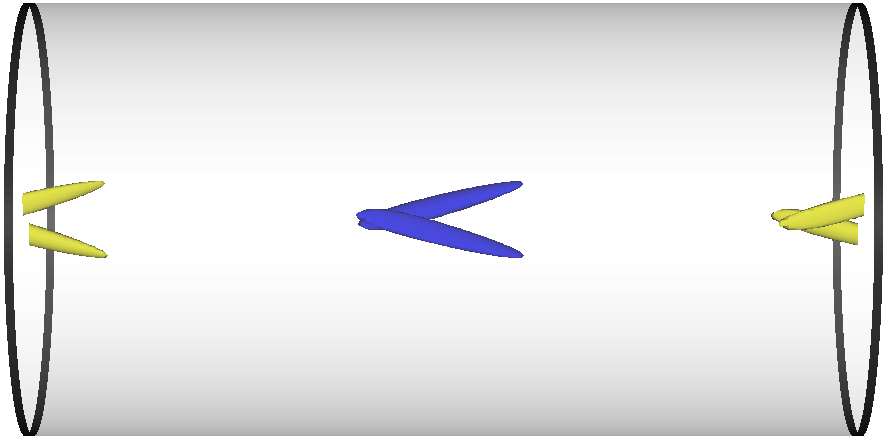} &
      \raisebox{0.16\linewidth}{(\textit{c})}\hspace{-0.6cm} &
      \includegraphics[height=0.15\linewidth,clip]{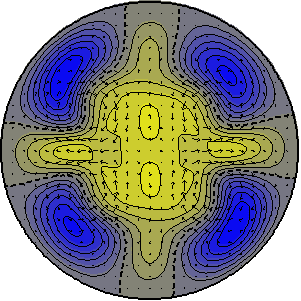} &
      \includegraphics[height=0.15\linewidth,clip]{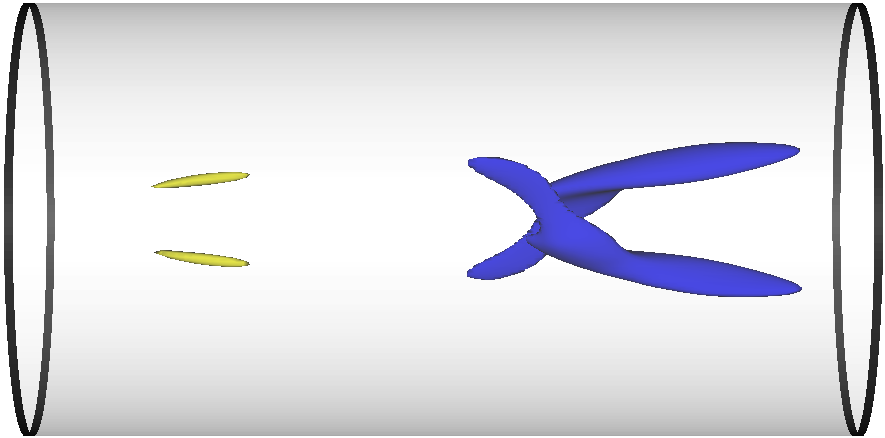}\\
      \raisebox{0.16\linewidth}{(\textit{b})}\hspace{-0.6cm} &
      \includegraphics[height=0.15\linewidth,clip]{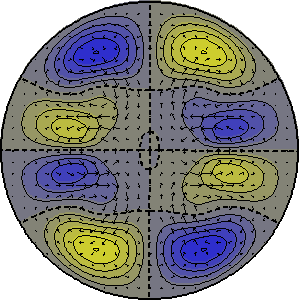} &
      \includegraphics[height=0.15\linewidth,clip]{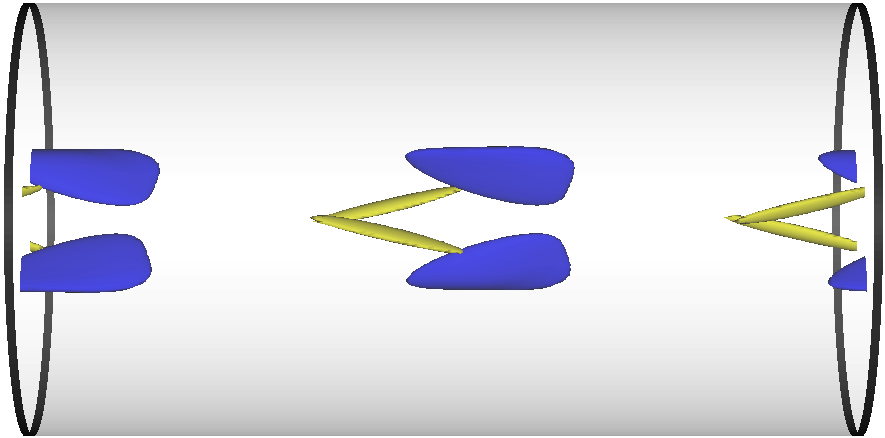} &
      &
      \includegraphics[height=0.15\linewidth,clip]{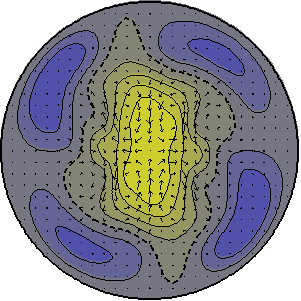} &
      \includegraphics[height=0.15\linewidth,clip]{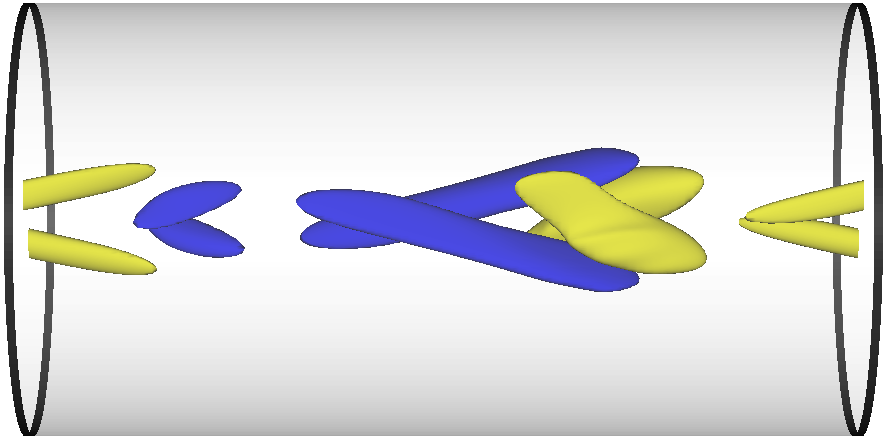}
    \end{tabular}
  \end{center}
  \caption{Unstable eigenmodes of $tw_3$ (see
    figure~\ref{fig:SAk1.63Re2215}\textit{a}) and $sw_1$ (see
    figure~\ref{fig:SAk1.63Re2215}\textit{b}) at
    $(\kappa,\Rey)=(1.63,2215)$. (\textit{a}) Symmetry-preserving real
    eigenvector of $tw_3$. (\textit{b}) Symmetry-breaking real
    eigenvector of $tw_3$. (\textit{c}) Real (top) and imaginary
    (bottom) parts of the unstable complex pair of $sw_1$. Shown are
    averaged cross-sectional axial velocity contours with in-plane
    velocity vectors and three-dimensional isosurfaces of axial
    vorticity (blue -dark- for negative, yellow -light- for
    positive). Axial vorticity has been scaled ($\omega_z(\Imag)=5
    \omega_z(\Real)$) to aid visualisation.}
  \label{fig:swev}
\end{figure}

Figures~\ref{fig:SAk1.63Re2215}(\textit{a}) and
\ref{fig:SAk1.63Re2215}(\textit{b}) depict the eigenspectra at
$(\Rey,\kappa)=(2215,1.63)$ of all four coexisting shift-reflect
travelling waves and of all three spiralling waves, respectively.
Apart from the degenerate eigenmodes corresponding to infinitesimal
translations and rotations shared by all waves (filled black square at
the origin), $tw_1$ (downward-pointing triangles in
figure~\ref{fig:SAk1.63Re2215}\textit{a}) has a single unstable real
eigenmode, a relict from the saddle-node at which the wave was created
together with $tw_2$. The upper-branch wave, $tw_4$ (upward-pointing triangles),
exhibits, after a long sequence of destabilisations and
restabilisations that go beyond the scope of this study, a total of
seven unstable directions: an unstable shift-reflect-symmetric complex
pair, and an unstable real eigenvalue and two complex pairs pointing
out of the symmetry subspace.  As shown in
figure~\ref{fig:stan}(\textit{a}), $tw_2$ has both an unstable real
eigenvalue and an unstable complex pair at $\Rey=2215$. The unstable
real eigenmode of $tw_2$ (shaded squares) is shift-reflect antisymmetric and
points in the direction of $sw_1$. Its symmetry-breaking nature is
clear from figure~\ref{fig:twev}(\textit{a}), where mean axial
velocity contours are shown to be exactly antisymmetric and axial
vorticity has a dominant sign, a feature that is imprinted upon
$sw_1$. Of course, shooting in the opposite direction (changing the
sign of the eigenmode), the shift-reflect-conjugate version of $sw_1$
can be found.
The leading complex pair of $tw_2$ are shift-reflect symmetric and
point in the direction of an invariant unstable manifold leading to a
modulated travelling wave that will be described later on. The real
and imaginary parts of the eigenmode are shown in
figure~\ref{fig:twev}(\textit{b}), where axial vorticity isosurfaces
of the real part are rescaled by a factor of $5$ to make real and
imaginary parts of comparable magnitude. The shift-reflect character of the
complex pair is clear from the figure.  The upper-middle-branch wave
$tw_3$ (circles) has an unstable shift-reflect-symmetric real
eigenvalue (figure~\ref{fig:swev}\textit{a}), presumably left from the
saddle-node at which the wave merges with $tw_2$ at higher $\Rey$, and
an additional symmetry-breaking real eigenvalue inherited from
$tw_2$. The associated eigenmode is shown in
figure~\ref{fig:swev}(\textit{b}) and preserves some of the features
of the unstable symmetry-breaking eigenmode of $tw_2$, such as the
streaks layout and the signature of the vortical structure, although
their respective symmetry-breaking unstable manifolds have deformed
and may already lead to different phase space regions.

Spiralling waves $sw_2$ (squares in
figure~\ref{fig:SAk1.63Re2215}\textit{b}) and $sw_3$ (circles)
exhibit an unstable real eigenvalue together with a complex pair and
just a complex pair, respectively. The real unstable eigenvalue of
middle-branch $sw_2$ is the remnant of saddle-node bifurcations
relating the wave with $sw_1$ and $sw_3$ at other values of $\Rey$ and
$\kappa$. The unstable manifold defined by it is therefore expected to
lead to $sw_1$ and $sw_3$ at either side of $sw_2$. The complex pair,
shared by all three spiralling waves, is at the origin of a branch of
modulated spiralling waves branching off $sw_1$ that will be described
later. This eigenmode has been represented in
figure~\ref{fig:swev}(\textit{c}).
This complex pair of eigenvectors is remarkably similar to that of
$tw_2$ (figure~\ref{fig:twev}\textit{b}). They look alike in all
respects except that the latter preserves the shift-reflect symmetry,
while the former does not. The fact that both Hopf lines of
figure~\ref{fig:stan}(\textit{b}) coalesce in a single point in
$(\kappa,\Rey)$ parameter space is not a coincidence. In fact, it is
to be expected that modulated travelling waves and modulated
spiralling waves are related by a pitchfork bifurcation of a symmetric
fixed cycle \cite[][]{Kuznetsov_B_95}.

All unstable eigenmodes of figure~\ref{fig:SAk1.63Re2215} have been
reported in the last two columns of table~\ref{tab:tsws}, either in
$ev \in S$ or $ev \in \bar{S}$, depending on whether they do
or do not belong to the shift-reflect subspace, respectively.


\section{Time-dependent solutions}\label{sec:timedep}

The Hopf bifurcations prepare the stage for the appearance of
time-dependent solutions. Only $2$-fold azimuthally-periodic flows
will be considered, so that time-dependent solutions that are stable
within this subspace can be computed through symmetry-restricted time
evolution. Unless otherwise stated, this is the only symmetry enforced
in the numerical representation, 
while all other symmetries are unconstrained and may be broken.

\subsection{Modulated travelling waves}

In the symmetry-preserving Hopf
bifurcation on the lower-middle travelling-wave branch ($tw_2$)
a shift-reflect-symmetric time-periodic solution appears which
is a modulated travelling wave, also referred to as a relative
periodic orbit. As $\Rey$ is increased, a complex pair of eigenvalues
crosses into the unstable half of the complex plane as shown in
figure~\ref{fig:stan}(\textit{a}) for $\kappa=1.63$ and extended for
all $\kappa$ in figure~\ref{fig:stan}(\textit{b}). 
In the $\kappa$-range of interest, 
the pitchfork bifurcation occurs before the Hopf bifurcation.
As a consequence, modulated waves emerging from the Hopf
instability, although supercritical, must be expected to inherit all
previous instabilities. Accordingly, at $\kappa=1.63$, emerging
modulated travelling waves are pitchfork unstable at onset.

Fortunately, the Hopf instability belongs precisely in the symmetry
subspace that the pitchfork instability breaks, so that the latter can
be suppressed by restricting time-evolution to the shift-reflect
subspace. It is precisely the fact that shift-reflect-restricted time
evolution unveils a stable branch of modulated travelling waves
pointing towards increasing $\Rey$, that indicates that the Hopf
bifurcation is supercritical in the sense that the first Lyapunov
coefficient is negative \cite[][]{Kuznetsov_B_95}.

Thus, a branch of shift-reflect modulated travelling waves ($mtw$) has
been unfolded at $\kappa=1.63$ and represented in
figure~\ref{fig:grpvsRe}(\textit{a}) (filled diamonds). The shape of
the continuation curve is suggestive of a turning point in a
fold-of-cycles at about $\Rey\simeq 2337$, implying the existence of a
saddle branch of cycles of larger amplitude. The consequences of this
were advanced in figure~\ref{fig:BifDia} and will be discussed
below. Let us first focus on the nodal branch and analyse one of the
solutions.

\begin{figure}
  \begin{center}
    \begin{tabular}{cccc}
      \raisebox{0.33\linewidth}{(\textit{a})}\hspace{-0.6cm} &
      \includegraphics[height=0.32\linewidth,clip]{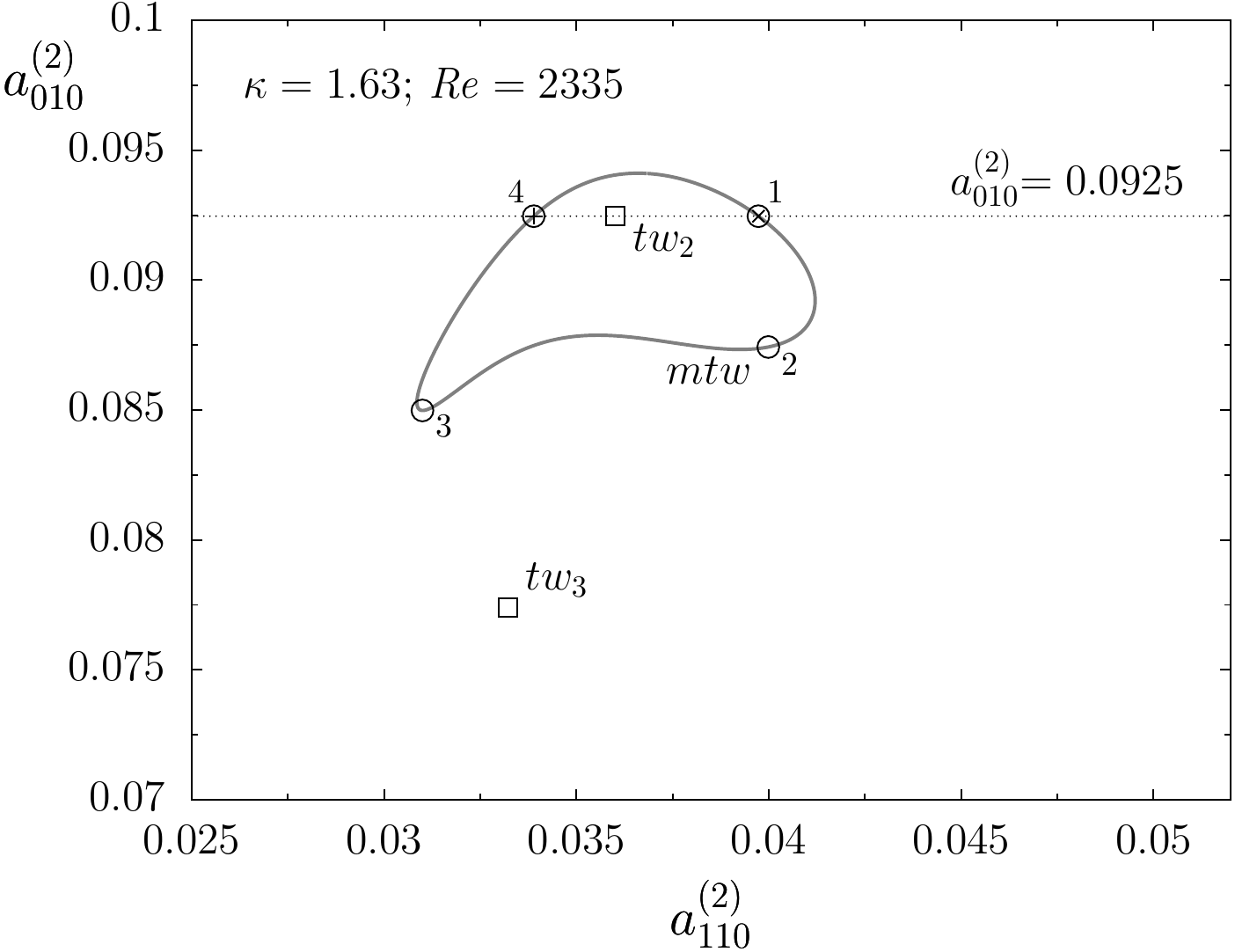} &
      \raisebox{0.33\linewidth}{(\textit{b})}\hspace{-0.6cm} &
      \includegraphics[height=0.31\linewidth,clip]{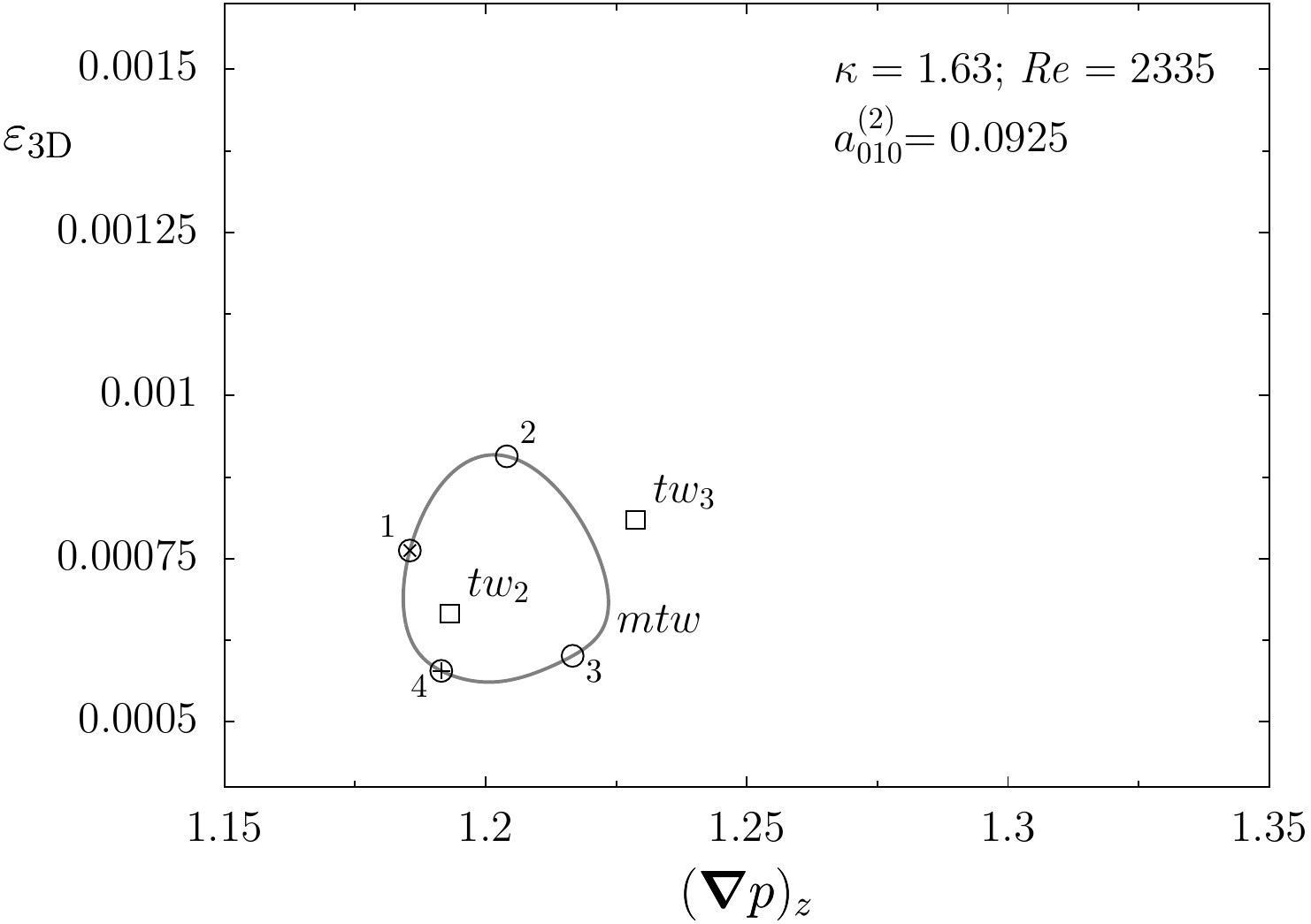}
    \end{tabular}
  \end{center}
  \caption{Phase map projections of a modulated travelling wave
    ($mtw$) at $(\kappa,\Rey)=(1.63,2335)$. (\textit{a})
    $a_{010}^{(2)}$ vs $a_{110}^{(2)}$. (\textit{b}) Three-dimensional
    energy ($\varepsilon_{\rm 3D}$) vs axial pressure gradient
    ($(\bnabla p)_z$). Middle-branch travelling waves ($tw_2$ and
    $tw_3$) have been labelled and marked with open squares. Positive
    and negative crossings of a Poincar\'e section defined by
    $a_{010}^{(2)}=a_{010}^{(2)}(tw_2)$ are indicated by plus signs
    and crosses, respectively. Numbered circles correspond to
    snapshots in figure~\ref{fig:SSTPOk1.63Re2335}.}
  \label{fig:PhMapTPOk1.63Re2335}
\end{figure}

Figure~\ref{fig:PhMapTPOk1.63Re2335} shows two alternative
axial-drift-independent phase map projections of a modulated
travelling wave solution at $(\kappa,\Rey)=(1.63,2335)$.
The solution describes a closed loop and can therefore be seen as a
relative periodic orbit. Travelling waves, which are degenerate or
relative equilibria, and solutions bifurcating from them have a pure
frequency associated to the advection speed (solid-body
translation). Time series of local quantities such as point velocities
or pressures are necessarily imprinted by this frequency, but the
modulation disappears by restating the problem in a comoving
frame. Global quantities such as modal energies or volume averaged
fields naturally overlook solid-body rotation and translation, making
them suitable for a decoupled analysis in the direction orthogonal to
the degenerate drift. Thus, as justified before, the bifurcation
analysis of travelling waves can be carried out analogously to that of
fixed points, as long as special care is taken in the neighbourhood of
homoclinic connections \cite[][]{Rand_ARMA_82,GoLeMe_JNS_00}. The Hopf
bifurcation adds a modulational frequency to the pure translational
frequency. As a result, global quantities cease to be constant and
oscillate with this frequency, while the representation of local
quantities would have made the solution appear as quasiperiodic.

At $\Rey=2335$, close to the presumed fold-of-cycles, the modulation
has grown large around the travelling wave ($tw_2$) from which it
originally bifurcated. To allow comparison with other solutions that
will be discussed later, a Poincar\'e section at
$a_{010}^{(2)}=a_{010}^{(2)}(tw_2)$ has been defined. The trajectory
pierces the Poincar\'e section twice, so it is convenient to
differentiate between positive and negative crossings (plus signs and
crosses, respectively, in figure~\ref{fig:PhMapTPOk1.63Re2335}). This
Poincar\'e section is special in the sense that it is independent of
the degenerate drift. Using global quantities is equivalent to working
in a reference frame moving with the wave. In this way, drifting waves
drop from the analysis, while modulated waves appear as equilibria of
the Poincar\'e application and doubly-modulated waves (or relative
tori) as discrete cycles.

\begin{figure}
  \begin{center}
    \begin{tabular}{cccc}
      \raisebox{0.30\linewidth}{(\textit{a})}\hspace{-0.6cm} &
      \includegraphics[height=0.27\linewidth,clip]{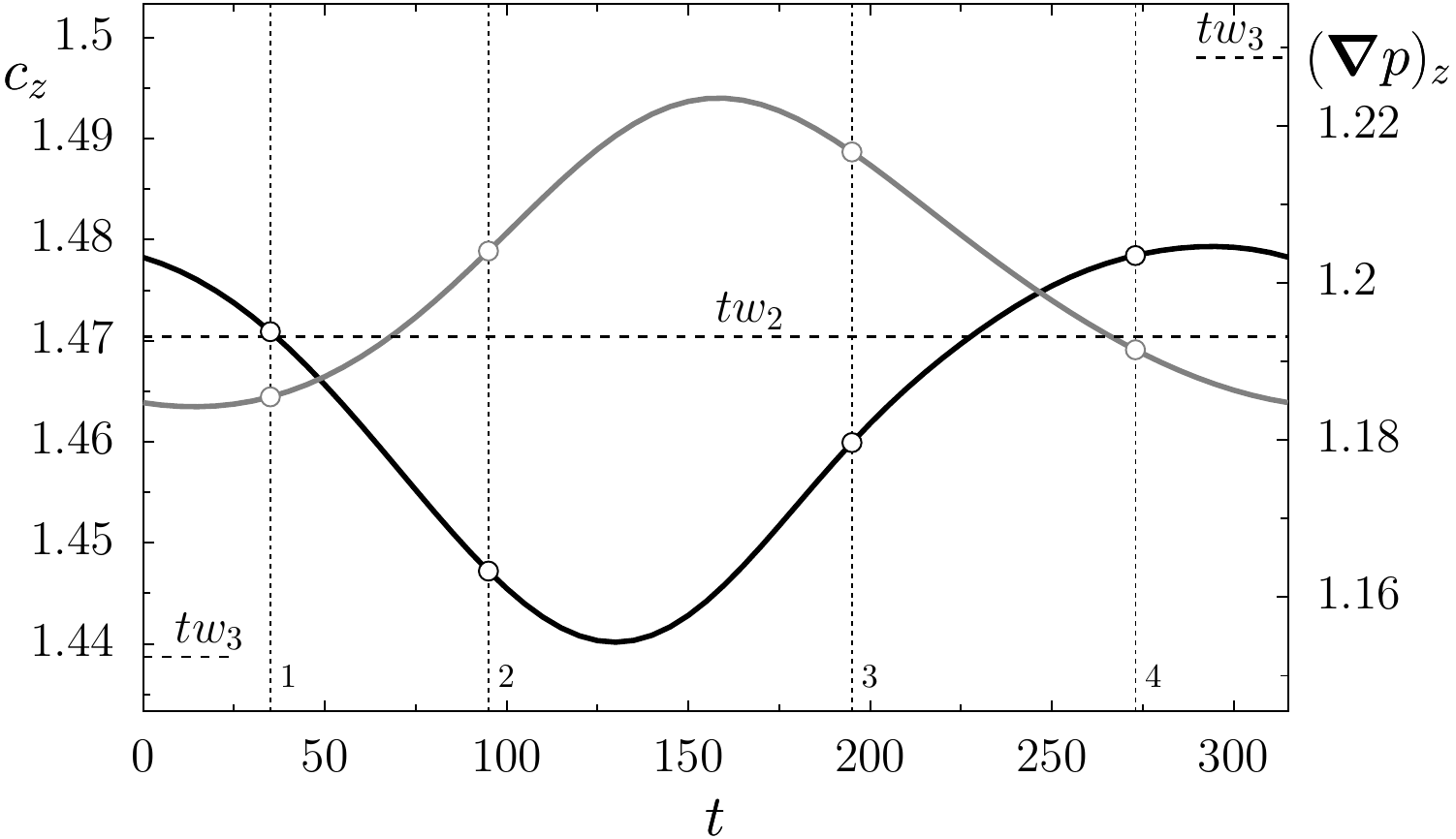} &
      \raisebox{0.30\linewidth}{(\textit{b})}\hspace{-0.6cm} &
      \includegraphics[height=0.28\linewidth,clip]{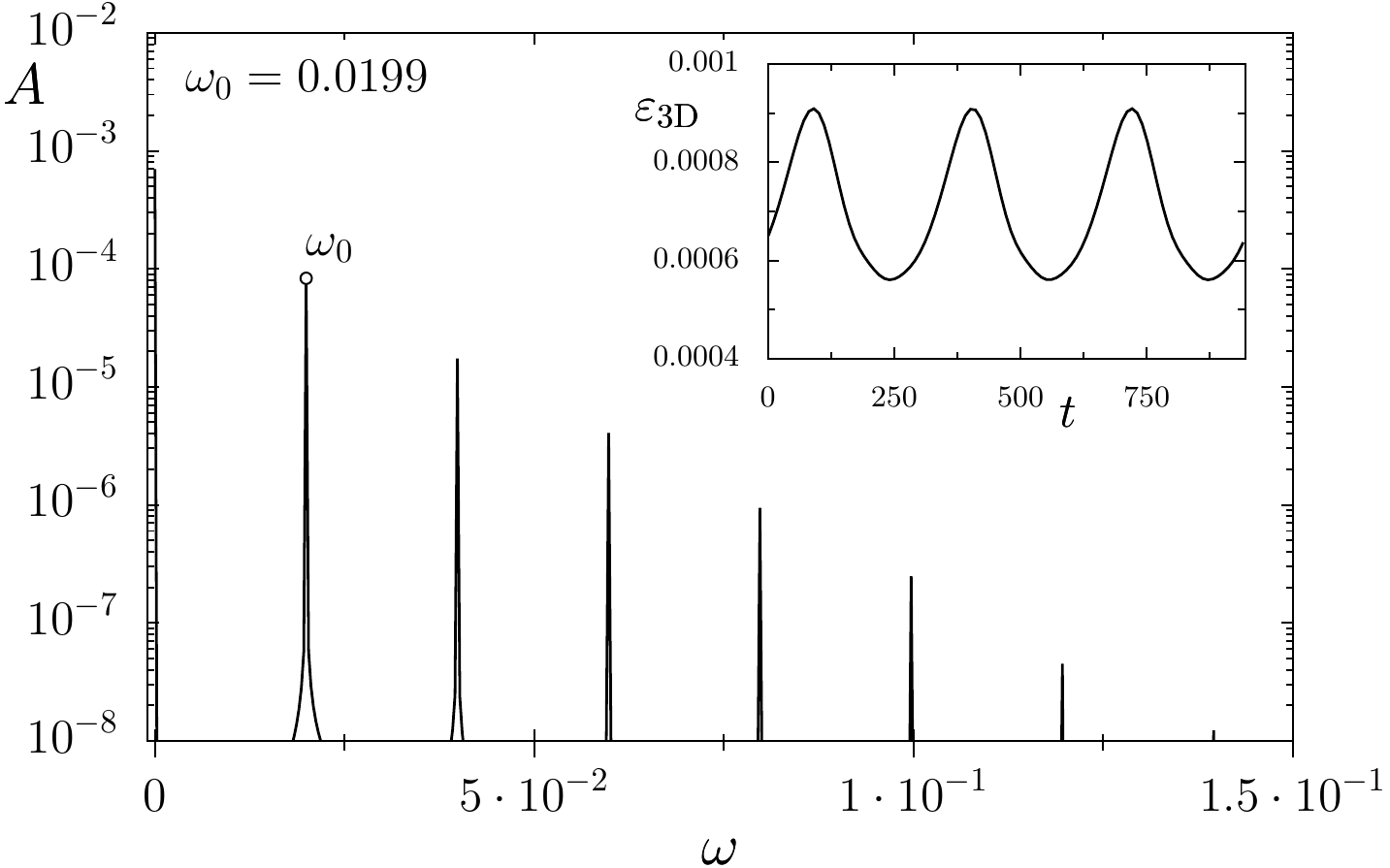}
    \end{tabular}
  \end{center}
  \caption{Modulated travelling wave ($mtw$) at
    $(\kappa,\Rey)=(1.63,2335)$. (\textit{a}) Axial phase speed
    ($c_z$, black line) and axial pressure gradient ($(\bnabla p)_z$,
    gray line) time-series. Dotted horizontal lines indicate the
    values for $tw_2$ and $tw_3$ while numbered vertical lines and
    open circles indicate snapshots in
    figure~\ref{fig:SSTPOk1.63Re2335}. (\textit{b}) Fourier transform
    of the non-axisymmetric streamwise-dependent modal energy contents
    ($\varepsilon_{\rm 3D}$). Part of the time signal is plotted in
    the inset.}
  \label{fig:TSFTTPOk1.63Re2335}
\end{figure}

The time dependence is clarified by
figure~\ref{fig:TSFTTPOk1.63Re2335}(\textit{a}), where axial phase
speed ($c_z$, black line) and mean axial pressure gradient ($(\bnabla
p)_z$, gray line) time-series along a full period of the solution have
been represented.
Comparing the time signals with the constant value for middle-branch
travelling waves ($tw_2$ and $tw_3$, horizontal dotted lines) it
becomes clear that the modulated wave oscillates around $tw_2$, and
seems to visit $tw_3$ once along every period, although
figure~\ref{fig:PhMapTPOk1.63Re2335} seems to discard too close a
visit. Both signals oscillate $\pm 1\%$ around a mean value that has
an offset when compared with $tw_2$. On average, the driving pressure
gradient is slightly higher and the axial drift rate slower than for
the travelling wave. This follows from the fact that we are no longer
close to the bifurcation point and nonlinear effects have long kicked
in. The Fourier transform of the energy contained in non-axisymmetric
streamwise-dependent modes ($\varepsilon_{\rm 3D}$) has been plotted
in figure~\ref{fig:TSFTTPOk1.63Re2335}(\textit{b}), with the time
signal shown in the inset frame. The spectrum reveals that the
solution has a strong mean component and a peak angular frequency at
$\omega_0=0.0199 \; (4U/D)$, corresponding to a period
$T_0=2\upi/\omega_0=315.2 D/(4 U)$. This period is extremely long when
compared with the streamwise advection time-scale, which is of order
$2 \upi /(\kappa c_z) \sim 3$. The signal is not strictly sinusoidal
and some energy is spread among a number of harmonics of $\omega_0$.

\begin{figure}
  \begin{center}
    \begin{tabular}{cccccc}
      \raisebox{0.16\linewidth}{($1$)}\hspace{-0.6cm} &
      \includegraphics[height=0.15\linewidth,clip]{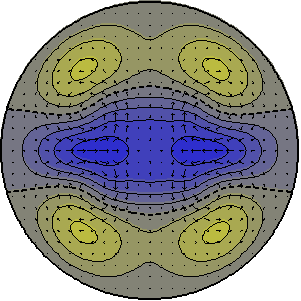} &
      \includegraphics[height=0.15\linewidth,clip]{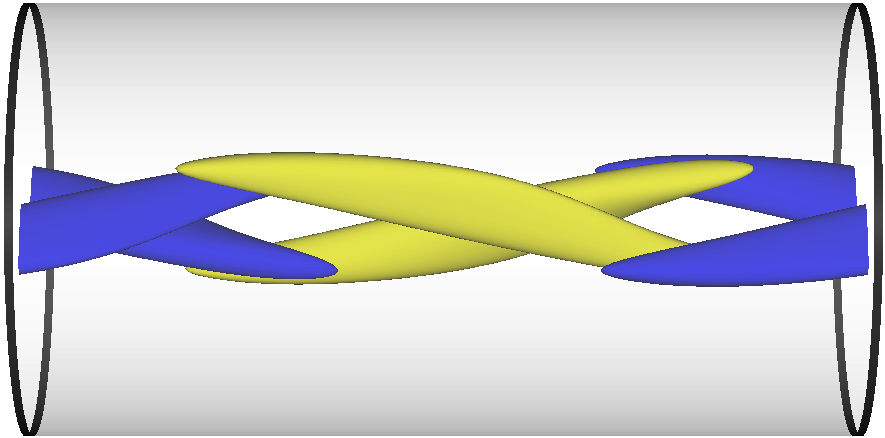} &
      \raisebox{0.16\linewidth}{($2$)}\hspace{-0.6cm} &
      \includegraphics[height=0.15\linewidth,clip]{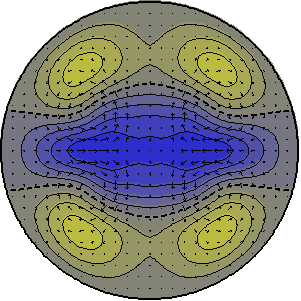} &
      \includegraphics[height=0.15\linewidth,clip]{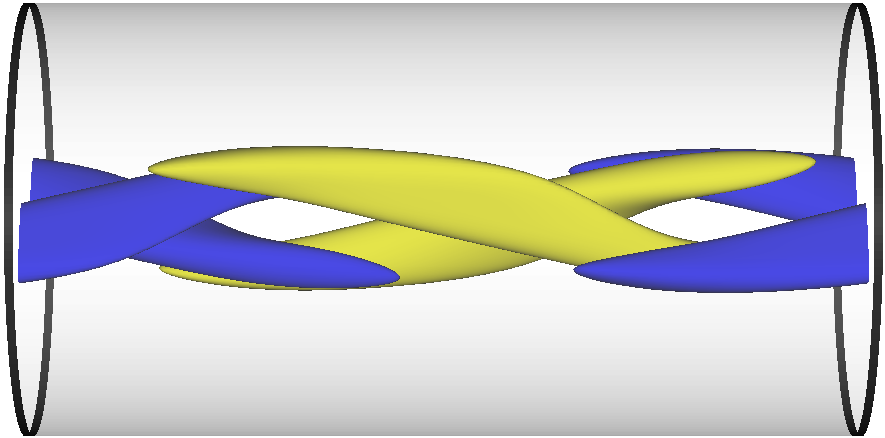}\\
      \raisebox{0.16\linewidth}{($3$)}\hspace{-0.6cm} &
      \includegraphics[height=0.15\linewidth,clip]{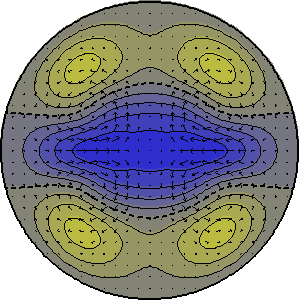} &
      \includegraphics[height=0.15\linewidth,clip]{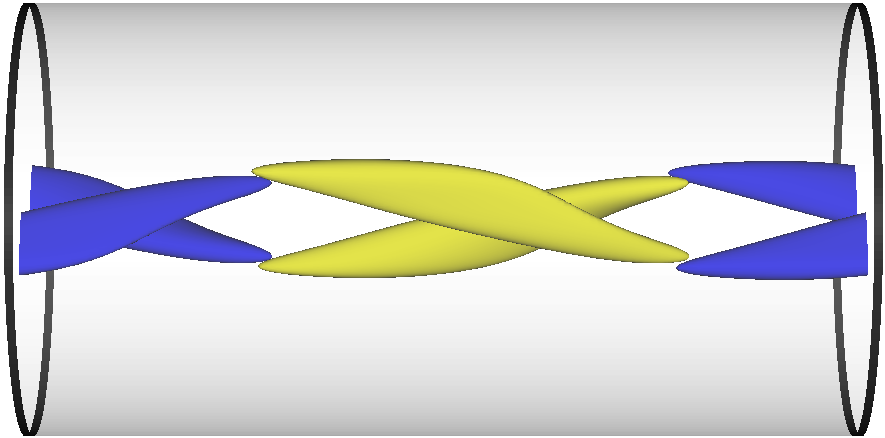} &
      \raisebox{0.16\linewidth}{($4$)}\hspace{-0.6cm} &
      \includegraphics[height=0.15\linewidth,clip]{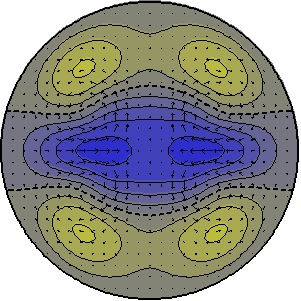} &
      \includegraphics[height=0.15\linewidth,clip]{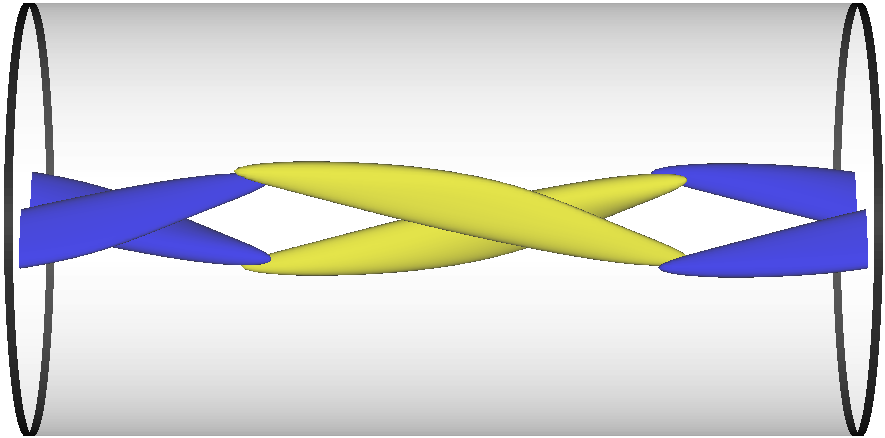}\\
    \end{tabular}
  \end{center}
  \caption{Modulated travelling wave at
    $(\kappa,\Rey)=(1.63,2335)$. Left: $z$-averaged cross-sectional
    axial velocity contours spaced at intervals of $\Delta \langle
    u_z\rangle_z = 0.1 U$. In-plane velocity vectors are also
    displayed. Right: axial vorticity isosurfaces at $\omega_z=\pm 1
    U/D$. Fluid flows rightwards. Blue (dark gray) for negative,
    yellow (light) for positive. ($1$) $t=35$, ($2$) $t=95$, ($3$)
    $t=195$ and ($4$) $t=273$ $D/(4U)$. To avoid drift due to
    streamwise advection, snapshots are taken in a comoving frame
    moving with the instantaneous advection speed from
    figure~\ref{fig:TSFTTPOk1.63Re2335}(\textit{a}). The snapshots
    have been indicated with empty circles in
    figures~\ref{fig:PhMapTPOk1.63Re2335} and
    \ref{fig:TSFTTPOk1.63Re2335}(\textit{a}).}
  \label{fig:SSTPOk1.63Re2335}
\end{figure}

To better convey the modulational character of the instability, a few
snapshots (conveniently marked in
figures~\ref{fig:PhMapTPOk1.63Re2335} and
\ref{fig:TSFTTPOk1.63Re2335}\textit{a}) of the flow field along a
cycle have been represented in figure~\ref{fig:SSTPOk1.63Re2335} (see
online movie).
It is clear from the snapshots that the modulated wave indeed
oscillates around $tw_2$ (figure~\ref{fig:tws}\textit{b}) and that, as
noted, preserves the shift-reflect symmetry. As expected from the mild
oscillation of all time signals, the modulation is not very prominent
and the snapshots look all fairly similar.


\subsection{Modulated spiralling waves}

The Hopf bifurcation on the branch of spiralling waves occurs at lower
$\Rey$ than on the branch of travelling waves
(figure~\ref{fig:stan}\textit{b}). Spiralling wave $sw_1$ loses
stability in a supercritical Hopf and a branch of stable modulated
spiralling waves emerges, pointing in the direction of increasing
$\Rey$. Three of such branches have been plotted in
figure~\ref{fig:grpvsRe} (filled circles, labelled $msw$) for varying
$\kappa$ (shading as explained in the legend).

\begin{figure}
  \begin{center}
    \includegraphics[height=0.5\linewidth,clip]{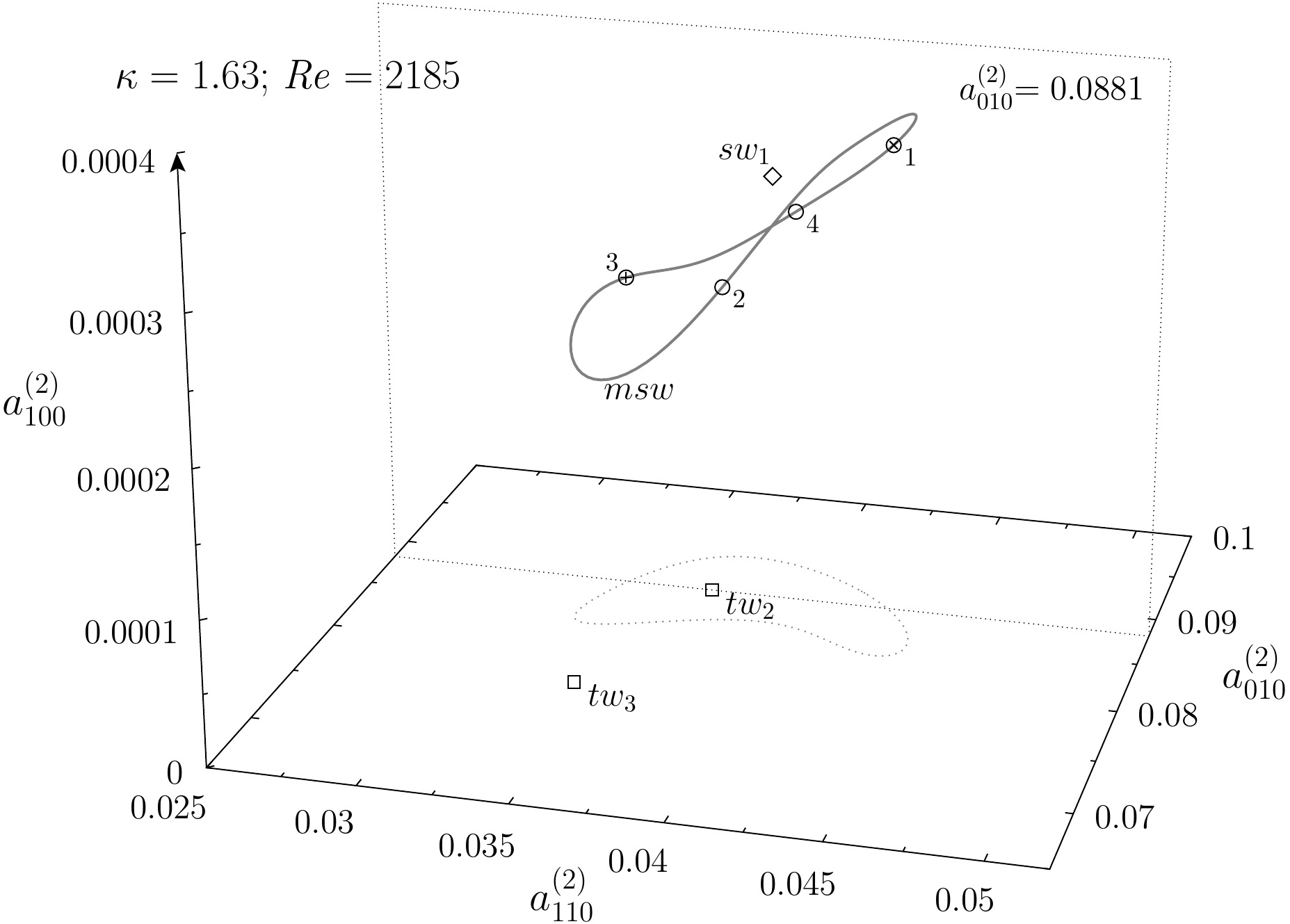}
  \end{center}
  \caption{Three-dimensional phase map projection
    $(a_{010}^{(2)},a_{110}^{(2)},a_{100}^{(2)})$ of a modulated
    spiralling wave ($msw$) at
    $(\kappa,\Rey)=(1.63,2185)$. Middle-branch travelling waves
    ($tw_2$ and $tw_3$) have been labelled and marked with open
    squares and spiralling wave ($sw_1$) with an open
    diamond. Positive and negative crossings of a Poincar\'e section
    defined by $a_{010}^{(2)}=a_{010}^{(2)}(tw_2)$ are indicated by
    plus signs and crosses, respectively. Numbered circles correspond
    to snapshots in figure~\ref{fig:SSRPOk1.63Re2185}.}
  \label{fig:PhMap3DRPOk1.63Re2185}
\end{figure}

Figure~\ref{fig:PhMap3DRPOk1.63Re2185} shows a three-dimensional phase
map projection on the space defined by
$(a_{010}^{(2)},a_{110}^{(2)},a_{100}^{(2)})$ of a modulated
spiralling wave ($msw$) at $(\kappa,\Rey)=(1.63,2185)$.
The wave describes a closed loop in phase space surrounding the
spiralling wave ($sw_1$) from which it bifurcated, and it does so at a
finite amplitude of $a_{100}^{(2)}$, away from the shift-reflect
subspace where this coefficient cancels exactly. The projection on the
shift-reflect subspace (light gray dotted line on the
$a_{100}^{(2)}=0$ plane) is reminiscent of
figure~\ref{fig:PhMapTPOk1.63Re2335}(\textit{a}), even if $\Rey$ is
different, which suggests that both waves are related by a pitchfork
of cycles.

\begin{figure}
  \begin{center}
    \begin{tabular}{cccc}
      \raisebox{0.33\linewidth}{(\textit{a})}\hspace{-0.6cm} &
      \includegraphics[height=0.32\linewidth,clip]{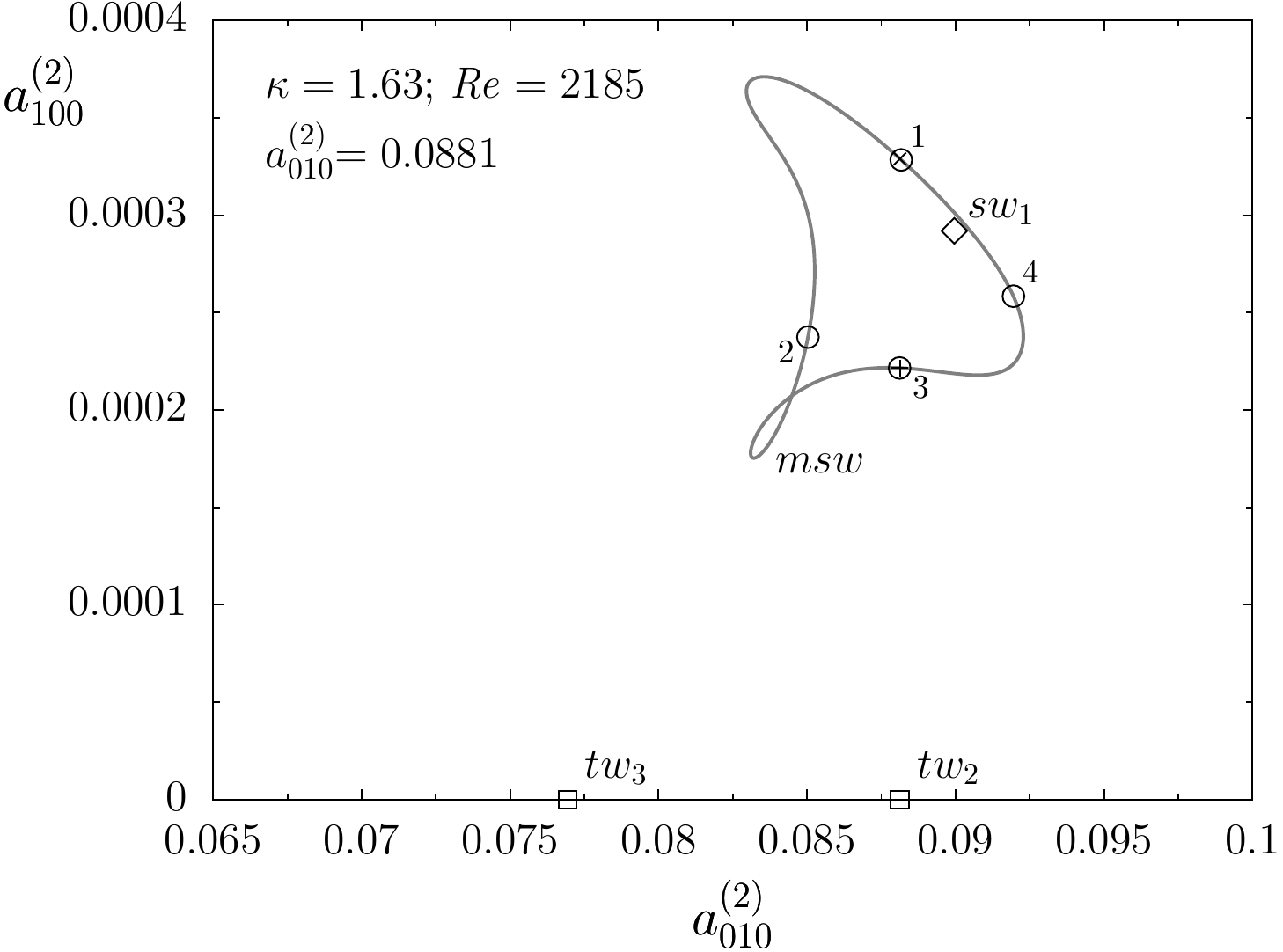} &
      \raisebox{0.33\linewidth}{(\textit{b})}\hspace{-0.6cm} &
      \includegraphics[height=0.31\linewidth,clip]{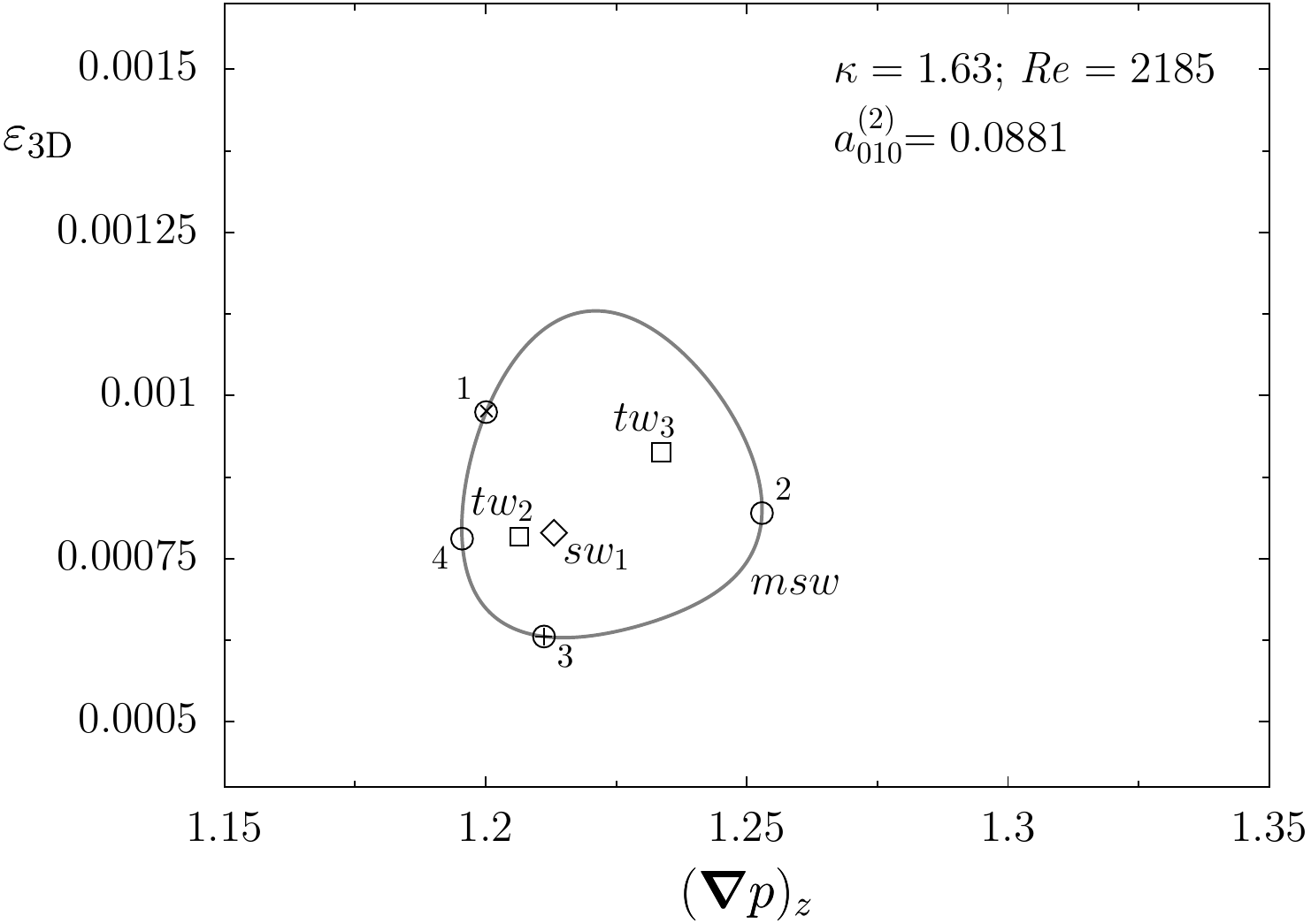}
    \end{tabular}
  \end{center}
  \caption{Phase map projections of a modulated spiralling wave
    ($msw$) at $(\kappa,\Rey)=(1.63,2185)$. (\textit{a})
    $a_{100}^{(2)}$ vs $a_{010}^{(2)}$. (\textit{b}) Three-dimensional
    energy ($\varepsilon_{\rm 3D}$) vs axial pressure gradient
    ($(\bnabla p)_z$). Middle-branch travelling waves ($tw_2$ and
    $tw_3$) have been labelled and marked with open squares and
    spiralling wave ($sw_1$) with an open diamond. Positive and
    negative crossings of a Poincar\'e section defined by
    $a_{010}^{(2)}=a_{010}^{(2)}(tw_2)$ are indicated by plus signs
    and crosses, respectively. Numbered circles correspond to
    snapshots in figure~\ref{fig:SSRPOk1.63Re2185}.}
  \label{fig:PhMapRPOk1.63Re2185}
\end{figure}

A couple of additional phase map projections have been represented in
figure~\ref{fig:PhMapRPOk1.63Re2185}.
Figure~\ref{fig:PhMapRPOk1.63Re2185}(\textit{a}) shows the projection
of figure~\ref{fig:PhMap3DRPOk1.63Re2185} on the
$(a_{100}^{(2)},a_{010}^{(2)})$ plane, while
figure~\ref{fig:PhMapRPOk1.63Re2185}(\textit{b}) represents the same
trajectory in the $(\varepsilon_{\rm 3D},(\bnabla p)_z)$
plane. Comparison of the latter with
figure~\ref{fig:PhMapTPOk1.63Re2335}(\textit{b}) further supports the
connection between $msw$ and $mtw$.

\begin{figure}
  \begin{center}
    \begin{tabular}{cccc}
      \raisebox{0.30\linewidth}{(\textit{a})}\hspace{-0.6cm} &
      \includegraphics[height=0.27\linewidth,clip]{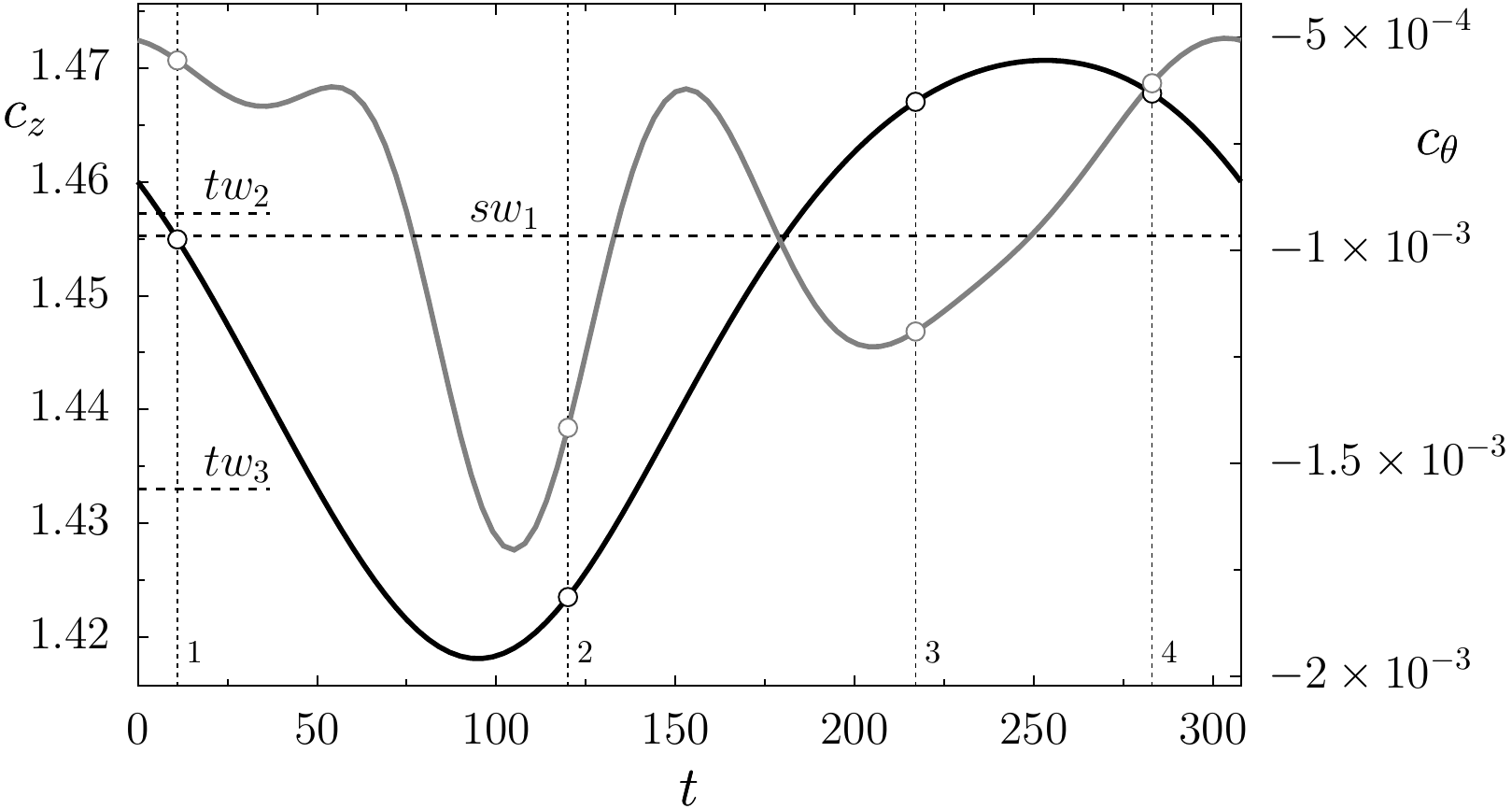} &
      \raisebox{0.30\linewidth}{(\textit{b})}\hspace{-0.6cm} &
      \includegraphics[height=0.28\linewidth,clip]{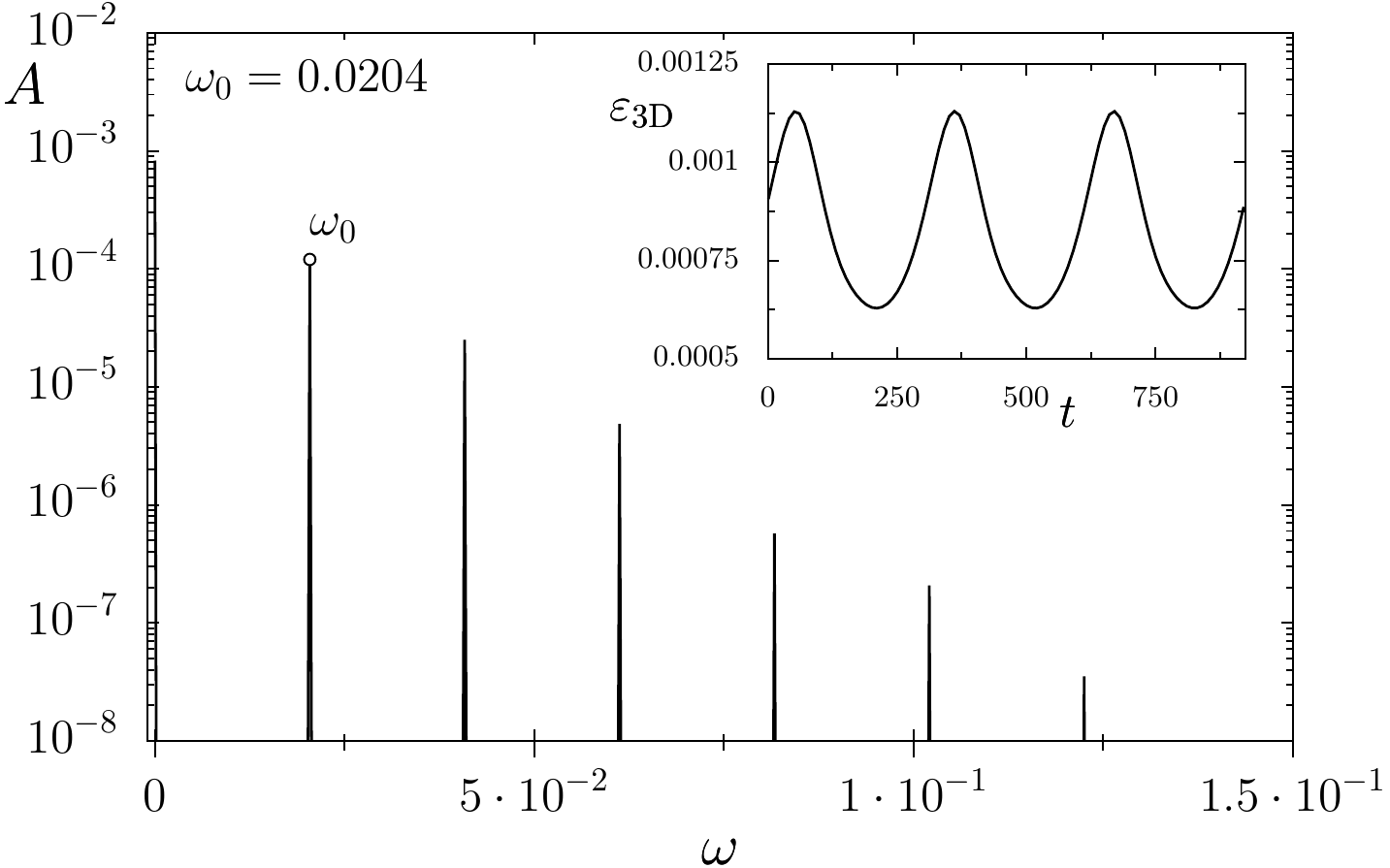}
    \end{tabular}
  \end{center}
  \caption{Modulated spiralling wave ($msw$) at
    $(\kappa,\Rey)=(1.63,2185)$. (\textit{a}) Axial ($c_z$, black
    line) and azimuthal ($c_{\theta}$, gray line) phase speed
    time-series. Dotted horizontal lines indicate the values for
    $sw_1$, $tw_2$ and $tw_3$ while numbered vertical lines and open
    circles indicate snapshots in
    figure~\ref{fig:SSRPOk1.63Re2185}. (\textit{b}) Fourier transform
    of the non-axisymmetric streamwise-dependent modal energy contents
    ($\varepsilon_{\rm 3D}$). Part of the time signal is plotted in
    the inset.}
  \label{fig:TSFTRPOk1.63Re2185}
\end{figure}

The variables used for phase map representation naturally filter the
frequencies associated to degenerate drifts. In this case, since we
are dealing with a modulated spiralling wave bifurcated from a
spiralling wave, not only the axial phase speed ($c_z$) is masked, but
also the azimuthal rotation rate ($c_{\theta}$). Together with the
modulational (Hopf) frequency, two additional frequencies associated
to drifts are involved, although only the former is relevant to the
dynamics. Both phase speeds have been plotted along a full period of
$msw$ in figure~\ref{fig:TSFTRPOk1.63Re2185}(\textit{a}).
They oscillate around the spiralling wave values with a certain offset
due to nonlinear
effects. Figure~\ref{fig:TSFTRPOk1.63Re2185}(\textit{b}) shows the
Fourier spectrum of the energy signal in the inset. As was already
observed for $mtw$, also $msw$ has a slow modulation angular frequency
of $\omega_0=0.0204 \; (4U/D)$, corresponding to a period $T_0=307.8
D/(4 U)$. The fact that both periods are of the same order reinforces
the suspicion that they are a consequence of the same mechanism.

\begin{figure}
  \begin{center}
    \begin{tabular}{cccccc}
      \raisebox{0.16\linewidth}{($1$)}\hspace{-0.6cm} &
      \includegraphics[height=0.15\linewidth,clip]{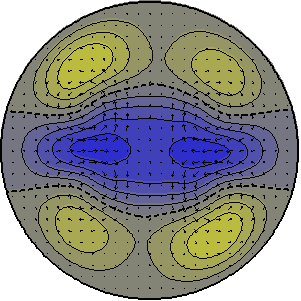} &
      \includegraphics[height=0.15\linewidth,clip]{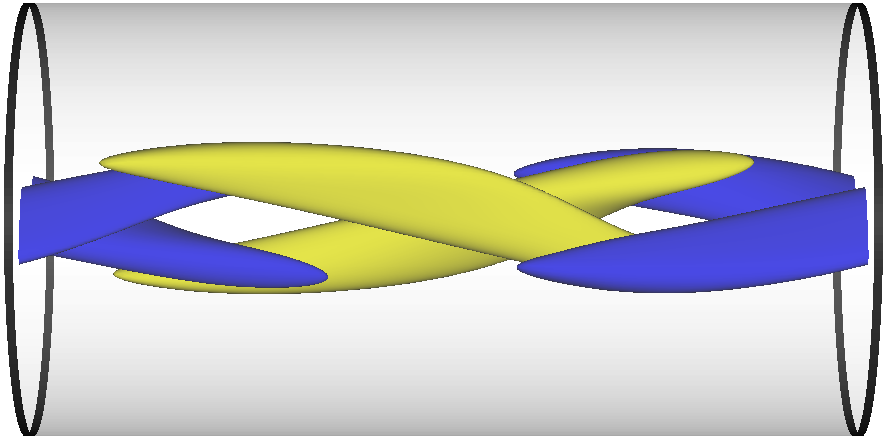} &
      \raisebox{0.16\linewidth}{($2$)}\hspace{-0.6cm} &
      \includegraphics[height=0.15\linewidth,clip]{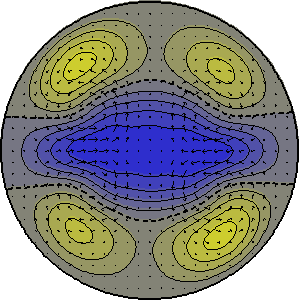} &
      \includegraphics[height=0.15\linewidth,clip]{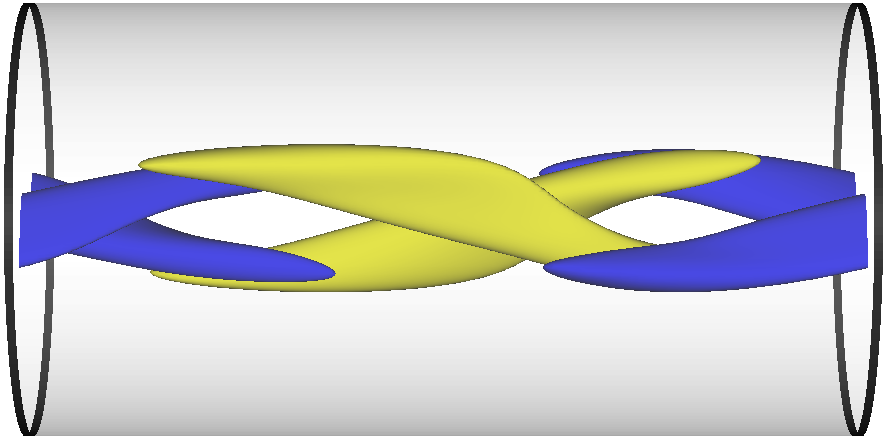}\\
      \raisebox{0.16\linewidth}{($3$)}\hspace{-0.6cm} &
      \includegraphics[height=0.15\linewidth,clip]{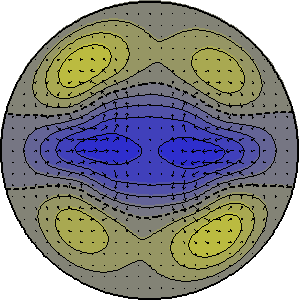} &
      \includegraphics[height=0.15\linewidth,clip]{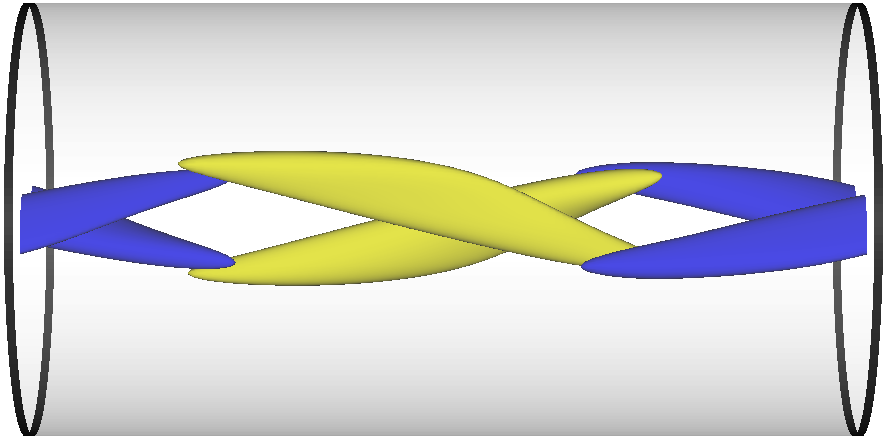} &
      \raisebox{0.16\linewidth}{($4$)}\hspace{-0.6cm} &
      \includegraphics[height=0.15\linewidth,clip]{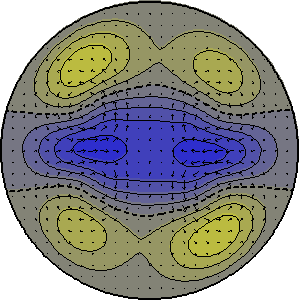} &
      \includegraphics[height=0.15\linewidth,clip]{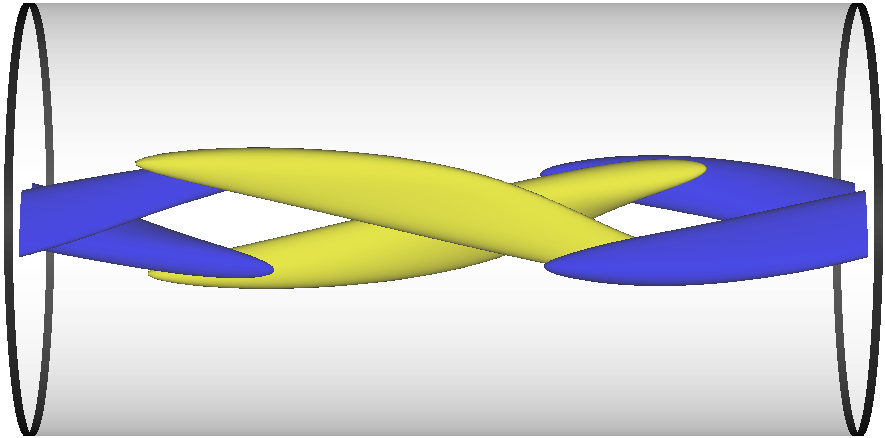}\\
    \end{tabular}
  \end{center}
  \caption{Modulated spiralling wave at
    $(\kappa,\Rey)=(1.63,2185)$. Left: $z$-averaged cross-sectional
    axial velocity contours spaced at intervals of $\Delta \langle
    u_z\rangle_z = 0.1 U$. In-plane velocity vectors are also
    displayed. Right: axial vorticity isosurfaces at $\omega_z=\pm 1
    U/D$. Fluid flows rightwards. Blue (dark gray) for negative,
    yellow (light) for positive. ($1$) $t=11$, ($2$) $t=120$, ($3$)
    $t=217$ and ($4$) $t=283$ $D/(4U)$. To avoid drift due to
    streamwise advection, snapshots are taken in a comoving frame
    spiralling with the instantaneous advection speeds from
    figure~\ref{fig:TSFTRPOk1.63Re2185}(\textit{a}). The snapshots
    have been indicated with circles in
    figures~\ref{fig:PhMap3DRPOk1.63Re2185},
    \ref{fig:PhMapRPOk1.63Re2185} and
    \ref{fig:TSFTRPOk1.63Re2185}(\textit{a}).}
  \label{fig:SSRPOk1.63Re2185}
\end{figure}

Snapshots corresponding to the time instants indicated with circles in
figures~\ref{fig:PhMap3DRPOk1.63Re2185}, \ref{fig:PhMapRPOk1.63Re2185}
and \ref{fig:TSFTRPOk1.63Re2185}(\textit{a}), have been plotted in
figure~\ref{fig:SSRPOk1.63Re2185} (see online movie).
As was the case for $mtw$ of figure~\ref{fig:SSTPOk1.63Re2335}, the
modulational character of $msw$ is very mild. Nevertheless, it clearly
orbits around $sw_1$ and the symmetry-breaking is fairly clear at all
times.

At the lower $\kappa$ explored (light solid circles in
figure~\ref{fig:grpvsRe}\textit{b}), the modulated spiralling waves
branch seems to undergo a fold-of-cycles. The bifurcation scenario
presented in figure~\ref{fig:BifDia} is not applicable to this lower
$\kappa$. At higher $\kappa$, instead, modulated waves lose stability
to doubly-modulated spiralling waves in a supercritical Neimark-Sacker
bifurcation.

\subsection{Doubly-modulated spiralling waves}

In an appropriately spiralling frame of reference, modulated
spiralling waves appear as periodic orbits. A Poincar\'e section can
be defined as in previous section to analyse the stability of the
orbit. When considering the Poincar\'e map associated to the relative
periodic orbit, the solution reduces to a simple equilibrium and
discrete-time bifurcation analysis can be applied.

For high enough $\kappa$, modulated spiralling waves undergo a
supercritical Neimark-Sacker bifurcation (Hopf of cycles) that occurs
away from strong resonances \cite[][]{Kuznetsov_B_95}. Bifurcated
waves are relative quasiperiodic solutions involving $2$ degenerate
and $2$ modulational frequencies, adding up to $4$ frequencies. The
two extrema (minimum and maximum) that were used to represent
modulated waves split in four to convey the existence of a modulation
of the modulation (figure~\ref{fig:grpvsRe}, filled circles, labelled
$m^2sw$). This is visible at the highest $\Rey$ values for
$\kappa=1.60$ (intermediate-gray) and more clearly, yet in a short
range of less than $10$ $\Rey$-units, for $\kappa=1.63$ (black).

Track of the doubly-modulated waves at $\kappa=1.60$ is lost shortly
after they bifurcate in what seems to be a fold-of-tori that will not
be pursued here. At this $\kappa$, the bifurcation diagram of
figure~\ref{fig:BifDia} would be complicated in ways that make it
inaccessible. The more extended existence of doubly-modulated waves at
$\kappa=1.63$ leaves enough room for them to evolve nonlinearly away
from modulated waves so that they can be properly analysed.

\begin{figure}
  \begin{center}
    \includegraphics[height=0.5\linewidth,clip]{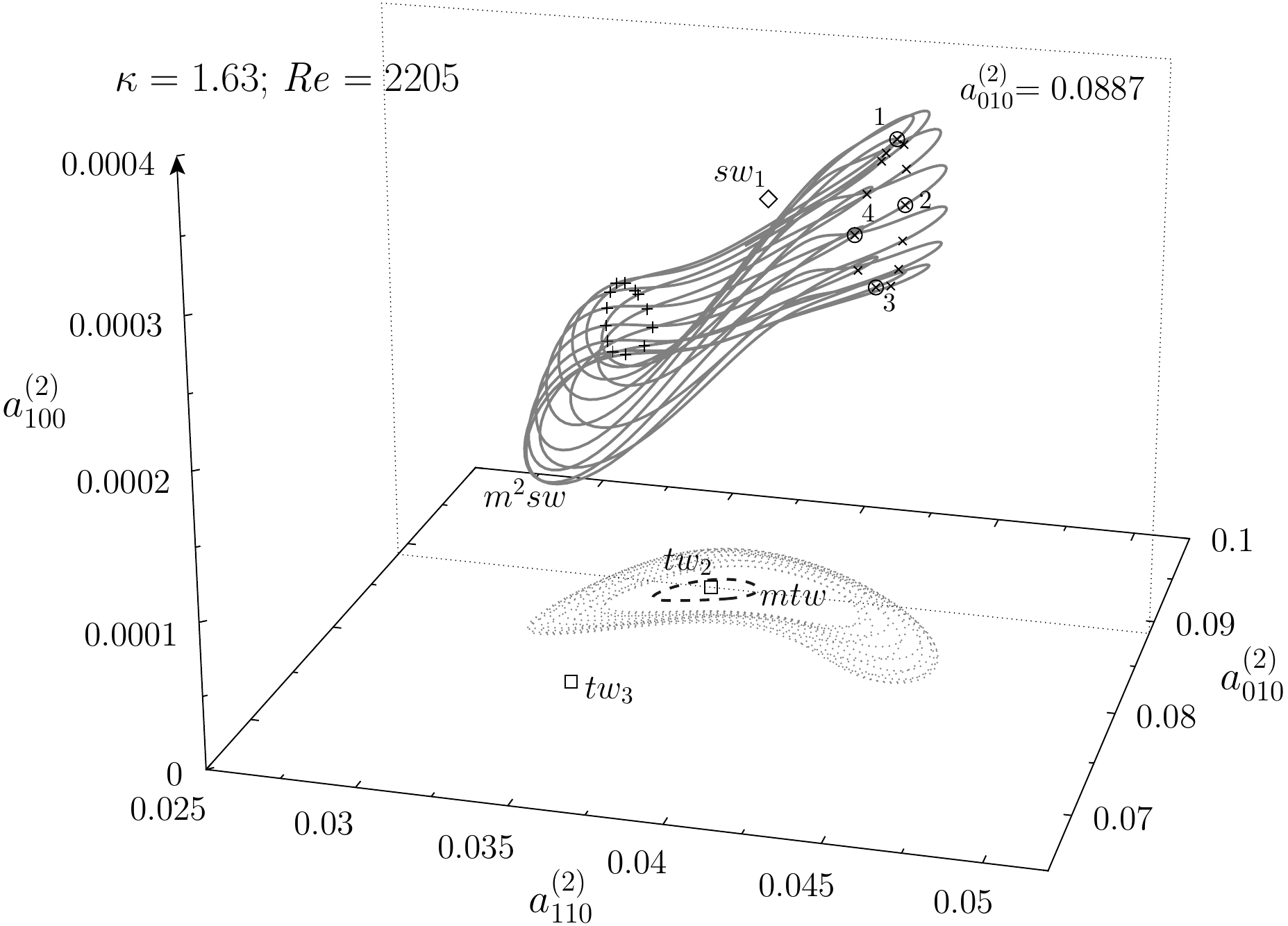}
  \end{center}
  \caption{Three-dimensional phase map projection
    $(a_{010}^{(2)},a_{110}^{(2)},a_{100}^{(2)})$ of a
    doubly-modulated spiralling wave ($m^2sw$) at
    $(\kappa,\Rey)=(1.63,2205)$. Middle-branch travelling waves
    ($tw_2$ and $tw_3$) have been labelled and marked with open
    squares and spiralling wave ($sw_1$) with an open
    diamond. Positive and negative crossings of a Poincar\'e section
    defined by $a_{010}^{(2)}=a_{010}^{(2)}(tw_2)$ are indicated by
    plus signs and crosses, respectively. Numbered circles correspond
    to snapshots in figure~\ref{fig:SSRPTk1.63Re2205}.}
  \label{fig:PhMap3DRPTk1.63Re2205}
\end{figure}

Figure~\ref{fig:PhMap3DRPTk1.63Re2205} shows a three-dimensional phase
map projection on the space defined by
$(a_{010}^{(2)},a_{110}^{(2)},a_{100}^{(2)})$ of a doubly-modulated
spiralling wave ($m^2sw$) at $(\kappa,\Rey)=(1.63,2205)$.
The modulated spiralling wave ($msw$) from which $m^2sw$ bifurcates
has gone unstable and is therefore inaccessible. Nevertheless, it can
be safely presumed that it looks similar to $msw$ as shown for
$\Rey=2185$ in figure~\ref{fig:SSRPOk1.63Re2185} and that it must be
contained within the region of phase space delimited by the invariant
torus on which $m^2sw$ runs. Looking at the Poincar\'e section, it is
easy to identify the drift undergone by the wave after every
flight. The drift goes in a circle, appearing as an invariant cycle on
the Poincar\'e map.

\begin{figure}
  \begin{center}
    \begin{tabular}{cccc}
      \raisebox{0.33\linewidth}{(\textit{a})}\hspace{-0.6cm} &
      \includegraphics[height=0.32\linewidth,clip]{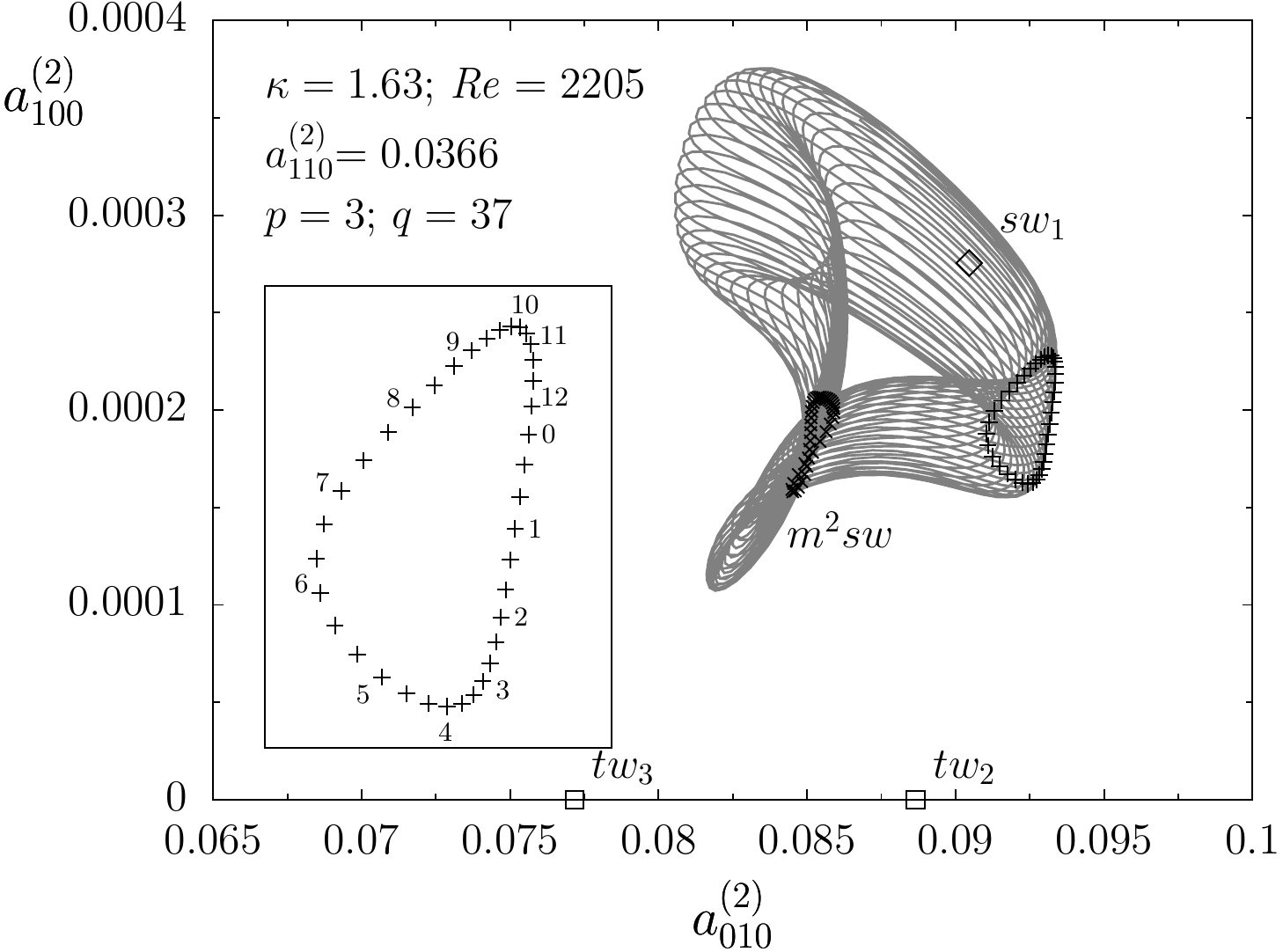} &
      \raisebox{0.33\linewidth}{(\textit{b})}\hspace{-0.6cm} &
      \includegraphics[height=0.31\linewidth,clip]{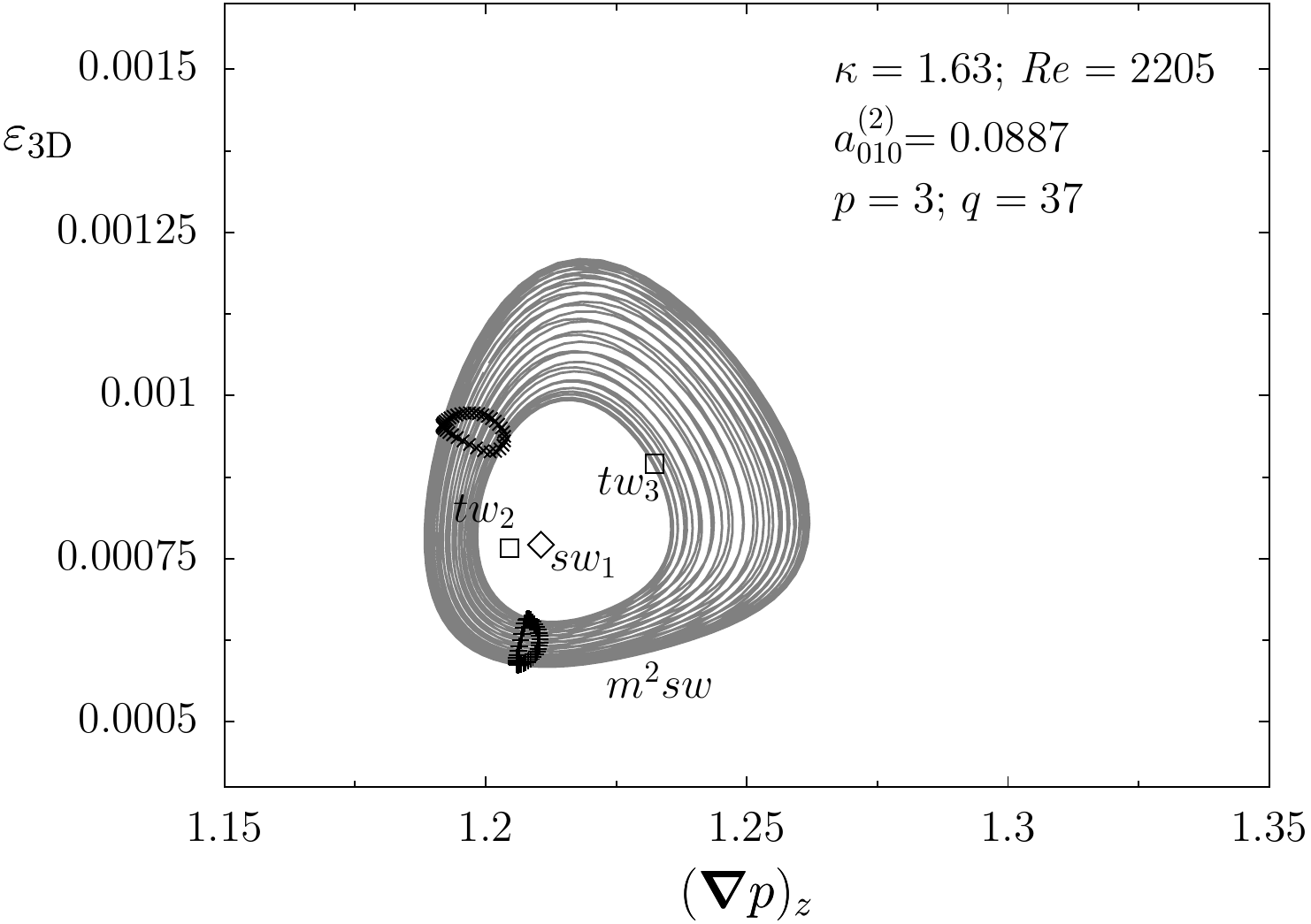}
    \end{tabular}
  \end{center}
  \caption{Phase map projections of a doubly-modulated spiralling wave
    ($m^2sw$) at $(\kappa,\Rey)=(1.63,2205)$. (\textit{a})
    $a_{100}^{(2)}$ vs $a_{010}^{(2)}$. The inset shows a Poincar\'e
    section defined by
    $a_{110}^{(2)}=a_{110}^{(2)}(tw_2)$. (\textit{b})
    Three-dimensional energy ($\varepsilon_{\rm 3D}$) vs axial
    pressure gradient ($(\bnabla p)_z$). The Poincar\'e section is
    defined by $a_{010}^{(2)}=a_{010}^{(2)}(tw_2)$. Middle-branch
    travelling waves ($tw_2$ and $tw_3$) have been labelled and marked
    with open squares and spiralling wave ($sw_1$) with an open
    diamond. Positive and negative crossings of Poincar\'e sections
    are indicated by plus signs and crosses, respectively.}
  \label{fig:PhMapRPTk1.63Re2205}
\end{figure}

Doubly-modulated waves will generally be dense-filling on the torus
(quasiperiodic), since no resonant conditions are enforced by any of
the symmetries of the problem and the new frequency does not
necessarily need to be commensurate with the previously existing
modulation frequency inherited from the modulated wave. However, as
the parameter is varied, Arnold tongues might be crossed and phase
locking occur producing parameter windows where periodic orbits
exist. The torus persists under small parameter variations but the
orbit structure will be dependent on whether the rotation number of
the map defined by Poincar\'e reduction is rational or irrational. As
a matter of fact, the wave shown in
figure~\ref{fig:PhMap3DRPTk1.63Re2205} exhibits a rational proportion
between the two frequencies, as can be seen in
figure~\ref{fig:PhMapRPTk1.63Re2205}.
The orbit structure and the Poincar\'e sections clearly show that the
wave is a $(3,37)$-cycle: the cycle makes $37$ revolutions along the
modulated wave trajectory (short modulation period) and $3$ around the
invariant curve on the Poincar\'e map (long modulation period) before
closing. The $37$th iterate of the Poincar\'e map is therefore a fixed
point. The existence of a stable $(3,37)$-cycle implies that an
unstable $(3,37)$-cycle must also exist
\cite[][]{Kuznetsov_B_95}. This is the only phase locking identified
in this study, meaning that all other doubly-modulated waves found are
dense or, at least, of extremely long period.

\begin{figure}
  \begin{center}
    \begin{tabular}{cccc}
      \raisebox{0.30\linewidth}{(\textit{a})}\hspace{-0.6cm} &
      \includegraphics[height=0.27\linewidth,clip]{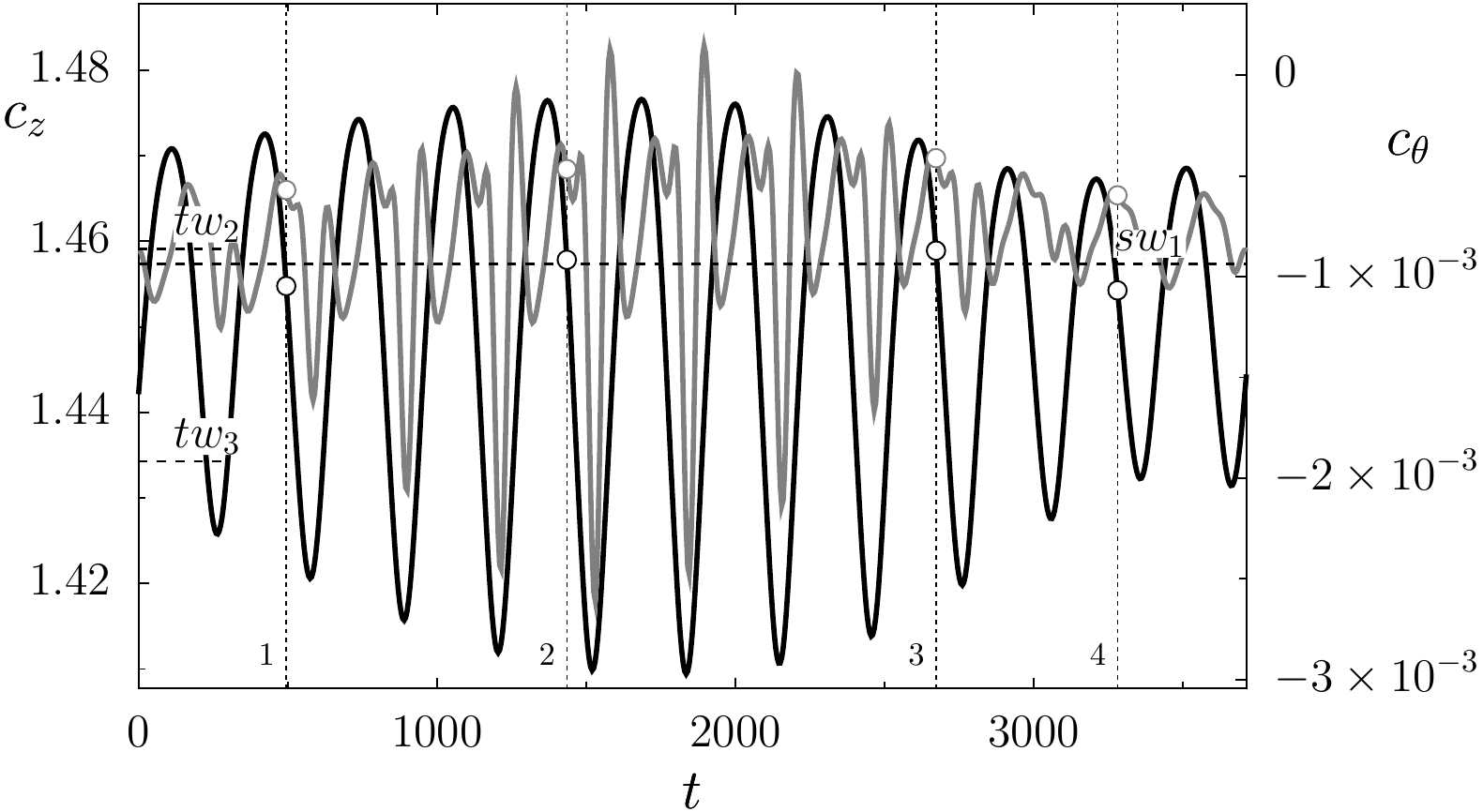} &
      \raisebox{0.30\linewidth}{(\textit{b})}\hspace{-0.6cm} &
      \includegraphics[height=0.28\linewidth,clip]{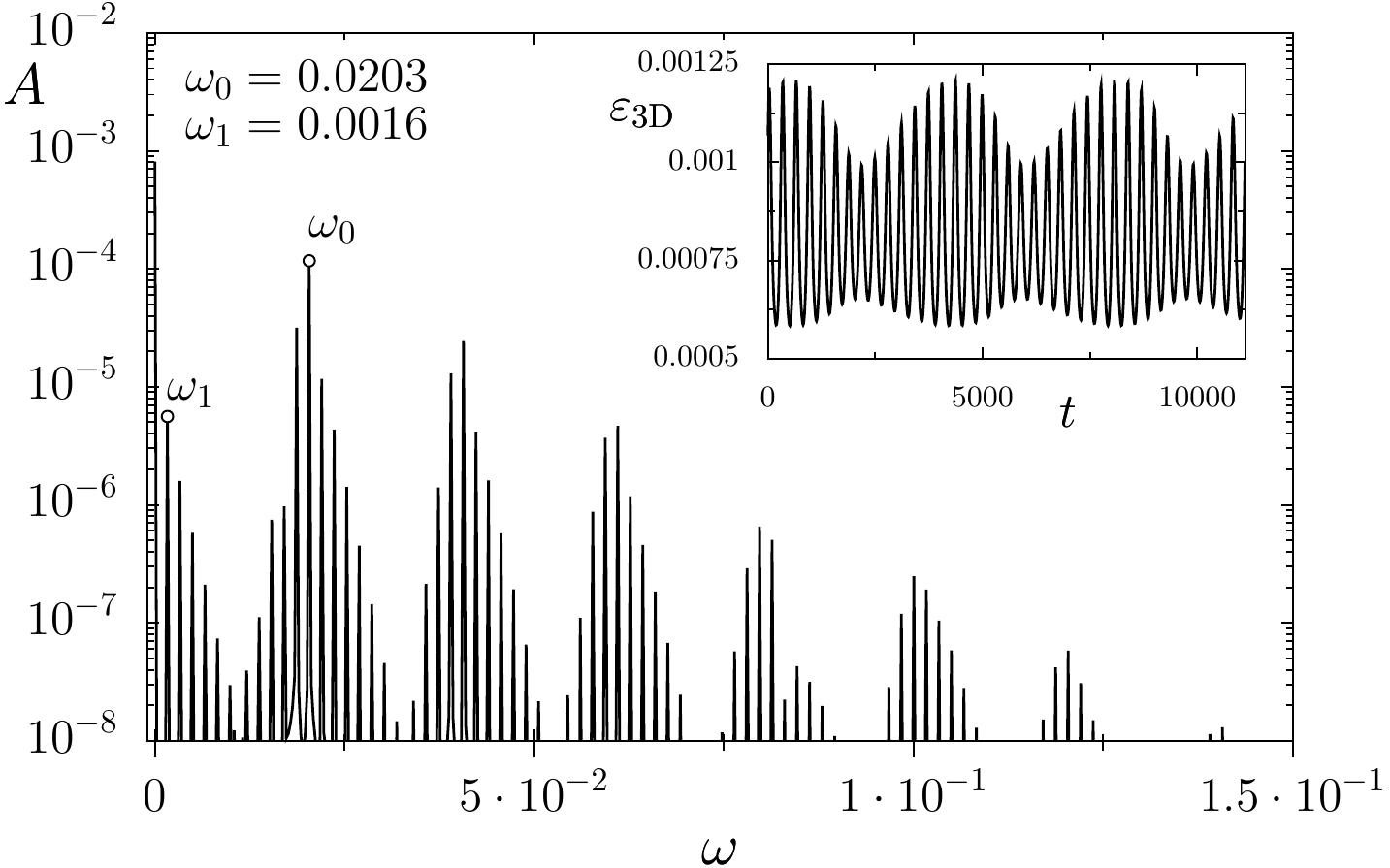}
    \end{tabular}
  \end{center}
  \caption{Doubly-modulated spiralling wave ($m^2sw$) at
    $(\kappa,\Rey)=(1.63,2205)$. (\textit{a}) Axial ($c_z$, black
    line) and azimuthal ($c_{\theta}$, gray line) phase speed
    time-series. Dotted horizontal lines indicate the values for
    $sw_1$, $tw_2$ and $tw_3$ while numbered vertical lines and open
    circles indicate snapshots in
    figure~\ref{fig:SSRPTk1.63Re2205}. (\textit{b}) Fourier transform
    of the non-axisymmetric streamwise-dependent modal energy contents
    ($\varepsilon_{\rm 3D}$). Part of the time signal is plotted in
    the inset.}
  \label{fig:TSFTRPTk1.63Re2205}
\end{figure}

The Neimark-Sacker modulational frequency is much higher than the
inherited Hopf frequency. Phase-speed time-series along a full cycle
around the Poincar\'e section for the wave at
$(\kappa,\Rey)=(1.63,2205)$ are shown in
figure~\ref{fig:TSFTRPTk1.63Re2205}(\textit{a}).
The signal is of about the same frequency as that in
figure~\ref{fig:TSFTRPOk1.63Re2185}(\textit{a}), but now, it is
slightly modulated with a much longer period. The spectrum of the
energy signal shown in figure~\ref{fig:TSFTRPTk1.63Re2205}(\textit{b})
features the same peaks, corresponding to the fast modulation
$\omega_0=0.0203$, as that of the modulated wave
(figure~\ref{fig:TSFTRPOk1.63Re2185}\textit{b}), except that
additional peaks at multiples of the slow frequency
$\omega_1=0.001649$ appear in the scene. The spectrum is still
discrete, as should be expected from a quasiperiodic signal, but now
two frequencies are identifiable. In this case, due to phase locking,
the ratio of frequencies is rational and coincides precisely with the
rotation number ($\rho=\omega_1/\omega_0=p/q=3/37$).

\begin{figure}
  \begin{center}
    \begin{tabular}{cccccc}
      \raisebox{0.16\linewidth}{($1$)}\hspace{-0.6cm} &
      \includegraphics[height=0.15\linewidth,clip]{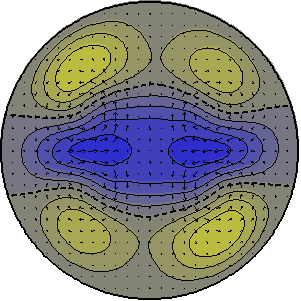} &
      \includegraphics[height=0.15\linewidth,clip]{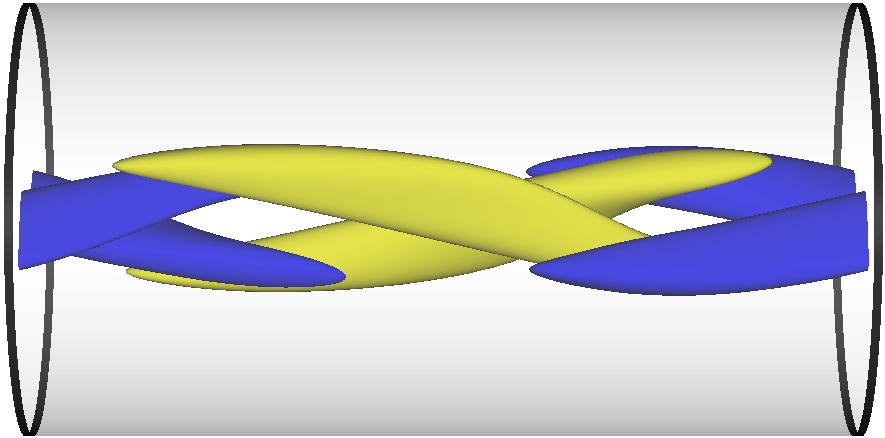} &
      \raisebox{0.16\linewidth}{($2$)}\hspace{-0.6cm} &
      \includegraphics[height=0.15\linewidth,clip]{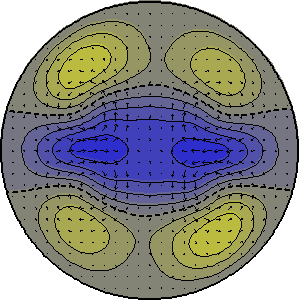} &
      \includegraphics[height=0.15\linewidth,clip]{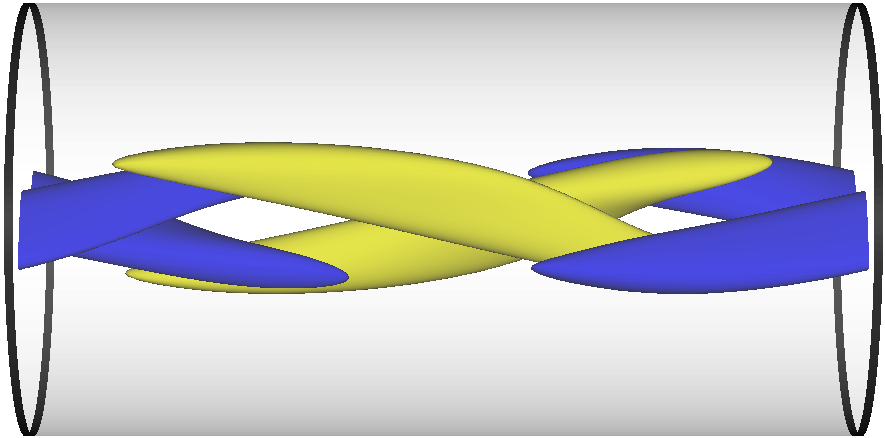}\\
      \raisebox{0.16\linewidth}{($3$)}\hspace{-0.6cm} &
      \includegraphics[height=0.15\linewidth,clip]{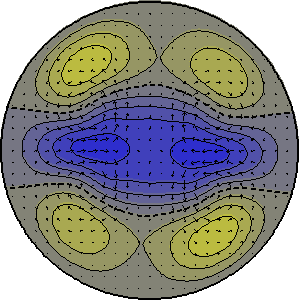} &
      \includegraphics[height=0.15\linewidth,clip]{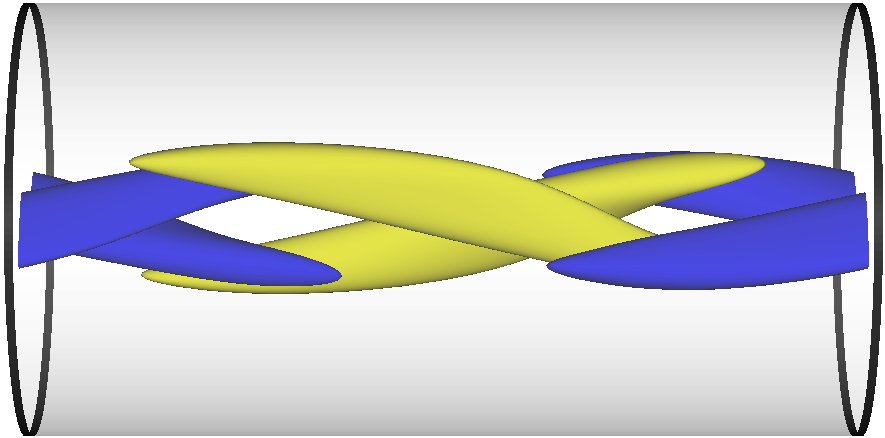} &
      \raisebox{0.16\linewidth}{($4$)}\hspace{-0.6cm} &
      \includegraphics[height=0.15\linewidth,clip]{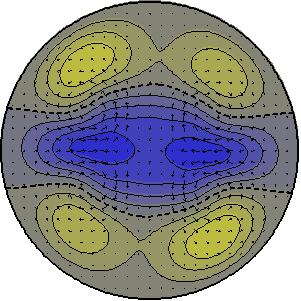} &
      \includegraphics[height=0.15\linewidth,clip]{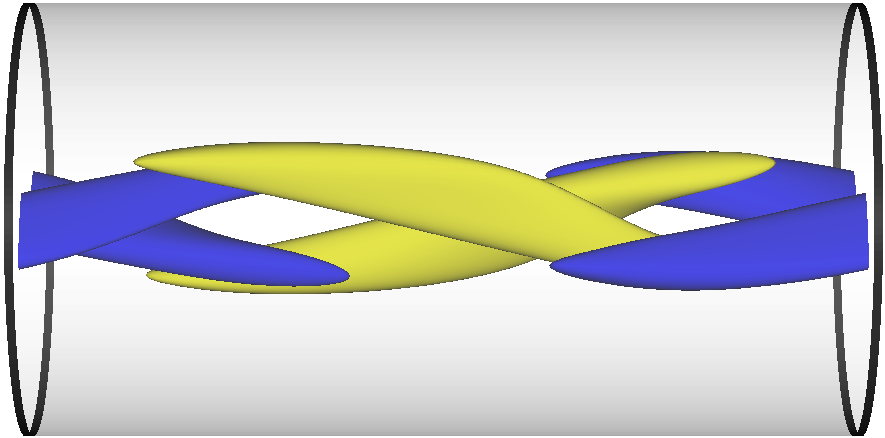}\\
    \end{tabular}
  \end{center}
  \caption{Doubly-modulated spiralling wave at
    $(\kappa,\Rey)=(1.63,2205)$. Left: $z$-averaged cross-sectional
    axial velocity contours spaced at intervals of $\Delta \langle
    u_z\rangle_z = 0.1 U$. In-plane velocity vectors are also
    displayed. Right: axial vorticity isosurfaces at $\omega_z=\pm 1
    U/D$. Fluid flows rightwards. Blue (dark gray) for negative,
    yellow (light) for positive. ($1$) $t=495$, ($2$) $t=1435$, ($3$)
    $t=2672$ and ($4$) $t=3280$ $D/(4U)$. To avoid drift due to
    streamwise advection, snapshots are taken in a comoving frame
    spiralling with the instantaneous advection speeds from
    figure~\ref{fig:TSFTRPTk1.63Re2205}(\textit{a}). The snapshots
    have been indicated with circles in
    figures~\ref{fig:PhMap3DRPTk1.63Re2205} and
    \ref{fig:TSFTRPTk1.63Re2205}(\textit{a}).}
  \label{fig:SSRPTk1.63Re2205}
\end{figure}

The slow period is very long ($T_1=2 \upi/\omega_1\simeq 3810.3 D/(4
U)$), which makes it difficult to select just a few snapshots that are
representative of the time dependence of the wave. Nevertheless, the
modulation along the short period has already been exemplified for a
modulated spiralling wave in figure~\ref{fig:SSRPOk1.63Re2185} and
should be fairly similar for a bifurcated doubly-modulated
wave. Therefore, we will focus here on the slow modulation effect by
representing the drift experienced by the wave every time the
Poincar\'e section is pierced. Figure~\ref{fig:SSRPTk1.63Re2205} (see
online movie) shows four such crossings of the Poincar\'e section
defined by $a_{010}^{(2)}=a_{010}^{(2)}(tw_2)$ as indicated in
figure~\ref{fig:PhMap3DRPTk1.63Re2205}.
The solution, which originally bifurcated at zero amplitude with
respect to the modulated spiralling wave, has barely left the linear
regime. Consequently, the modulation is extremely mild and only a
close inspection reveals that there is any modulation at all. The four
snapshots should be compared with figure~\ref{fig:SSRPOk1.63Re2185}($1$)
and seen as a modulation of it, since it represents the unique point
at which $msw$ crosses the Poincar\'e section.

Increasing $\Rey$ at $\kappa=1.63$ results in a significant swell of
the invariant torus and an ulterior catastrophic transition into a
chaotic attractor.

\subsection{Mildly chaotic spiralling waves}

Doubly-modulated travelling waves cease to exist abruptly and the new
dynamics are organised along pseudoperiodic trajectories. These
solutions largely preserve some of the features of the original torus
at some stages but depart frantically at some other
stages. Reattachment to the remnants of the torus never occurs at the
same point, introducing a mild degree of chaoticity.

\begin{figure}
  \begin{center}
    \includegraphics[height=0.5\linewidth,clip]{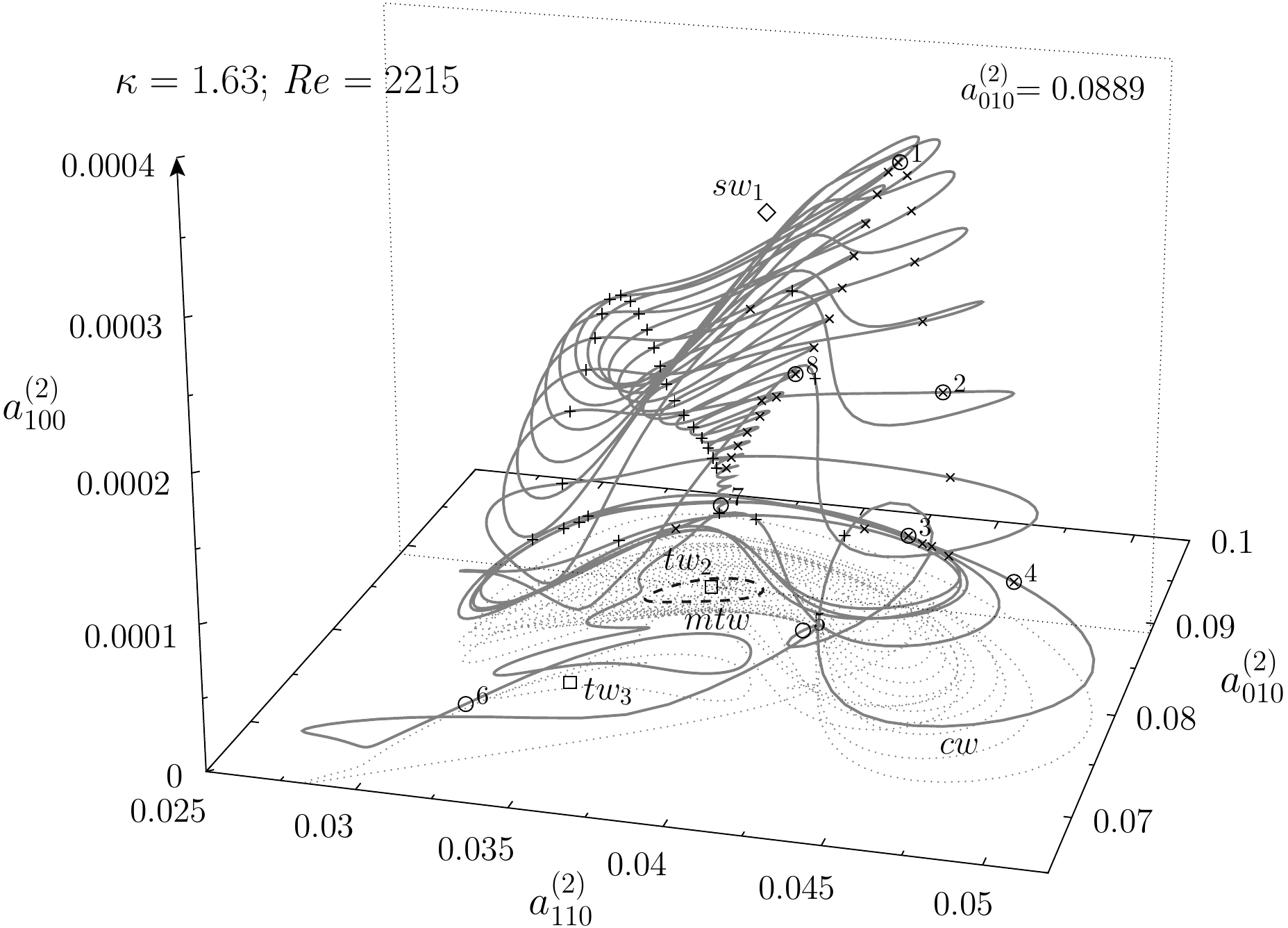}
  \end{center}
  \caption{Three-dimensional phase map projection
    $(a_{010}^{(2)},a_{110}^{(2)},a_{100}^{(2)})$ of a chaotic
    spiralling wave ($cw$) at $(\kappa,\Rey)=(1.63,2215)$ along a full
    pseudo-period. Middle-branch travelling waves ($tw_2$ and $tw_3$)
    have been labelled and marked with open squares and spiralling
    wave ($sw_1$) with an open diamond. A coexisting modulated
    travelling wave ($mtw$, black dashed line) has also been
    represented. Positive and negative crossings of a Poincar\'e
    section defined by $a_{010}^{(2)}=a_{010}^{(2)}(tw_2)$ are
    indicated by plus signs and crosses, respectively. Numbered
    circles correspond to snapshots in
    figure~\ref{fig:SSChaosk1.63Re2215}.}
  \label{fig:PhMap3DChaosk1.63Re2215}
\end{figure}

A full pseudo-period of a time-evolving chaotic spiralling wave ($cw$)
at $(\kappa,\Rey)=(1.63,2215)$ has been represented in
figure~\ref{fig:PhMap3DChaosk1.63Re2215}.
For a while, the orbit runs on the remnants of the invariant torus
that existed at lower $\Rey$
(figure~\ref{fig:PhMap3DRPTk1.63Re2205}). As it goes around, though,
the flow is unable to close as before, and the trajectory is thrown
away and gets hooked in a seemingly periodic motion as if it was
captured by the stable manifold of some periodic orbit. All this
happens at very low $a_{100}^{(2)}$, which is indicative of a close
approach to the shift-reflect subspace. After some turns, the
trajectory departs again with a violent thrust and then reconnects
back onto what seems to be the unstable manifold of $tw_2$. Finally,
following this manifold, the trajectory makes its trip back onto the
torus-dominated region and the process starts anew.

\begin{figure}
  \begin{center}
    \begin{tabular}{cccc}
      \raisebox{0.33\linewidth}{(\textit{a})}\hspace{-0.6cm} &
      \includegraphics[height=0.32\linewidth,clip]{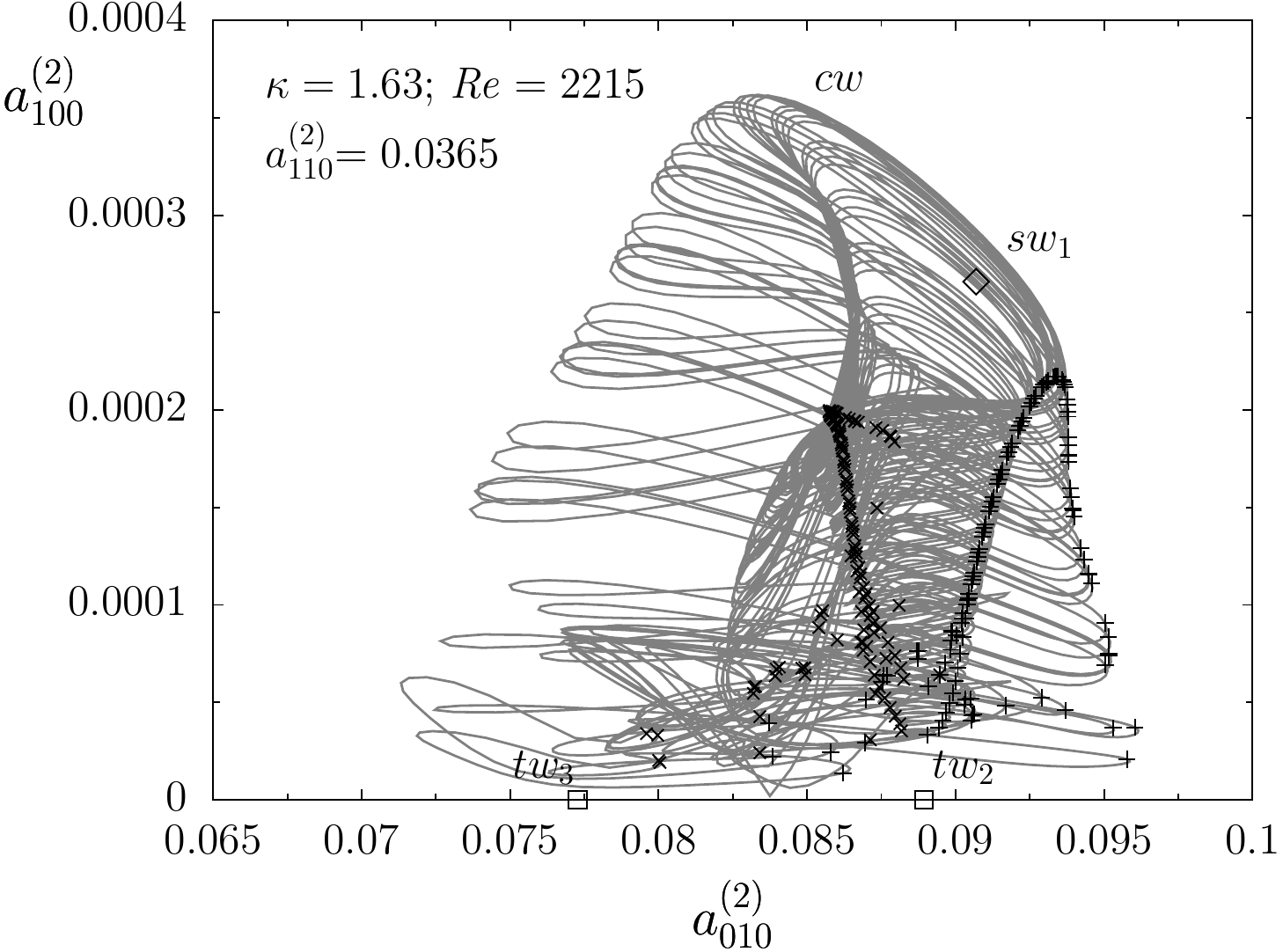} &
      \raisebox{0.33\linewidth}{(\textit{b})}\hspace{-0.6cm} &
      \includegraphics[height=0.31\linewidth,clip]{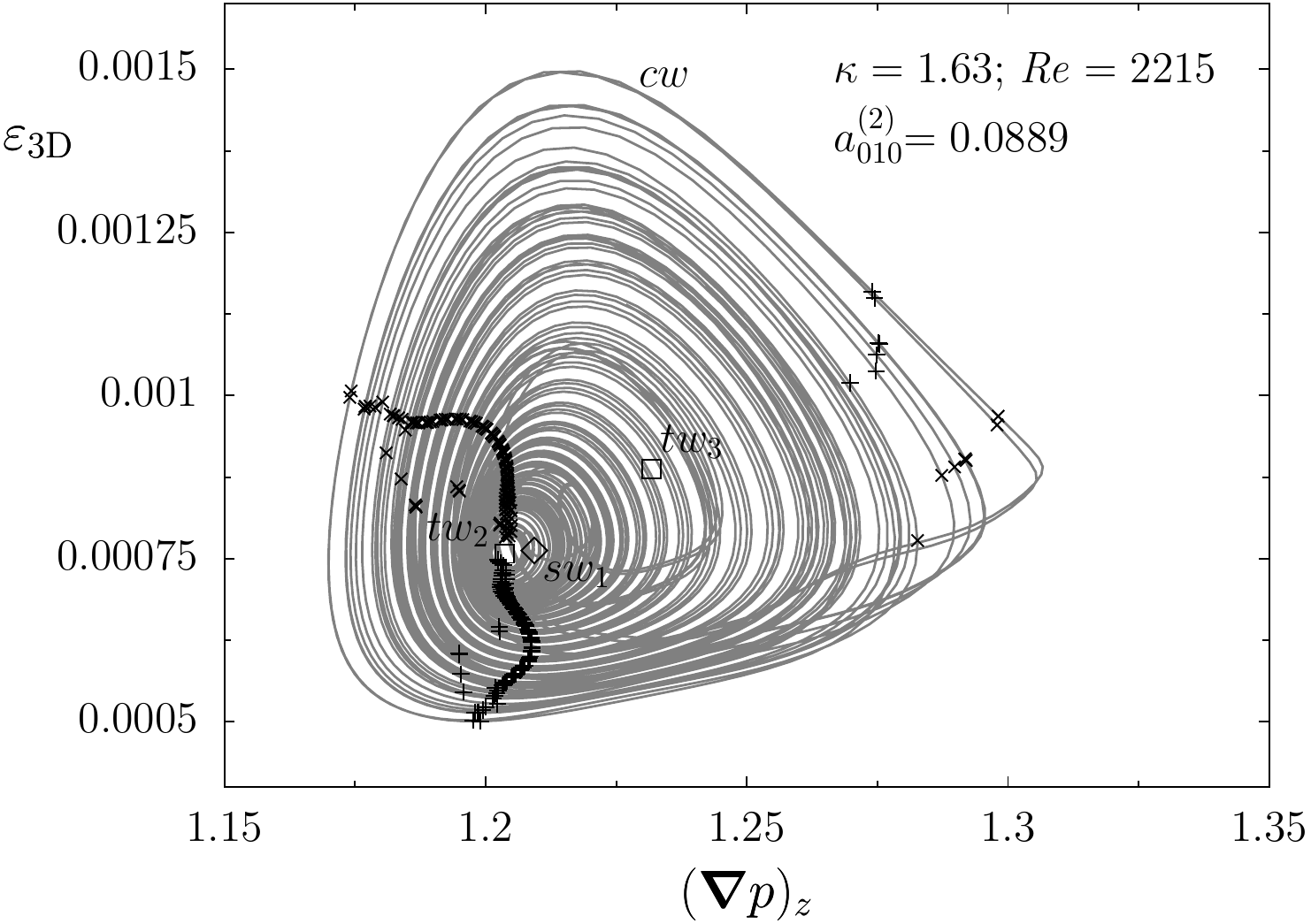}
    \end{tabular}
  \end{center}
  \caption{Phase map projections of a chaotic spiralling wave ($cw$)
    at $(\kappa,\Rey)=(1.63,2215)$ along several
    pseudo-periods. (\textit{a}) $a_{100}^{(2)}$ vs
    $a_{010}^{(2)}$. Marks indicate crossings of a Poincar\'e section
    defined by $a_{110}^{(2)}=a_{110}^{(2)}(tw_2)$. (\textit{b})
    Three-dimensional energy ($\varepsilon_{\rm 3D}$) vs axial
    pressure gradient ($(\bnabla p)_z$). The Poincar\'e section is
    defined by $a_{010}^{(2)}=a_{010}^{(2)}(tw_2)$. Middle-branch
    travelling waves ($tw_2$ and $tw_3$) have been labelled and marked
    with open squares and spiralling wave ($sw_1$) with an open
    diamond. Positive and negative crossings of Poincar\'e sections
    are indicated by plus signs and crosses, respectively.}
  \label{fig:PhMapChaosk1.63Re2215}
\end{figure}

Several pseudo-periods of the wave have been represented in a couple
of phase map projections (figure~\ref{fig:PhMapChaosk1.63Re2215}) to
illustrate its mildly chaotic character.
The Poincar\'e map in
figure~\ref{fig:PhMapChaosk1.63Re2215}(\textit{a}) tends to be dense
and deterministic at the upper tip, where the solution runs on the
torus (to be compared with
figure~\ref{fig:PhMapRPTk1.63Re2205}\textit{a}). Meanwhile, the lower
side features a chaotic cloud of random crossings. Furthermore, the
return funnel that gets the trajectory back on the torus exhibits a
variable width that can get as narrow as in
figure~\ref{fig:PhMap3DChaosk1.63Re2215}. The $\varepsilon_{\rm 3D}$
vs $(\bnabla p)_z$ phase map also gives a clear view on the fate of
the torus. While at $\Rey=2205$
(figure~\ref{fig:PhMapRPTk1.63Re2205}\textit{b}) the doughnut shape
with closed $mtw$-centred orbits on the Poincar\'e sections was clearly
identifiable, at $\Rey=2215$
(figure~\ref{fig:PhMapChaosk1.63Re2215}\textit{b}), the hole has been
disrupted and long excursions to high three-dimensional energy values
and large axial pressure gradients occur.

\begin{figure}
  \begin{center}
    \begin{tabular}{cccc}
      \raisebox{0.30\linewidth}{(\textit{a})}\hspace{-0.6cm} &
      \includegraphics[height=0.27\linewidth,clip]{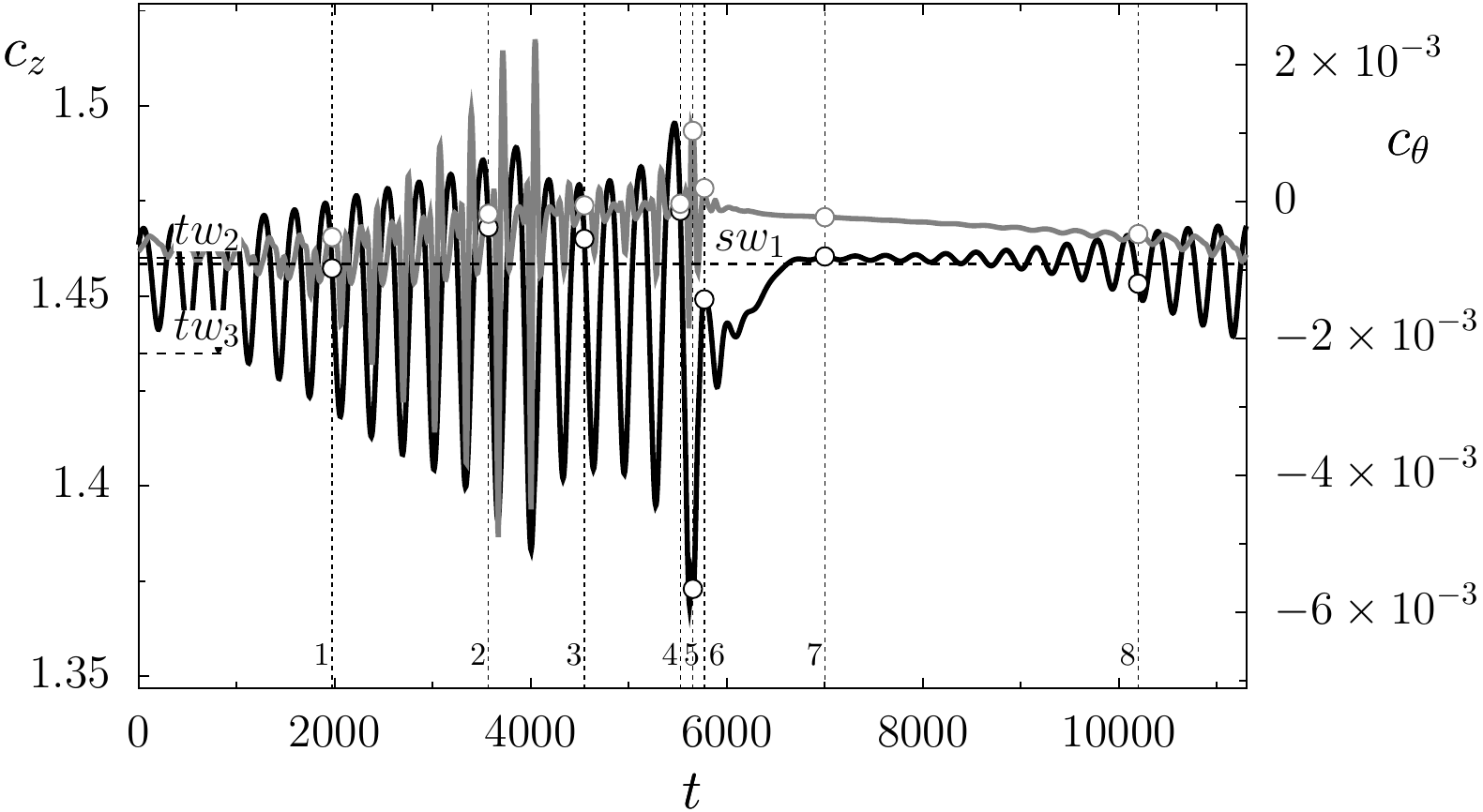} &
      \raisebox{0.30\linewidth}{(\textit{b})}\hspace{-0.6cm} &
      \includegraphics[height=0.28\linewidth,clip]{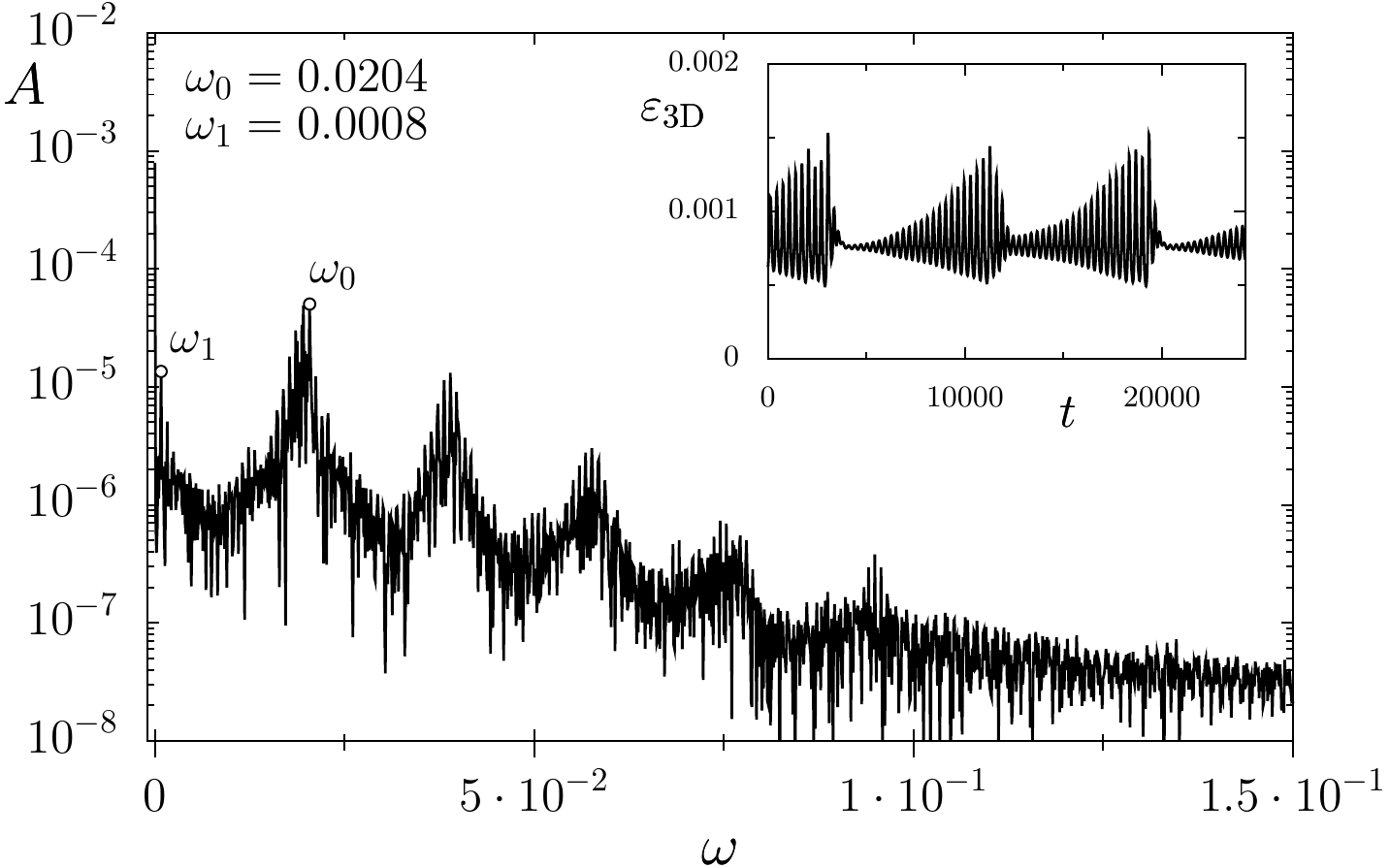}
    \end{tabular}
  \end{center}
  \caption{Chaotic spiralling wave ($cw$) at
    $(\kappa,\Rey)=(1.63,2215)$. (\textit{a}) Axial ($c_z$, black
    line) and azimuthal ($c_{\theta}$, gray line) phase speed
    time-series. Dotted horizontal lines indicate the values for
    $sw_1$, $tw_2$ and $tw_3$ while numbered vertical lines and open
    circles indicate snapshots in
    figure~\ref{fig:SSChaosk1.63Re2215}. (\textit{b}) Fourier
    transform of the non-axisymmetric streamwise-dependent modal
    energy contents ($\varepsilon_{\rm 3D}$). Part of the time signal
    is plotted in the inset.}
  \label{fig:TSFTChaosk1.63Re2215}
\end{figure}

Time signals preserve the fast frequency associated to the old Hopf
instability, while the slow frequency resulting from the
Neimark-Sacker instability is exchanged for a longer period modulation
that is no longer constant. Phase speed time-series for a
pseudo-period of the chaotic wave at $(\kappa,\Rey)=(1.63,2215)$ have
been drawn in figure~\ref{fig:TSFTChaosk1.63Re2215}(\textit{a}).
Both signals undergo a steady amplitude growth before becoming
somewhat chaotic. Phase speed modulations get fairly large just before
the chaotic transient, with the azimuthal phase speed ($c_{\theta}$)
even changing sign, which means that at some intervals the wave
becomes retrograde. Towards the end of the chaotic transient, the
azimuthal phase speed vanishes as the wave approaches the
shift-reflect subspace and, shortly after, the axial phase speed
($c_z$) evolves towards lower-middle-branch travelling-wave values
($tw_2$). From there on, the wave evolves as if it was following the
combined pitchfork-Hopf-unstable manifold of $tw_2$ and the process
restarts. It is remarkable how, despite the close visit to the
shift-reflect subspace, and even if the azimuthal speed changes sign,
the wave always makes its way back to the same side of phase
space. There are no reversals, contrary to what happens for the
Lorenz attractor \cite*[][]{Strogatz_B_94}, and the
shift-reflect-conjugate side of phase space is never
visited. Therefore, a shift-reflect-symmetric chaotic attractor
spiralling with opposite rotation rate exists and the two attractors
do not interact.

The inset of figure~\ref{fig:TSFTChaosk1.63Re2215}(\textit{b})
illustrates the variable nature of the long modulational period. The
spectral energy density of the signal reflects this variability. There
still are clear peaks at the high frequency $\omega_0=0.0204
\; (4U/D)$ and its harmonics, largely preserved from modulated and
doubly-modulated spiralling waves. The long period modulation, though,
is no longer represented by a discrete peak, but by a certain
dispersion around $\omega_1 \simeq 0.0008 \; (4U/D)$. The
pseudo-period is much longer than for doubly-modulated waves due to
the fact that the wave spends large amounts of time in a region of
phase space not visited by $m^2sw$. The effect of the dispersion is
that the wave ceases to be quasiperiodic and the spectrum becomes
continuous.

\begin{figure}
  \begin{center}
    \begin{tabular}{cccccc}
      \raisebox{0.16\linewidth}{($1$)}\hspace{-0.6cm} &
      \includegraphics[height=0.15\linewidth,clip]{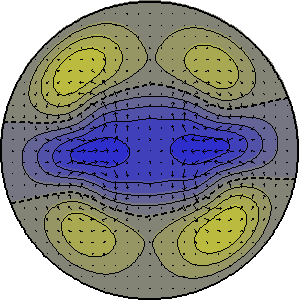} &
      \includegraphics[height=0.15\linewidth,clip]{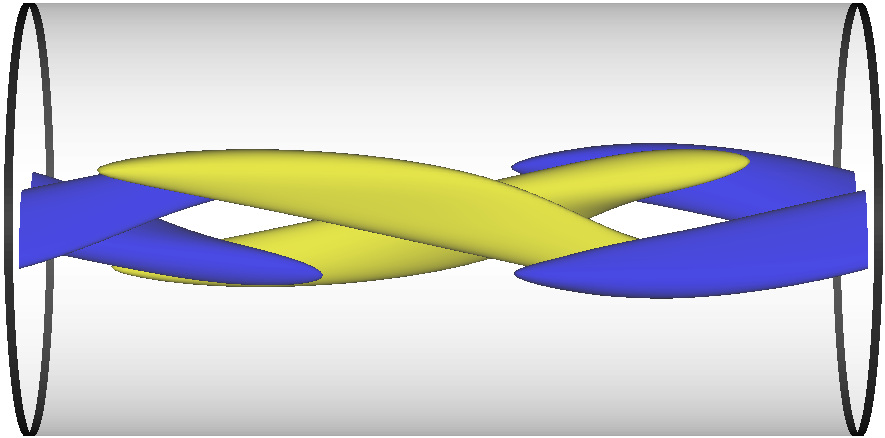} &
      \raisebox{0.16\linewidth}{($2$)}\hspace{-0.6cm} &
      \includegraphics[height=0.15\linewidth,clip]{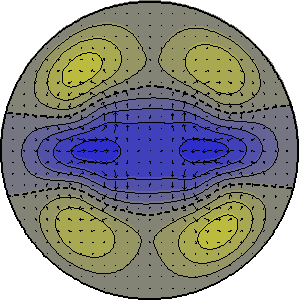} &
      \includegraphics[height=0.15\linewidth,clip]{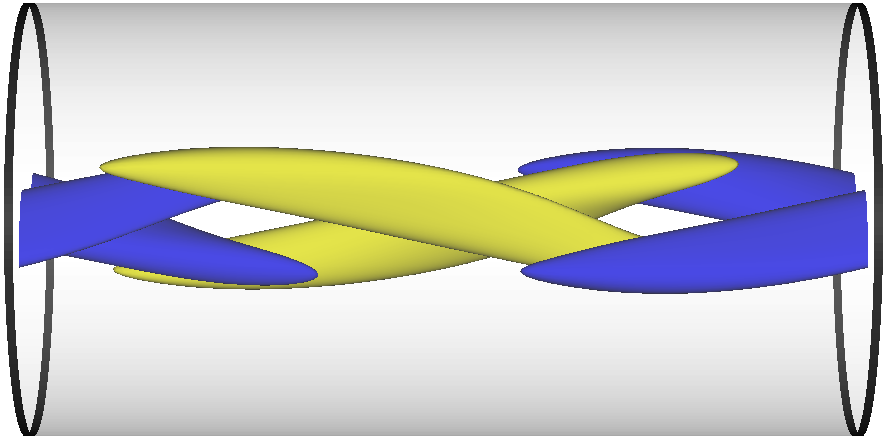}\\

      \raisebox{0.16\linewidth}{($3$)}\hspace{-0.6cm} &
      \includegraphics[height=0.15\linewidth,clip]{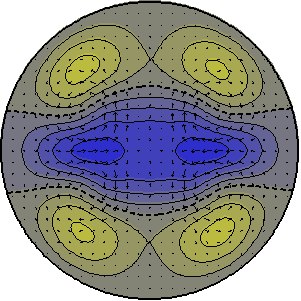} &
      \includegraphics[height=0.15\linewidth,clip]{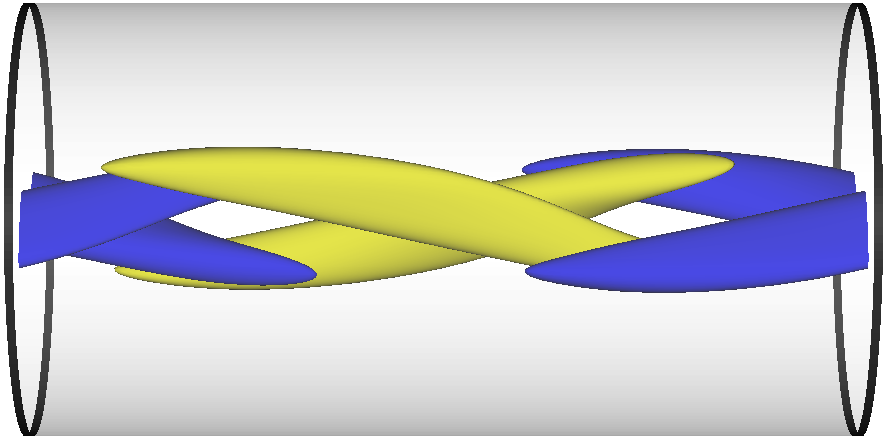} &
      \raisebox{0.16\linewidth}{($4$)}\hspace{-0.6cm} &
      \includegraphics[height=0.15\linewidth,clip]{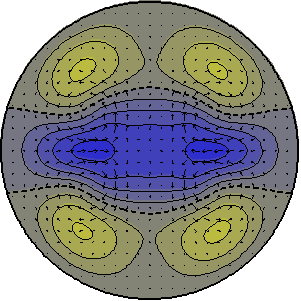} &
      \includegraphics[height=0.15\linewidth,clip]{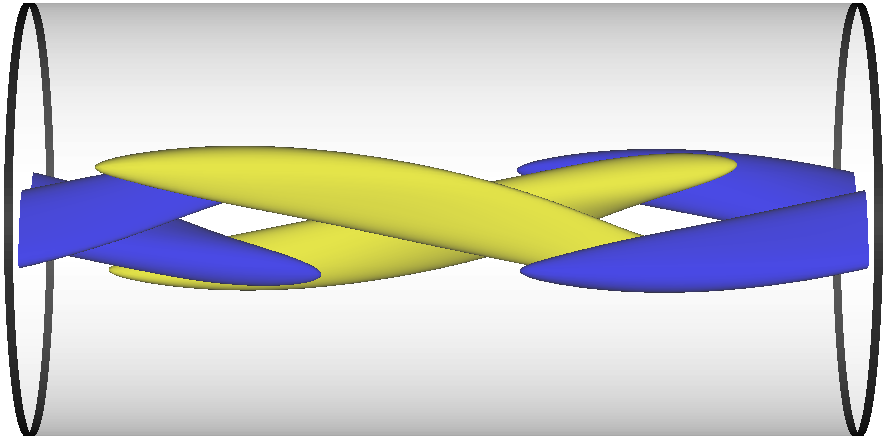}\\

      \raisebox{0.16\linewidth}{($5$)}\hspace{-0.6cm} &
      \includegraphics[height=0.15\linewidth,clip]{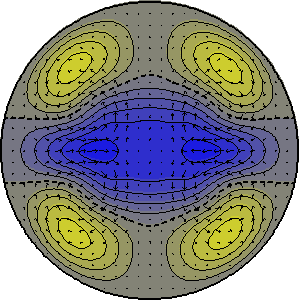} &
      \includegraphics[height=0.15\linewidth,clip]{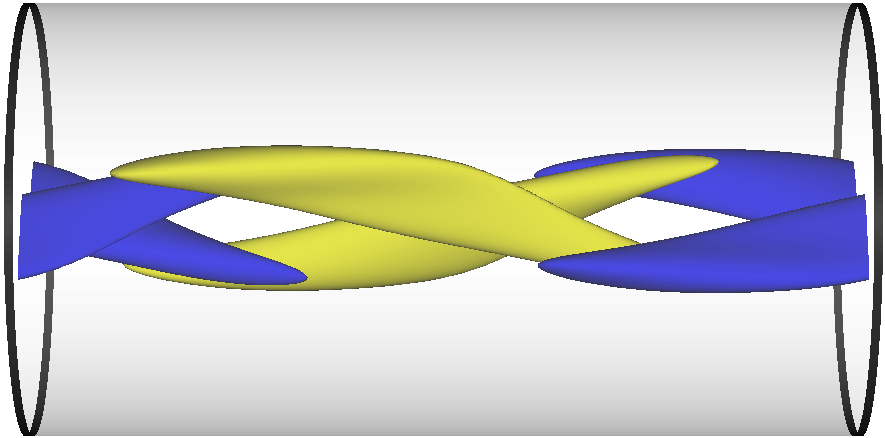} &
      \raisebox{0.16\linewidth}{($6$)}\hspace{-0.6cm} &
      \includegraphics[height=0.15\linewidth,clip]{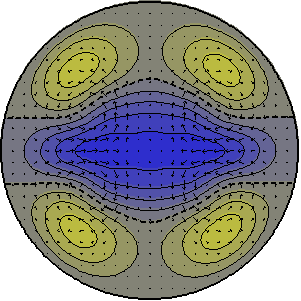} &
      \includegraphics[height=0.15\linewidth,clip]{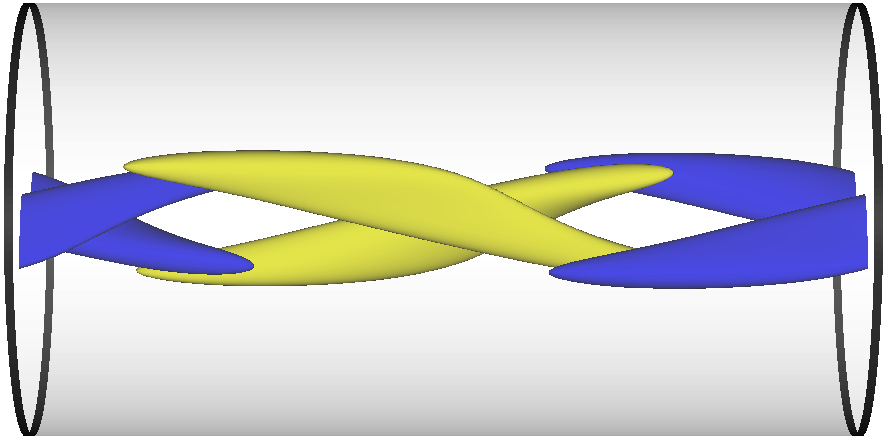}\\

      \raisebox{0.16\linewidth}{($7$)}\hspace{-0.6cm} &
      \includegraphics[height=0.15\linewidth,clip]{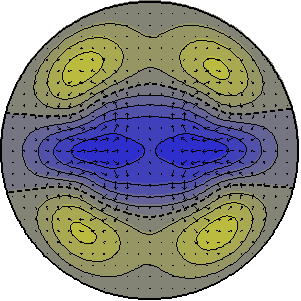} &
      \includegraphics[height=0.15\linewidth,clip]{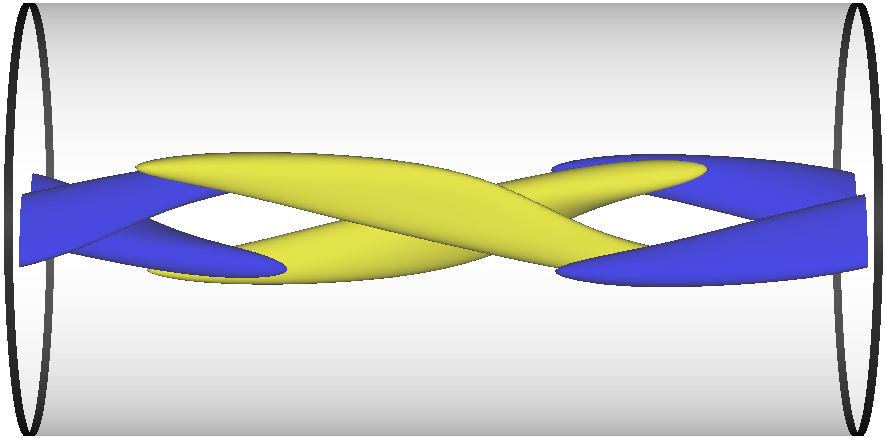} &
      \raisebox{0.16\linewidth}{($8$)}\hspace{-0.6cm} &
      \includegraphics[height=0.15\linewidth,clip]{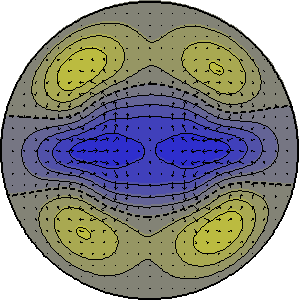} &
      \includegraphics[height=0.15\linewidth,clip]{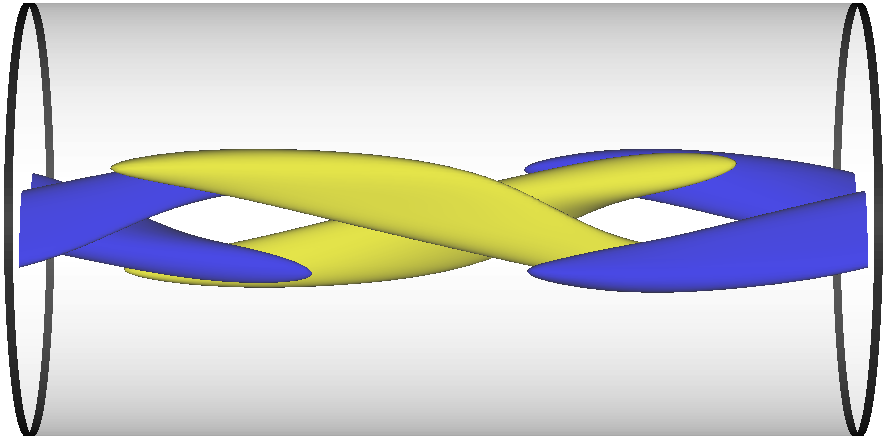}\\
    \end{tabular}
  \end{center}
  \caption{Chaotic spiralling wave at
    $(\kappa,\Rey)=(1.63,2215)$. Left: $z$-averaged cross-sectional
    axial velocity contours spaced at intervals of $\Delta \langle
    u_z\rangle_z = 0.1 U$. In-plane velocity vectors are also
    displayed. Right: axial vorticity isosurfaces at $\omega_z=\pm 1
    U/D$. Fluid flows rightwards. Blue (dark gray) for negative,
    yellow (light) for positive. ($1$) $t=1974$, ($2$) $t=3569$, ($3$)
    $t=4547$, ($4$) $t=5526$ , ($5$) $t=5654$ , ($6$) $t=5771$ , ($7$)
    $t=7000$ and ($8$) $t=10196$ $D/(4U)$. To avoid drift due to
    streamwise advection, snapshots are taken in a comoving frame
    spiralling with the instantaneous advection speeds from
    figure~\ref{fig:TSFTChaosk1.63Re2215}(\textit{a}). The snapshots
    have been indicated with circles in
    figures~\ref{fig:PhMap3DChaosk1.63Re2215} and
    \ref{fig:TSFTChaosk1.63Re2215}(\textit{a}).}
  \label{fig:SSChaosk1.63Re2215}
\end{figure}

To portray the time-dependence of the flow field, several snapshots of
the chaotic spiralling wave along a pseudo-period have been gathered
in figure~\ref{fig:SSChaosk1.63Re2215}.
To reduce the number of snapshots to a minimum and at the same time
allow comparison with $m^2sw$, flow fields have been represented
either on the same Poincar\'e section as in
figure~\ref{fig:SSRPTk1.63Re2205} (snapshots $1$-$4$ and $8$) or along
the violent escape in the vicinity of the shift-reflect subspace
(snapshots $5$-$7$).

Snapshots ($1$), ($2$) and ($8$) clearly correspond to time-instants
at which the trajectory runs on the torus. It is not surprising that
they look similar to the corresponding snapshots ($1$, $2$ and $4$ in
figure~\ref{fig:SSRPTk1.63Re2205}) of $m^2sw$. Snapshots ($3$) and
($4$), notwithstanding the fact that they correspond to Poincar\'e
crossings, exemplify new dynamics. They are in a region of phase space
away from where the torus existed and very close to the shift-reflect
subspace, as is apparent from the clearly identifiable quasi-symmetry
planes of the cross-sections. Snapshot ($3$) belongs in the region
where $cw$ seems to get hooked on a periodic orbit, while snapshot
($4$), taken at the time the violent excursion starts, seems to
confirm it, as it holds great similarity with ($3$). The axial
pressure gradient peak in
figure~\ref{fig:PhMapChaosk1.63Re2215}(\textit{b}) is represented by
snapshot ($5$), where the streaks are clearly very strong and the
shift-reflect symmetry fairly clear. Snapshot ($6$) shows the closest
visit to the shift-reflect subspace and is, at the same time, a
disputably close visit to $tw_3$ (figure~\ref{fig:tws}\textit{c}) as
indicated by both cross-sectional axial velocity contours and
axial-vorticity isosurfaces. ($4$)-($6$) occurs in a small fraction
of the full trajectory, which ultimately justifies the description of
the escape as violent. The shift-reflect symmetry starts to be
disrupted again in snapshot ($7$) and the wave evolves slowly as if
departing away from $tw_2$ along its unstable pitchfork-Hopf
manifold. The flow field in ($7$) is clearly half way between $tw_2$
(figure~\ref{fig:tws}\textit{b}) and $sw_1$
(figure~\ref{fig:sws}\textit{a}). At ($8$) the symmetry has been
completely broken and the flow field reattaches to the remnants of the
invariant torus.

\subsection{Phase-locked wave}

As in many systems \cite[see, \eg logistic map,][]{Strogatz_B_94}
where a chaotic attractor develops, there exist parameter intervals
where the solution locks onto a stable periodic orbit. These are
called periodic windows, and the chaotic set described above is no
exception.

\begin{figure}
  \begin{center}
    \includegraphics[height=0.5\linewidth,clip]{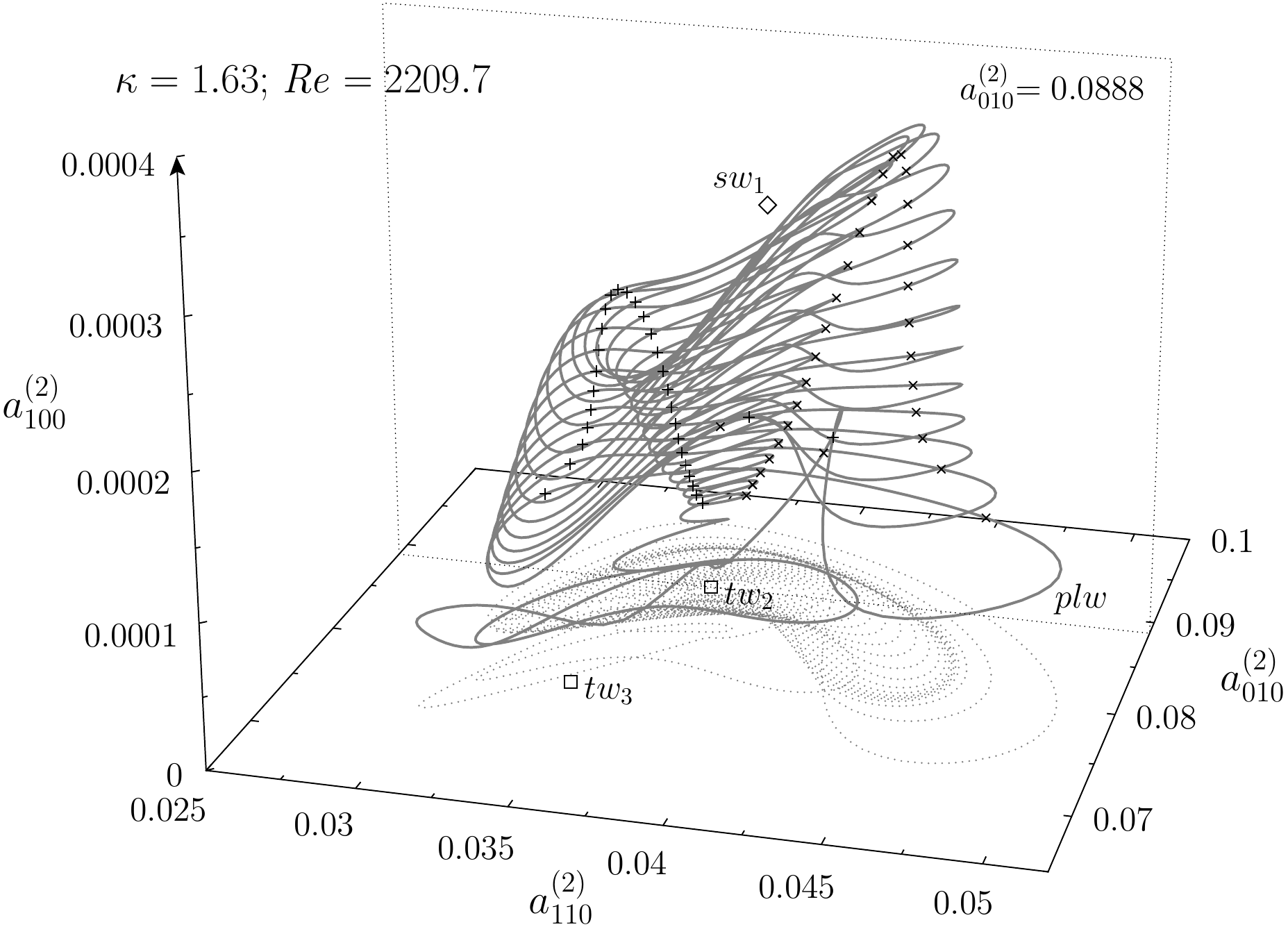}
  \end{center}
  \caption{Three-dimensional phase map projection
    $(a_{010}^{(2)},a_{110}^{(2)},a_{100}^{(2)})$ of a phase-locked
    spiralling wave ($plw$) at $(\kappa,\Rey)=(1.63,2209.7)$ along a
    full period. Middle-branch travelling waves ($tw_2$ and $tw_3$)
    have been labelled and marked with open squares and spiralling
    wave ($sw_1$) with an open diamond. Positive and negative
    crossings of a Poincar\'e section defined by
    $a_{010}^{(2)}=a_{010}^{(2)}(tw_2)$ are indicated by plus signs
    and crosses, respectively.}
  \label{fig:PhMap3DRTPOk1.63Re2209.7}
\end{figure}

At $\Rey=2209.7$, very slightly above torus breakdown and creation of
the chaotic set, and for a tiny interval, the flow becomes
periodic. The periodic trajectory, which looks very similar to the
pseudoperiodic trajectory at $\Rey=2215$, has been represented in
figure~\ref{fig:PhMap3DRTPOk1.63Re2209.7}.

\begin{figure}
  \begin{center}
    \begin{tabular}{cccc}
      \raisebox{0.30\linewidth}{(\textit{a})}\hspace{-0.6cm} &
      \includegraphics[height=0.27\linewidth,clip]{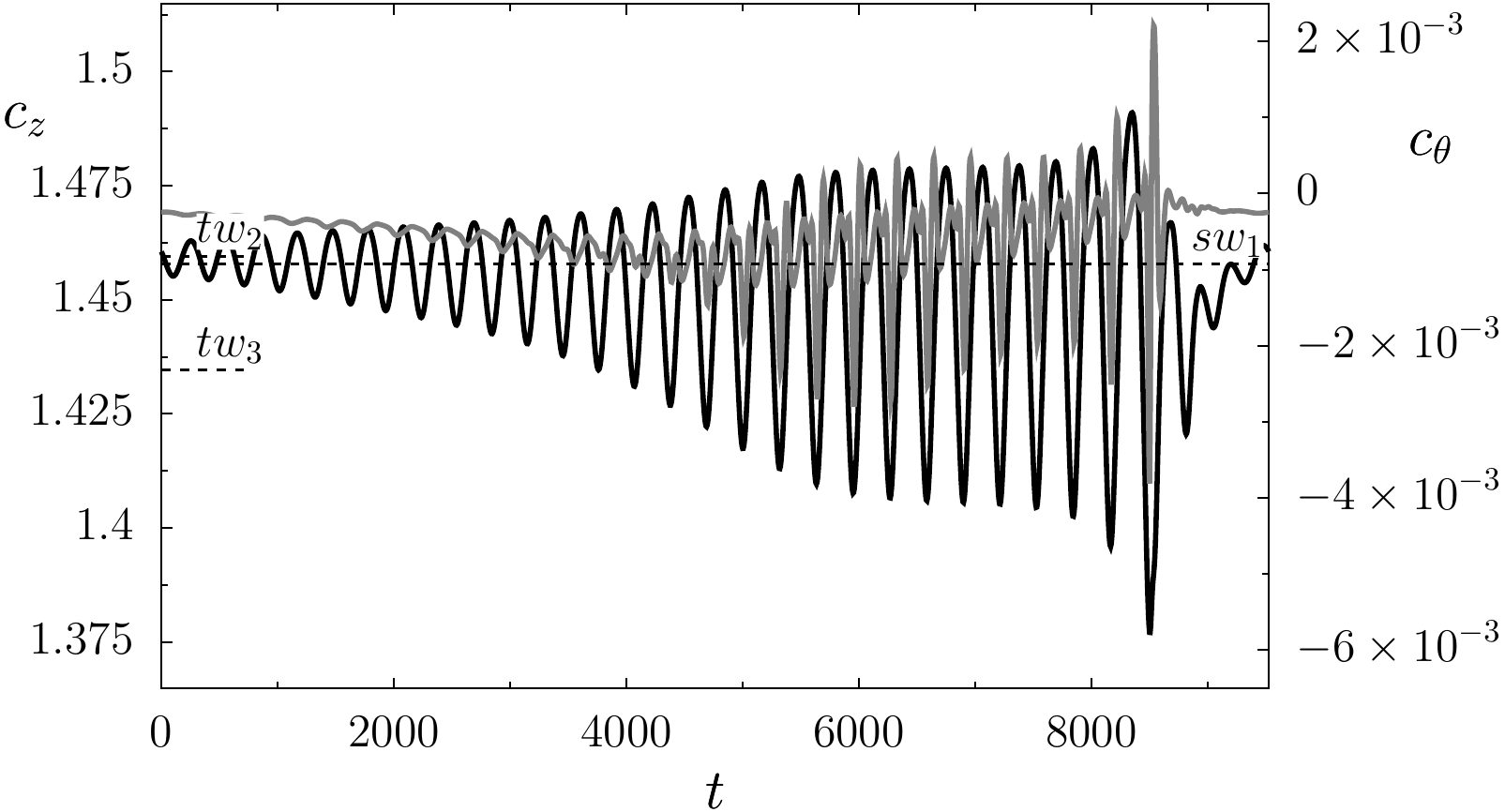} &
      \raisebox{0.30\linewidth}{(\textit{b})}\hspace{-0.6cm} &
      \includegraphics[height=0.28\linewidth,clip]{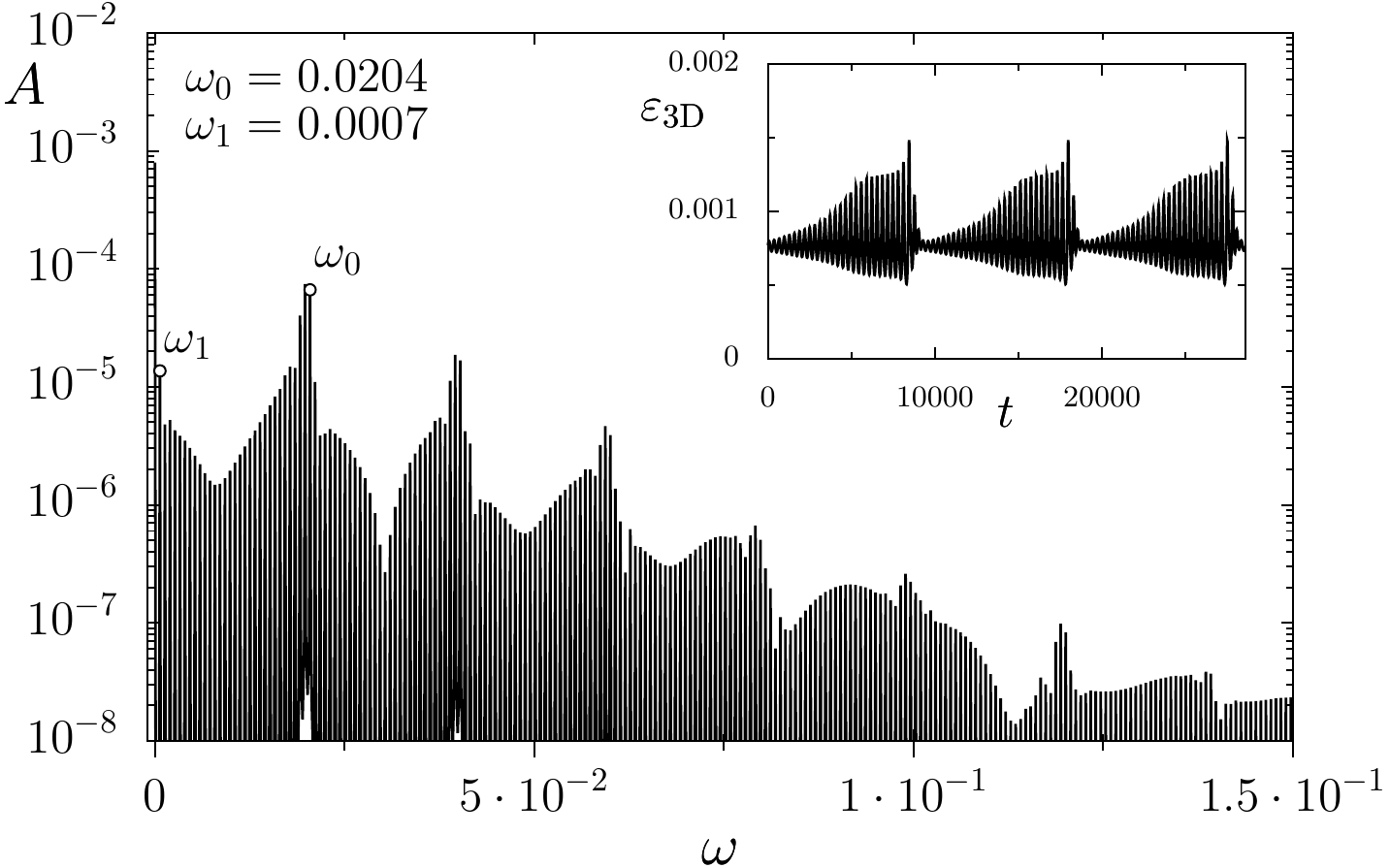}
    \end{tabular}
  \end{center}
  \caption{Phase-locked spiralling wave ($plw$) at
    $(\kappa,\Rey)=(1.63,2209.7)$. (\textit{a}) Axial ($c_z$, black
    line) and azimuthal ($c_{\theta}$, gray line) phase speed
    time-series. Dotted horizontal lines indicate the values for
    $sw_1$, $tw_2$ and $tw_3$. (\textit{b}) Fourier transform of the
    non-axisymmetric streamwise-dependent modal energy contents
    ($\varepsilon_{\rm 3D}$). Part of the time signal is plotted in
    the inset.}
  \label{fig:TSFTRTPOk1.63Re2209.7}
\end{figure}

Phase speeds have been plotted in
figure~\ref{fig:TSFTRTPOk1.63Re2209.7}(\textit{a}) for comparison with
$cw$.
Chaos disappears but the invariant torus is not reestablished and the
shift-reflect escape is preserved. The slowly evolving interval that
we identified as an approach to the unstable manifold of $tw_2$ is no
longer as clear, and the return funnel is wide from the beginning, so
that amplitude modulation is present all along. Despite the close
resemblance of all time-signals to some of the pseudo-periods at
$\Rey=2215$, time-stamps are now strictly periodic and the Fourier
transform of three-dimensional energy time-series is again discrete as
was the case at $\Rey=2205$. The fast frequency $\omega_0=0.0204
\; (4U/D)$ is clearly identifiable with the Hopf instability, while
the slow frequency $\omega_1=0.0006596 \; (4U/D)$ is inherited from
$cw$ but now takes a sharp value and fits exactly an integer number of
times in $\omega_0$.

We do not provide snapshots of the flow structures at selected times
along a period because of the close resemblance they bear with those
shown for $\Rey=2215$ in figure~\ref{fig:SSChaosk1.63Re2215}.

\subsection{Turbulent transients}

\begin{figure}
  \begin{center}
    \begin{tabular}{cccc}
      \raisebox{0.33\linewidth}{(\textit{a})}\hspace{-0.6cm} &
      \includegraphics[height=0.32\linewidth,clip]{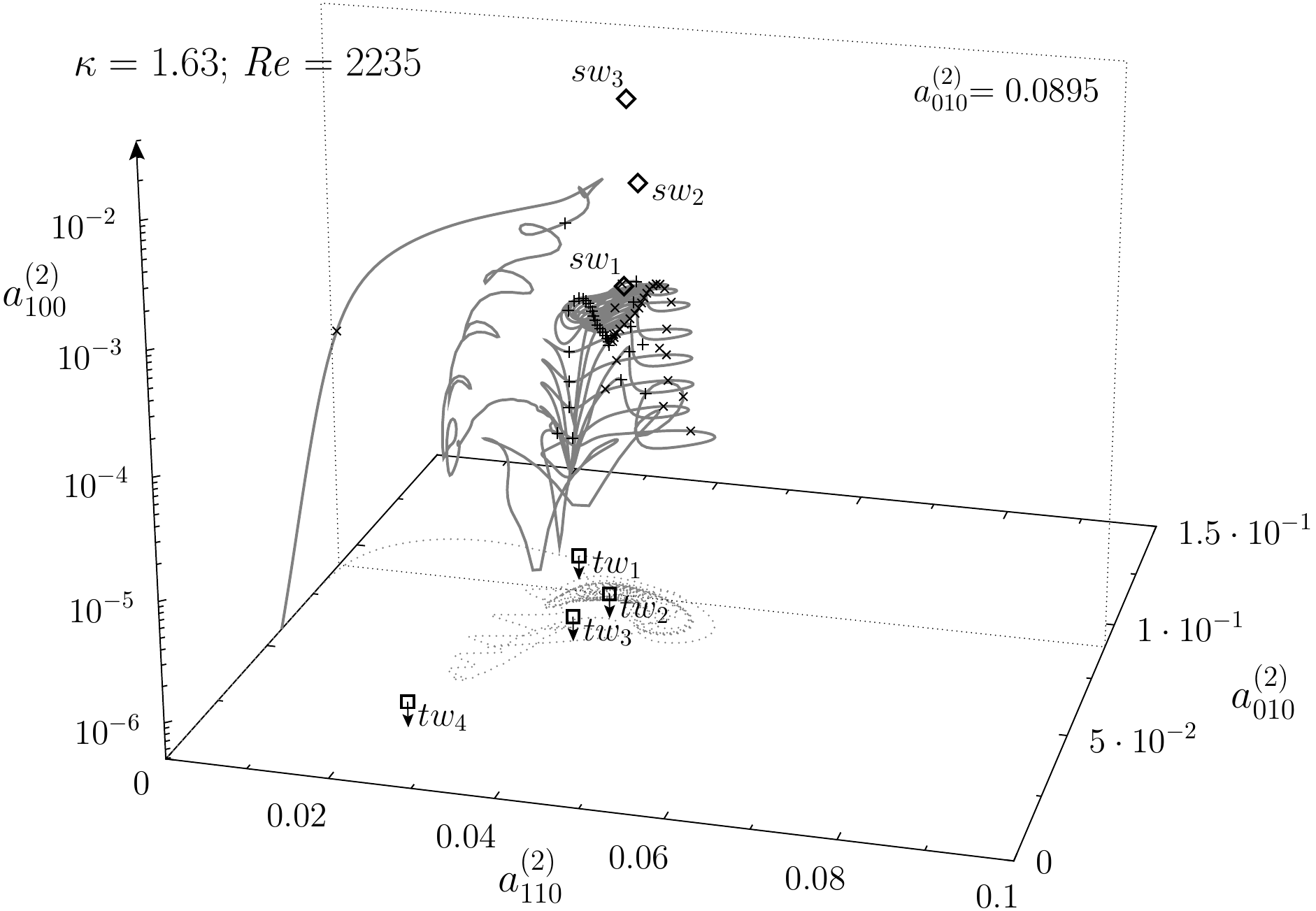} &
      \raisebox{0.33\linewidth}{(\textit{b})}\hspace{-0.6cm} &
      \includegraphics[height=0.32\linewidth,clip]{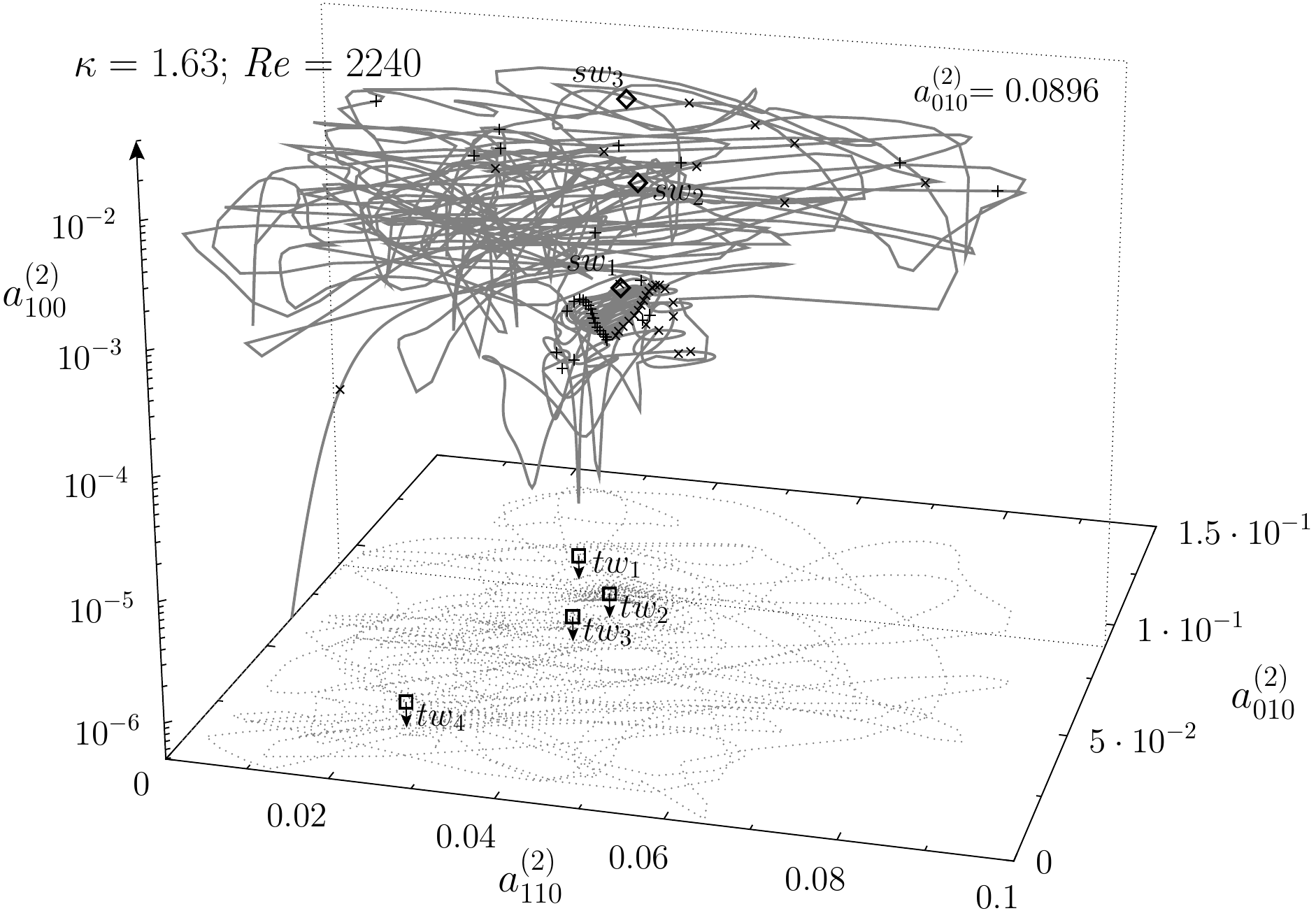}
    \end{tabular}
  \end{center}
  \caption{Three-dimensional phase map projection
    $(a_{010}^{(2)},a_{110}^{(2)},a_{100}^{(2)})$ of two escapes from
    the chaotic set. (\textit{a})
    $(\kappa,\Rey)=(1.63,2235)$. (\textit{b})
    $(\kappa,\Rey)=(1.63,2240)$. Travelling waves ($tw_1$, $tw_2$,
    $tw_3$ and $tw_4$) have been labelled and marked with open
    squares and spiralling waves ($sw_1$, $sw_2$ and $sw_3$) with
    open diamonds. Positive and negative crossings of a Poincar\'e
    section defined by $a_{010}^{(2)}=a_{010}^{(2)}(tw_2)$ are
    indicated by plus signs and crosses, respectively.}
  \label{fig:PhMap3DTurbk1.63Re2230s}
\end{figure}

At somewhat larger $\Rey$ the chaotic set ceases to be an attractor
and cannot hold trajectories
indefinitely. Figure~\ref{fig:PhMap3DTurbk1.63Re2230s} exemplifies two
departures from the chaotic set at $\kappa=1.63$ and $\Rey=2235$ and
$2240$.
Logarithmic scale has been used for the vertical axis to aid
visualisation. The dense structure in the middle of the figure is the
remains of the chaotic attractor, which preserves the same structure
as for lower $\Rey$, except that it is no longer an attractor. For
both $\Rey$ shown, the solution departs in pretty much the same
direction towards the region in phase space where the strongly
spiralling waves $sw_2$ and $sw_3$ are. At $\Rey=2235$
(figure~\ref{fig:PhMap3DTurbk1.63Re2230s}\textit{a}), the trajectory
does not reach the waves and relaminarisation follows
shortly. Instead, at $\Rey=2240$
(figure~\ref{fig:PhMap3DTurbk1.63Re2230s}\textit{b}), the flow reaches
the location where spiralling waves live and is kicked towards
turbulence. The solution stays turbulent for around $1000 D/(4 U)$ and
finally relaminarises, oddly enough, along a path that seems close to
that followed at $\Rey=2235$.

Full understanding of the actual nature of the $2$-fold
azimuthally-symmetric turbulent saddle is beyond the scope of this
study and the issue will not be pursued further. It is nevertheless
pertinent to stress that turbulence does occur within the azimuthal
subspace and even when computations are further restricted to preserve
the shift-reflect symmetry.

\section{Formation and destruction of the chaotic attractor}\label{sec:discu}

Most of the transitions between states reported in this study, once
the drifts due to axial and azimuthal degeneracy have been
appropriately tackled, are of a local nature and conspicuously well
understood. This is not the case of the transition from
doubly-modulated waves to mildly chaotic waves. There are several
known paths whereby deterministic solutions revolving around an
invariant torus become chaotic when the torus disappears. Some of them
have been described in the context of turbulent
transition. \cite*{Lan_DANCCCP_44} suggested quasiperiodic motion
resulting from an infinite cascade of bifurcating incommensurable
frequencies that ultimately lead to turbulence. \cite*{RuTa_CMP_71}
and \cite*{NeRuTa_CMP_71} modified the scenario to include
dissipative systems, such as viscous fluids, that do not, in general,
have quasiperiodic motions. Their route to chaos involved successive
bifurcations from a stable equilibrium into a stable limit cycle
followed by transition to a stable $2$-torus and then to chaos. The
scenario, though, does not give a full description of all possible
bifurcation scenarios leading to the destruction of a $2$-torus and
does not provide conditions under which chaos may or may not
follow. Some scenarios have been analysed by generalising Floquet
theory in systems without symmetry \cite*[][]{ChIo_ARMA_79} and some
theorems on torus breakdown have been formulated
\cite*[][]{AnSaCh_ITCS_93}. These theorems, known under the name
Afraimovich-Shilnikov theorems, suggest three distinct scenarios: (i)
breakdown due to some ordinary bifurcation of phase-locked limit
cycles such as period doubling (flip) or Neimark-Sacker; (ii) sudden
transition to chaos due to the appearance of a homoclinic connection;
(iii) breakdown following from a gradual loss of smoothness. We will
argue that our one-dimensional path follows (ii).

\begin{figure}
  \begin{center}
    \begin{tabular}{cccc}
      \raisebox{0.31\linewidth}{(\textit{a})}\hspace{-0.6cm} &
      \includegraphics[height=0.29\linewidth,clip]{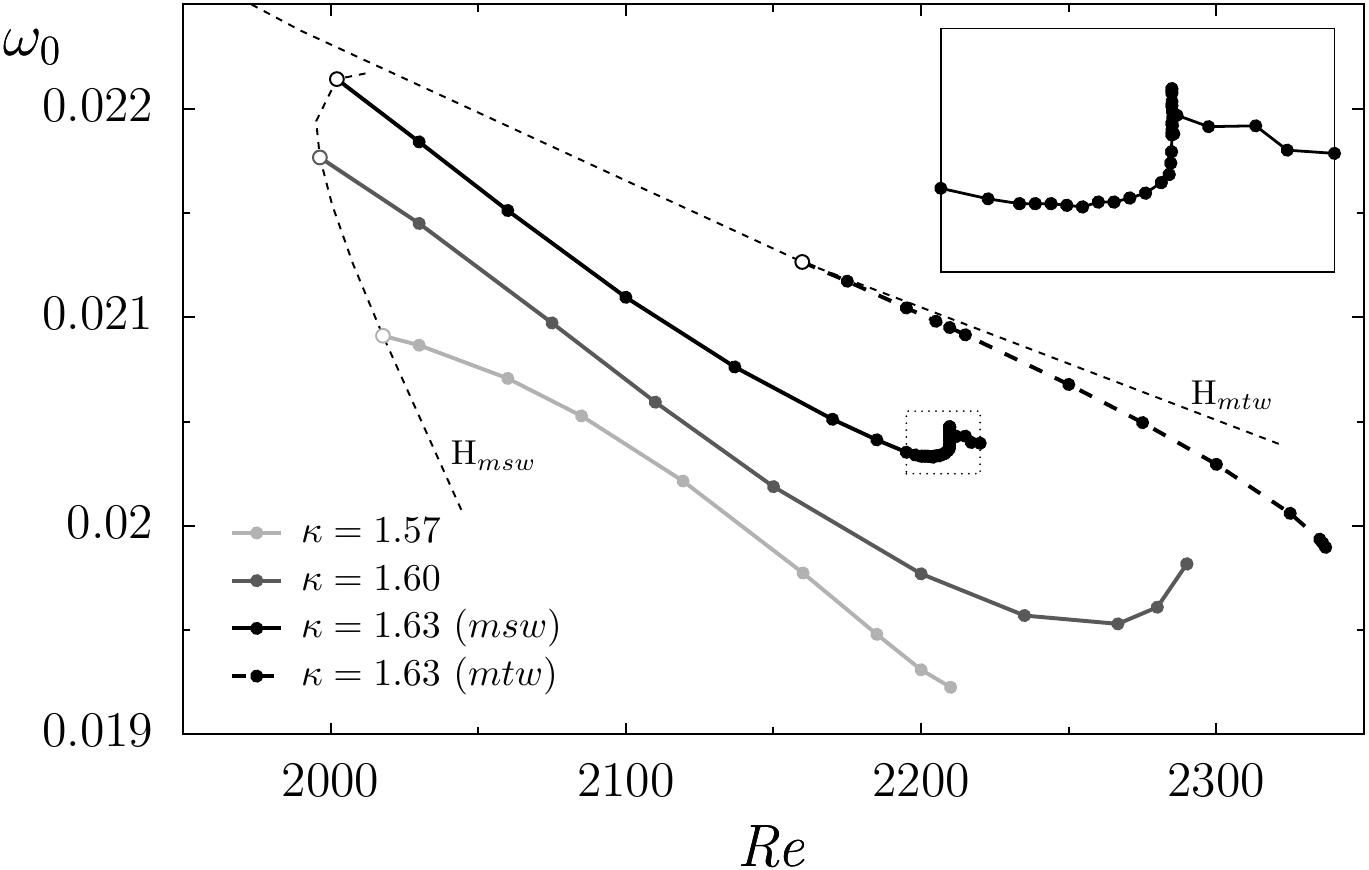} &
      \raisebox{0.31\linewidth}{(\textit{b})}\hspace{-0.6cm} &
      \includegraphics[height=0.29\linewidth,clip]{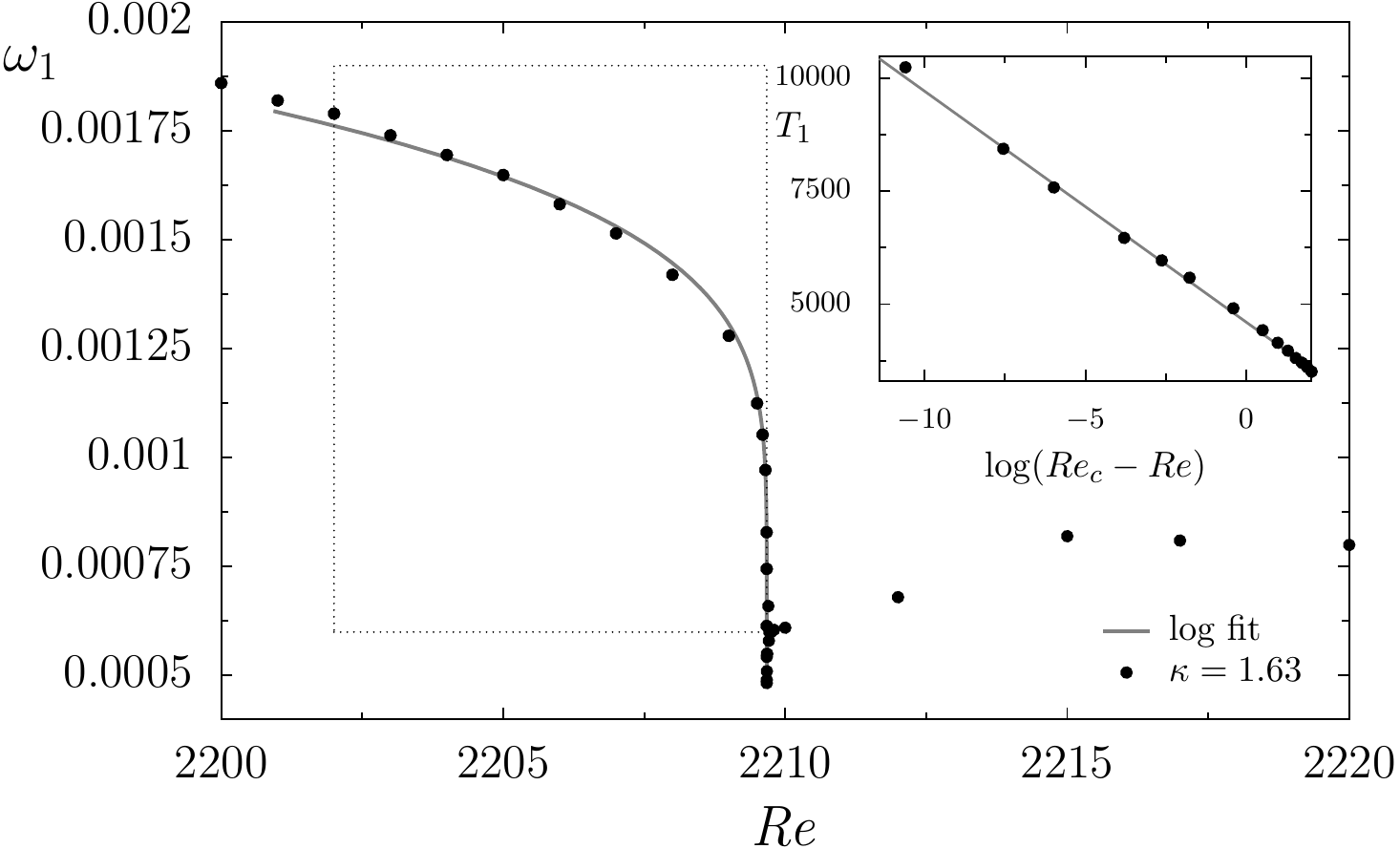}
    \end{tabular}
  \end{center}
  \caption{(\textit{a}) Fast angular frequency ($\omega_0$) of all
    time-dependent solutions as a function of $\Rey$ and $\kappa$.
    Solid lines for spiralling waves and dashed for
    travelling. Shading for varying $\kappa$ as indicated in the
    legend. Hopf bifurcation curves are indicated with dotted lines
    and labelled (H$_{mtw}$ and H$_{msw}$). The inset shows a close-up
    of the small dotted square. (\textit{b}) Slow frequency
    ($\omega_1$) of doubly-modulated waves at $\kappa=1.63$. The
    dotted-square-region is zoomed-in in the inset with a coordinate
    transformation ($T_1=2\upi/\omega_1$ vs $\log(\Rey_c-\Rey)$).}
  \label{fig:freqvsRe}
\end{figure}

Very valuable information can be retrieved from the frequency of
quasiperiodic solutions as they approach the bifurcation point. As
shown in figure~\ref{fig:freqvsRe}(\textit{a}), the fast angular
frequency ($\omega_0$) evolves smoothly through the Neimark-Sacker
bifurcation from modulated to doubly-modulated waves. So much so, that
the bifurcation point for $\kappa=1.63$ (black solid line) at
$\Rey\sim 2199$ is not discernible and doubly-modulated waves inherit
this Hopf frequency directly from modulated waves.
As quasiperiodic motion evolves, the frequency grows slightly but
remains bounded and stabilises at a fairly constant value (on average)
for chaotic waves after the global bifurcation has taken place. In
contrast, the slow frequency ($\omega_1$) experiences substantial
variation across the bifurcation, as is evident from
figure~\ref{fig:freqvsRe}(\textit{b}). Since the two frequencies
associated to degenerate drifts ($c_z$ and $c_{\theta}$) are constant
and $\omega_0$ is nearly constant, while the period
$T_1=2\upi/\omega_1$ becomes unbounded, we will represent the
$4$-torus as a limit cycle and relative limit cycles as equilibria. We
can do this because constant and nearly constant frequencies do not
play a role in the dynamics near bifurcation points.

Infinite-period bifurcations are commonly associated with homoclinic
or heteroclinic behaviour. The two most typical cases are the
saddle-node and the saddle-loop homoclinic bifurcations, depending on
whether the bifurcating orbit becomes homoclinic to a non-hyperbolic
or to an hyperbolic equilibrium, respectively
\cite[][]{Strogatz_B_94,Kuznetsov_B_95}. One of the mensurable
properties that allows to discriminate between the two is the way in
which the period diverges. When a saddle-node develops on a limit cycle,
the period diverges according to:
\begin{equation}
  T_1 = \frac{2\upi}{\omega_1} \sim \frac{C}{\sqrt{\Rey_c-\Rey}}+D,
  \label{eq:Tsn}
\end{equation}
where $\Rey_c$ is the critical $\Rey$, and $C$ and $D$ are fitting
parameters. Meanwhile, when a cycle grows to collide with a hyperbolic
saddle and then disappear in a saddle-loop bifurcation, the period
diverges logarithmically, like \cite*[][]{Ga_JPC_90}
\begin{equation}
  T_1 = \frac{2\upi}{\omega_1} \sim -\lambda^{-1}\log(\Rey_c-\Rey)+D,
  \label{eq:Tsl}
\end{equation}
where the factor $\lambda$ is the eigenvalue of the unstable direction
of the hyperbolic fixed point and $\log$ represents the natural
logarithm.

The behaviour of the period near a saddle-loop bifurcation may be
extended to more complex scenarios such as Shilnikov bifurcations in
n-dimensional dynamical systems or even tori collision with hyperbolic
limit cycles \cite*[][]{MaLoSh_PD_01}, while the saddle-node
homoclinic trend is also observed in the case of orbits homoclinic to
nonhyperbolic equilibria or to other nonhyperbolic cycles \cite*[\eg a
  blue sky catastrophe, see][but also in the case when the invariant
  set forms a torus]{MeMeBaRa_PRL_04}.

The frequency-range of the doubly-modulated-waves is in very good
agreement with a logarithmic fit (\ref{eq:Tsl}) (gray solid line in
figure~\ref{fig:freqvsRe}\textit{b}, see the inset). As a result from
the fit, an estimated value of $\Rey_c=2209.67253$ is obtained for the
critical bifurcation point and $\lambda=1.959\times 10^{-3}
(4U/D)$. This suggests that the torus might be becoming homoclinic to
a hyperbolic saddle cycle with an unstable multiplier of
$\mu=\mathrm{e}^{\lambda T_0}\simeq 1.8253$ ($T_0=307.2 D/(4U)$ is the
fast period of trajectories on the torus, which should coincide, in
the limit, with the period of the saddle cycle) as the bifurcation is
approached.

\begin{figure}
  \begin{center}
    \begin{tabular}{cccc}
      \raisebox{0.36\linewidth}{(\textit{a})}\hspace{-0.6cm} &
      \includegraphics[height=0.38\linewidth,clip]{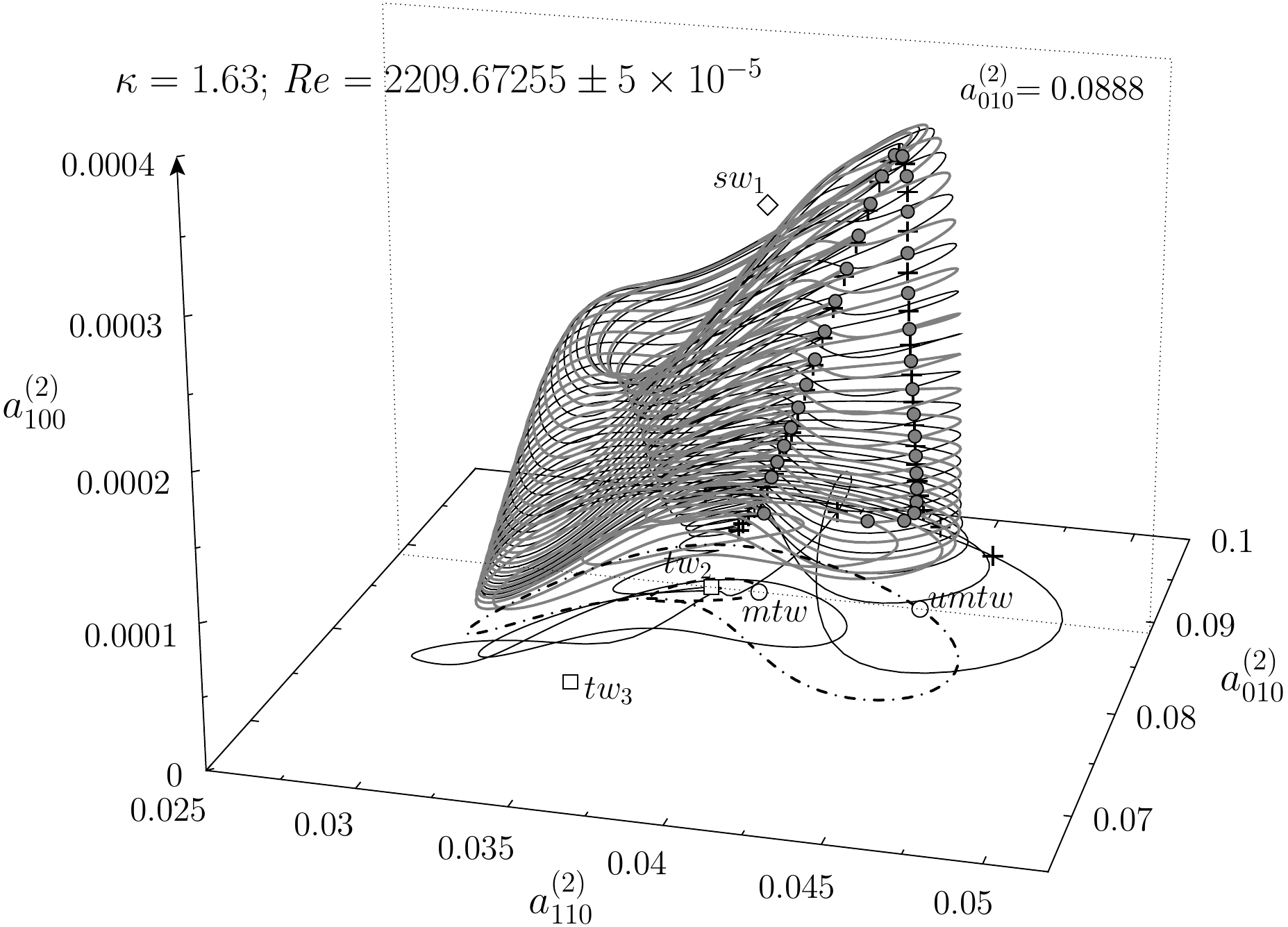} &
      \raisebox{0.36\linewidth}{(\textit{b})}\hspace{-0.6cm} &
      \includegraphics[height=0.35\linewidth,clip]{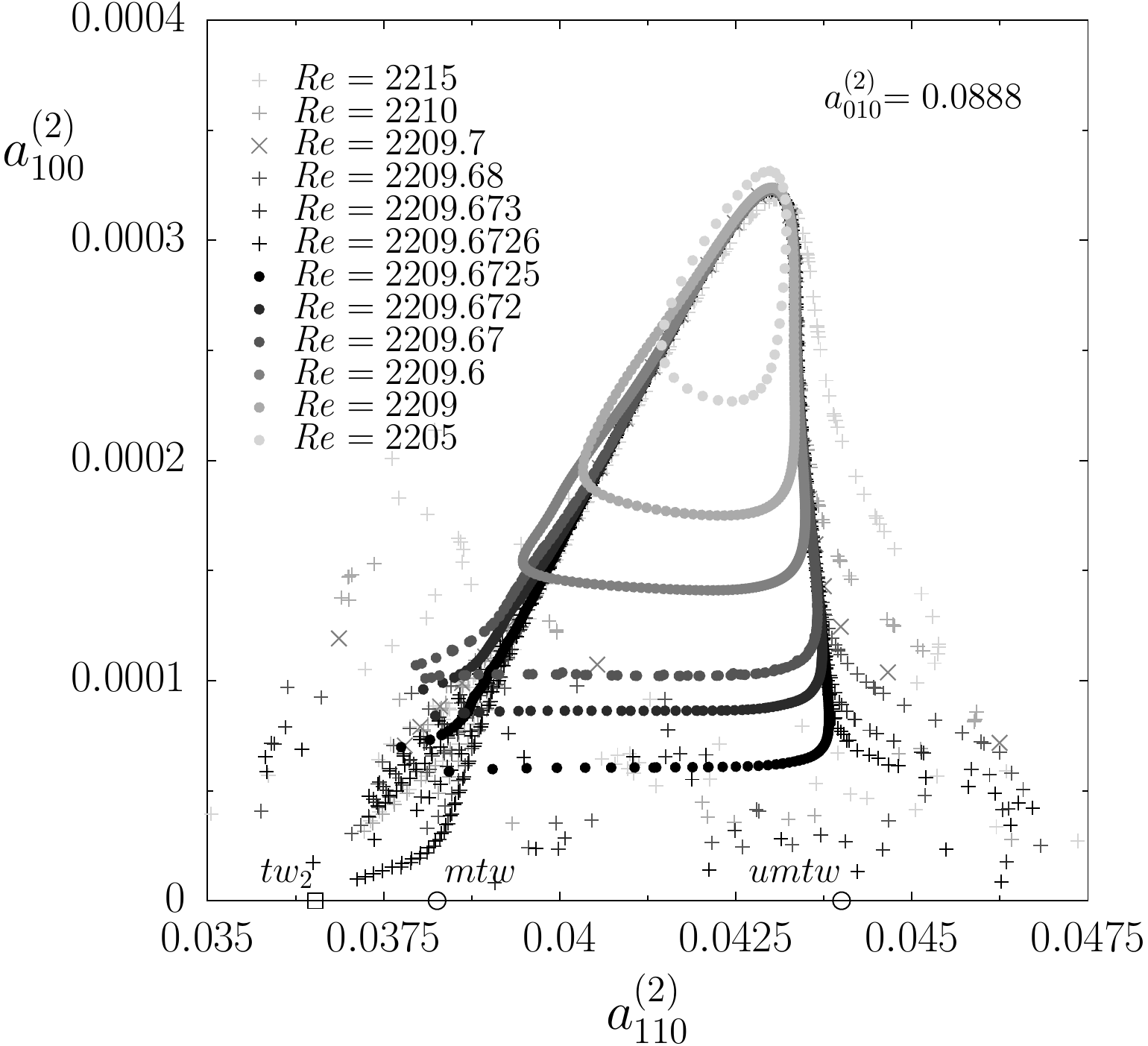}
    \end{tabular}
  \end{center}
  \caption{(\textit{a}) Three-dimensional phase map projection
    $(a_{010}^{(2)},a_{110}^{(2)},a_{100}^{(2)})$ just before (gray,
    circles for Poincar\'e crossings) and after (black, plus signs)
    transition. (\textit{b}) Poincar\'e section showing solutions
    before (filled circles) and after (plus signs and crosses for
    phase-locked wave) transition.  Travelling waves ($tw_2$ and
    $tw_3$) have been labelled and marked with open squares and
    spiralling wave ($sw_1$) with an open diamond. The shift-reflect
    stable ($mtw$) and unstable ($umtw$) modulated waves (dashed and
    dashdotted lines in \textit{a}, open circles in \textit{b}) are
    also shown.}
  \label{fig:PhMapTrans}
\end{figure}

The evolution of the frequency suggests an approach to a homoclinic
or, as we will see, a heteroclinic cyclic connection involving
hyperbolic orbits. To elucidate with which solution or solutions the
torus collides, figure~\ref{fig:PhMapTrans} depicts trajectories
corresponding to refined approaches to the transition point from
either side of the bifurcation.
Figure~\ref{fig:PhMapTrans}(\textit{a}) shows a three-dimensional
phase map representation of the stable solution just before
(doubly-modulated wave, gray solid line) and after (mildly chaotic
wave, black solid line) transition. For a large fraction of the time,
both solutions run together, as is clear from the upper part of the
plot. However, as the solutions spiral down the outer side of the
torus and approach the shift-reflect subspace, the doubly-modulated
wave closes on the torus from below, while the chaotic wave is kicked
away and frenzies in the vicinity of the shift-reflect subspace for a
while before joining the torus back at the inner side to spiral
up. The divergence of the trajectories is further evidenced when
comparing the crossings on the Poincar\'e section (open circles for
the quasiperiodic wave, plus signs for the chaotic wave). The
Poincar\'e map associated to the doubly-modulated wave describes a
triangle-shaped cycle, while the chaotic wave only preserves the upper
part of the triangle and breaks down at the lower side. The solutions,
as shown in the figure, only revolve once around the torus, but they
fill the trajectories on the Poincar\'e map densely when followed for
sufficiently long times.

Figure~\ref{fig:PhMapTrans}(\textit{b}) shows these solutions on the
Poincar\'e section, together with an array of solutions at varying
distance from either side of the bifurcation point. As transition is
approached from $\Rey<\Rey_c$ (filled circles getting darker), the
cycle gets more and more triangle-shaped, starting from the
phase-locked pseudo-circular shape at $\Rey=2205$. The cycles run
clockwise. The upper corner remains fixed, while the lower side gets
aligned with the shift-reflect subspace, represented by the horizontal
axis, and approaches it in a blow-up fashion (note that solutions are
not $\Rey$-equispaced). For each quasiperiodic wave, a fairly
equidistant chaotic wave at the other side of the transition point has
been represented (plus signs, crosses for the phase-locked wave at
$\Rey=2209.7$). The right side of the chaotic wave closes on the right
side of the quasiperiodic wave like a zipper as bifurcation is
approached. After departing from the right vertex of the triangle,
chaos follows and reattachment to the left side of the torus is less
neat.

The trend followed by the right vertex of the triangle suggests that a
shift-reflect saddle cycle (a fixed point of the Poincar\'e map) might
be approached. Less clear, due to the chaotic dynamics, but still
plausible, is the conjecture that the left vertex approaches the
modulated travelling wave ($mtw$). This modulated travelling wave, a
stable node in its shift-reflect subspace, was presumed to undergo a
fold-of-cycles at around $\Rey\simeq 2337$, thus pairing with a
shift-reflect saddle cycle of larger amplitude. This would be the
saddle cycle approached by the right vertex. If collision of the torus
with the two limit cycles occurred simultaneously, as suggested by
figure~\ref{fig:PhMapTrans}(\textit{b}) and supported by the
fold-pitchfork bifurcation scenario for maps \cite*[see][for the
  closely related fold-flip bifurcation]{KuMeVV_IJBC_04}, a non-robust
heteroclinic cycle would be created at $\Rey_c$ leaving no
attractors. No full description of the fold-pitchfork bifurcation for
maps is available in the literature. It would correspond to the
analysis of a 1:1 resonance \cite[][]{Kuznetsov_B_95} in the presence
of $Z_2$ symmetry and the details are too intricate to be discussed
here and shall be treated elsewhere.

Using shift-reflect symmetry-restricted time-evolution and
edge-tracking refinement \cite[][]{SkYoEc_PRL_06,ScEcYo_PRL_07}, where
the laminar and turbulent asymptotic states have been replaced by two
clearly distinct paths towards the stable modulated wave, we have
numerically computed the conjectured unstable modulated wave ($umtw$)
with sufficient accuracy. It has a single real multiplier outside of
the unit circle corresponding to a shift-reflect eigenmode. Its
location is perfectly congruent with a collision with the outer
surface of the torus (right vertex) as the global bifurcation point is
approached (see figure~\ref{fig:PhMapTrans}).

It can be shown that the $mtw$ branch effectively experiences a
fold-of-cycles (a shift-reflect real multiplier crosses the unit
circle) and that $umtw$ belongs precisely to the saddle branch created
at the fold as schematically shown in figure~\ref{fig:BifDia}.

\begin{figure}
  \begin{center}
    \begin{tabular}{cccc}
      \raisebox{0.33\linewidth}{(\textit{a})}\hspace{-0.6cm} &
      \includegraphics[height=0.32\linewidth,clip]{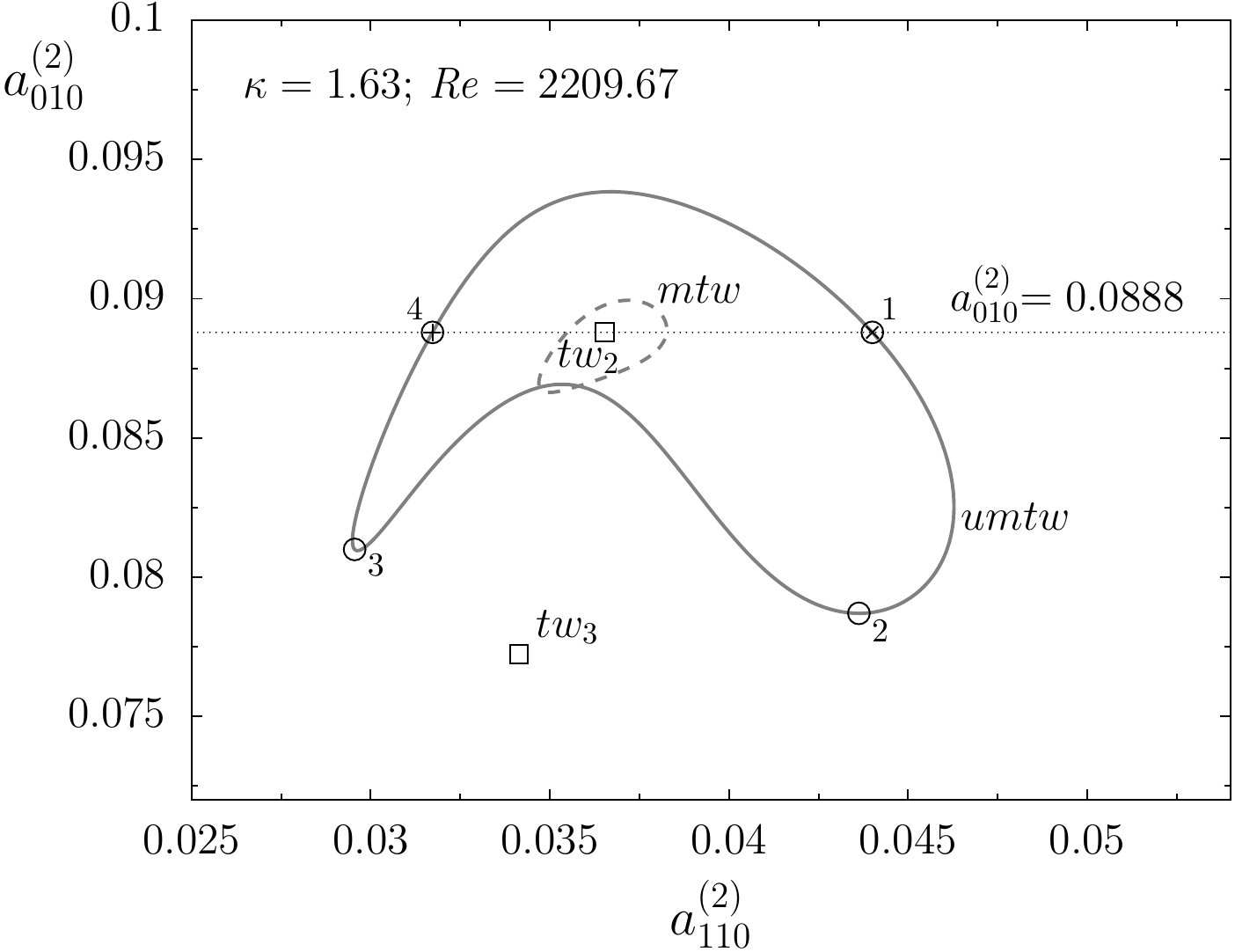} &
      \raisebox{0.33\linewidth}{(\textit{b})}\hspace{-0.6cm} &
      \includegraphics[height=0.31\linewidth,clip]{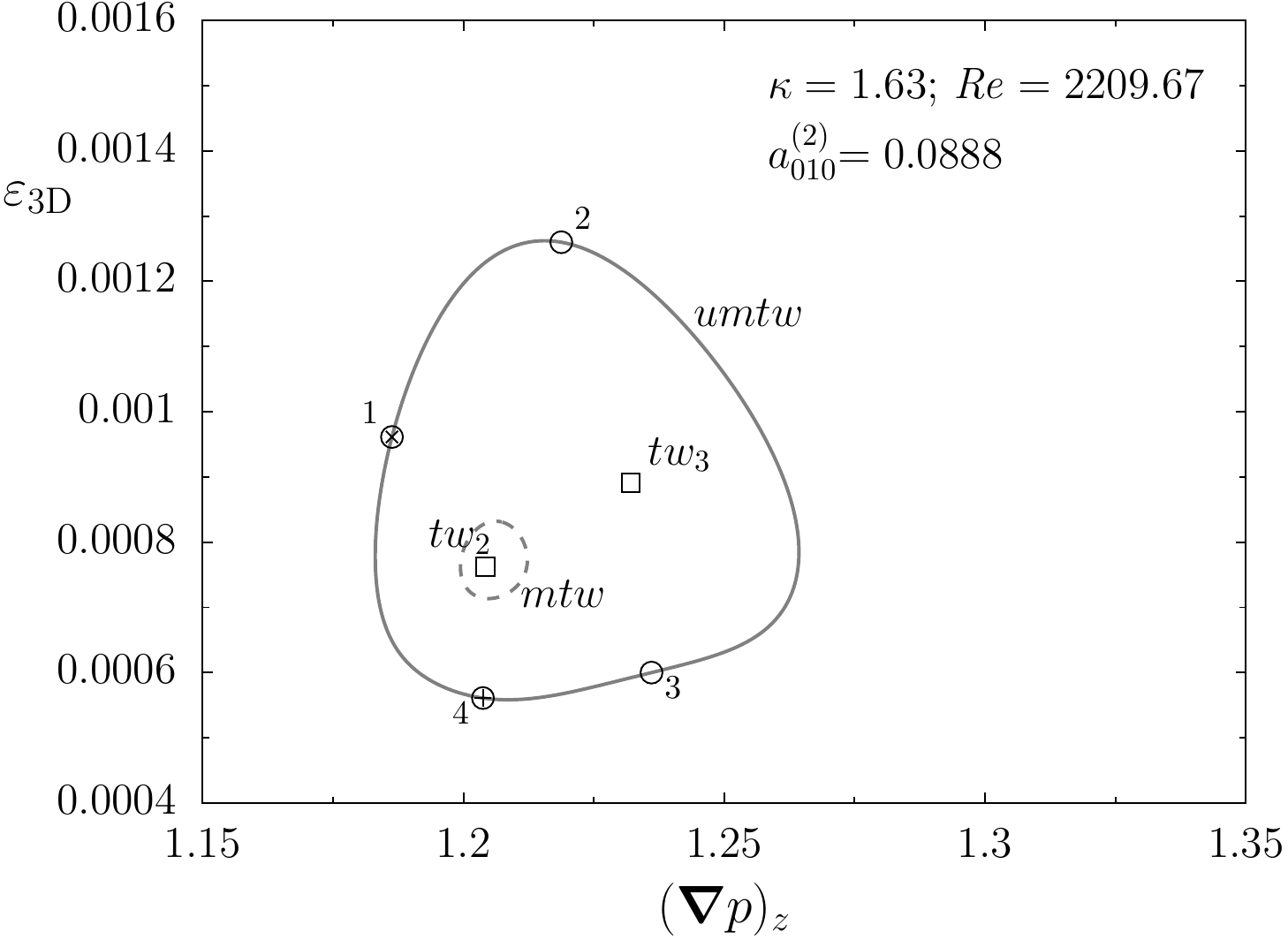}
    \end{tabular}
  \end{center}
  \caption{Phase map projections of an unstable modulated travelling
    wave ($umtw$) at $(\kappa,\Rey)=(1.63,2209.67)$. (\textit{a})
    $a_{010}^{(2)}$ vs $a_{110}^{(2)}$. (\textit{b}) Three-dimensional
    energy ($\varepsilon_{\rm 3D}$) vs axial pressure gradient
    ($(\bnabla p)_z$). Middle-branch travelling waves ($tw_2$ and
    $tw_3$) have been labelled and marked with open squares.  The
    modulated travelling wave $mtw$ at the same parameter values is
    indicated with a dashed line. Positive and negative crossings of a
    Poincar\'e section defined by $a_{010}^{(2)}=a_{010}^{(2)}(tw_2)$
    are indicated by plus signs and crosses, respectively. Numbered
    circles correspond to snapshots in
    figure~\ref{fig:SSUTPOk1.63Re2209.67}.}
  \label{fig:PhMapUTPOk1.63Re2209.67}
\end{figure}

Figure~\ref{fig:PhMapUTPOk1.63Re2209.67} shows two alternative
axial-drift-independent phase map projections of the unstable
modulated travelling wave solution at $(\kappa,\Rey)=(1.63,2209.67)$,
close to the bifurcation point.
It is a straightforward observation to identify $mtw$ and $umtw$ as
bounding solutions for the asymptotic shift-reflect projection of the
doubly-modulated wave $m^2sw$ as it approaches the bifurcation point
(compare the three solutions in
figure~\ref{fig:PhMapTrans}). Especially $umtw$, since $mtw$ seems to
be affected by manifold tangles. This correspondence gets even clearer
when comparing figure~\ref{fig:PhMapUTPOk1.63Re2209.67}(\textit{b})
and figure~\ref{fig:PhMapRPTk1.63Re2205}(\textit{b}). The projection
of the doubly-modulated wave seems to grow in the region bounded by
$mtw$ at the inside and $umtw$ at the outside, the heteroclinic
bifurcation corresponding to the simultaneous collision with both
modulated waves.

\begin{figure}
  \begin{center}
    \begin{tabular}{cccc}
      \raisebox{0.30\linewidth}{(\textit{a})}\hspace{-0.6cm} &
      \includegraphics[height=0.27\linewidth,clip]{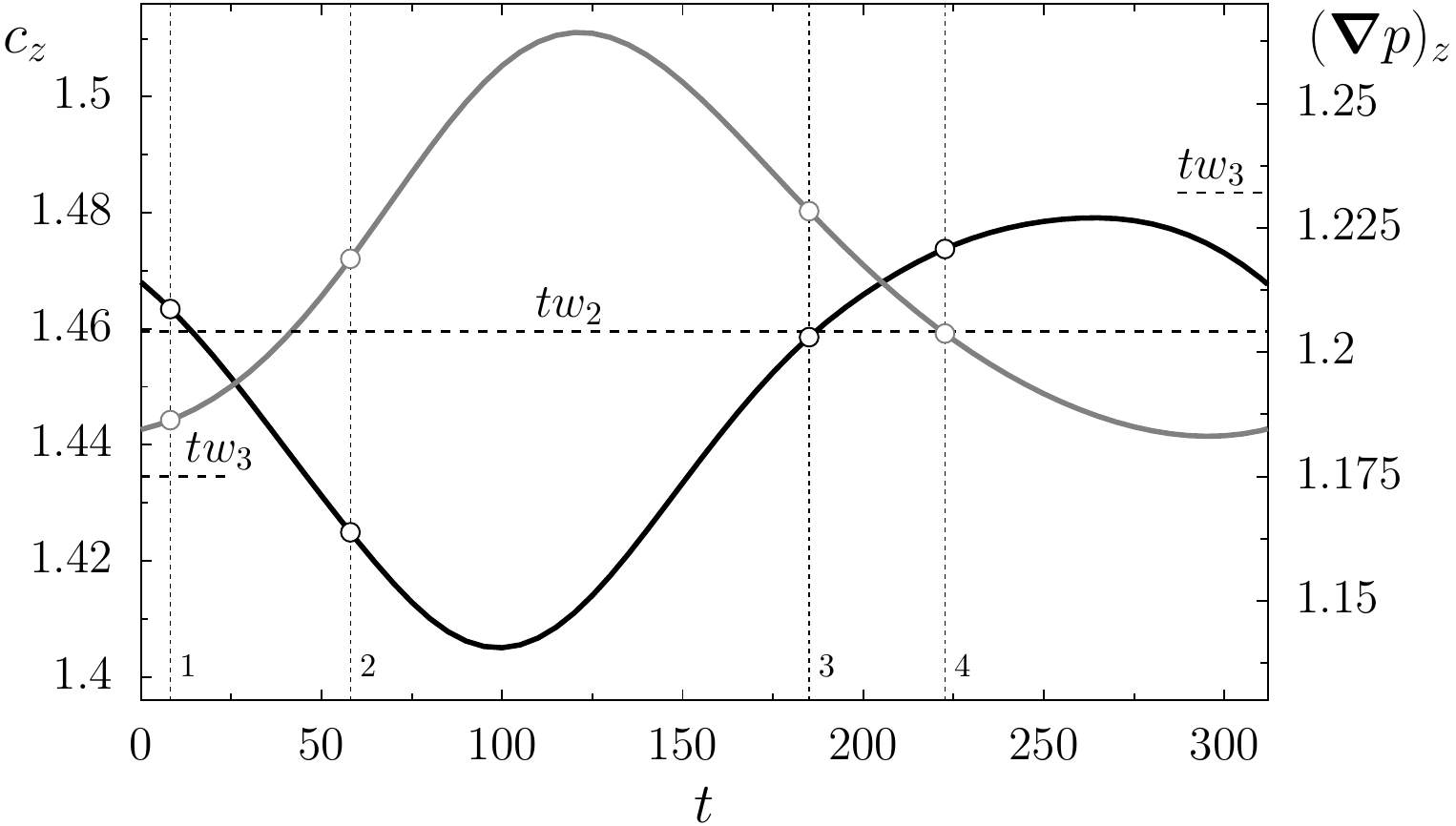} &
      \raisebox{0.30\linewidth}{(\textit{b})}\hspace{-0.6cm} &
      \includegraphics[height=0.28\linewidth,clip]{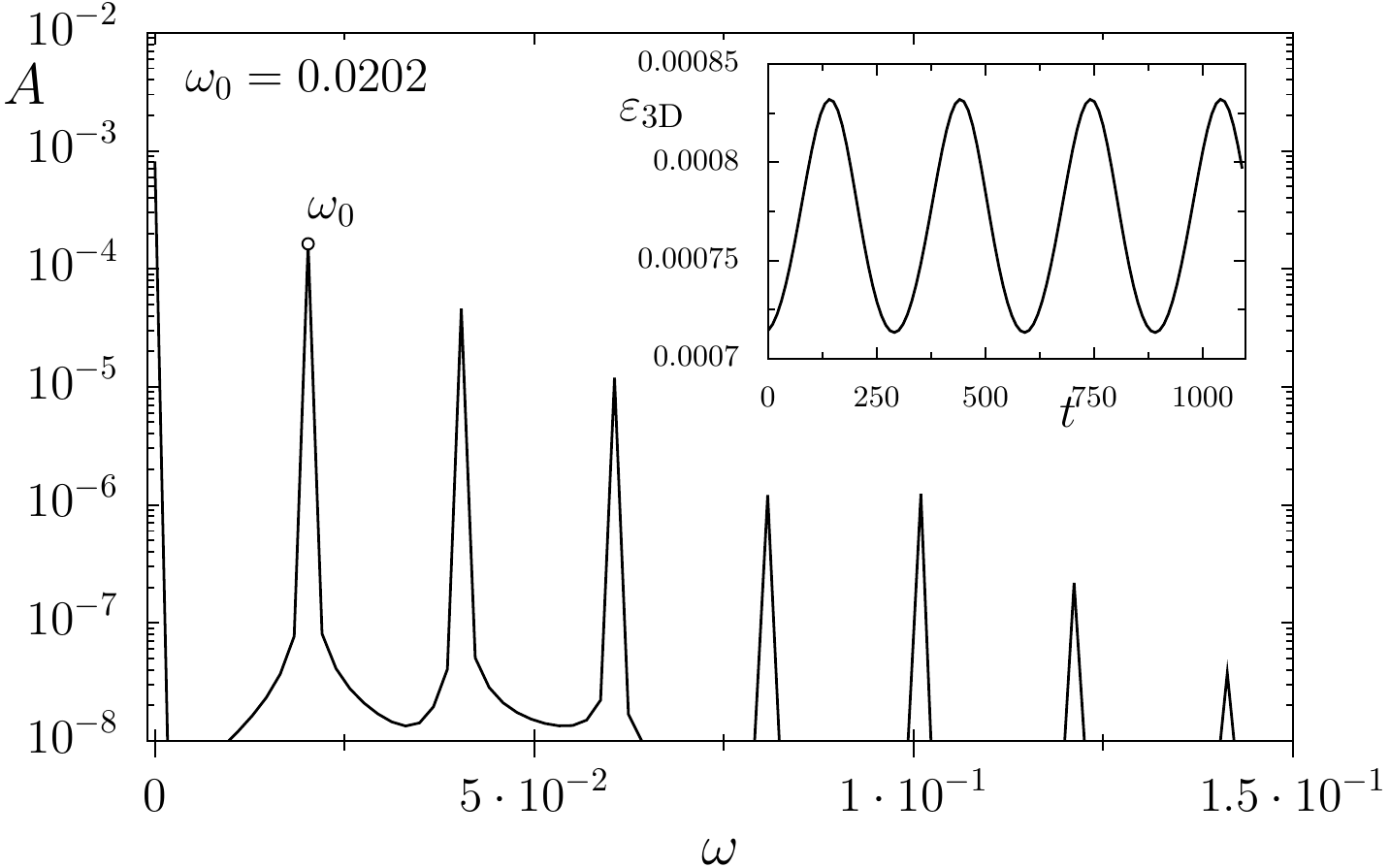}
    \end{tabular}
  \end{center}
  \caption{Unstable modulated travelling wave ($umtw$) at
    $(\kappa,\Rey)=(1.63,2209.67)$. (\textit{a}) Axial phase speed
    ($c_z$, black line) and axial pressure gradient ($(\bnabla p)_z$,
    gray line) time-series. Dotted horizontal lines indicate the
    values for $tw_2$ and $tw_3$ while numbered vertical lines and
    open circles indicate snapshots in
    figure~\ref{fig:SSUTPOk1.63Re2209.67}. (\textit{b}) Fourier transform
    of the non-axisymmetric streamwise-dependent modal energy contents
    ($\varepsilon_{\rm 3D}$). Part of the time signal is plotted in
    the inset.}
  \label{fig:TSFTUTPOk1.63Re2209.67}
\end{figure}

The time evolution of $umtw$ is characterised in
figure~\ref{fig:TSFTUTPOk1.63Re2209.67}(\textit{a}) through the
representation of the axial phase speed ($c_z$, black line) and the
mean axial pressure gradient ($(\bnabla p)_z$, gray line) time-series
along a full period of the solution.
Comparing the time signals with those for $mtw$ in
figure~\ref{fig:TSFTTPOk1.63Re2335}(\textit{a}) it becomes relatively
clear that $umtw$ results from an amplification of $mtw$ once it has
gone unstable at the fold. The Fourier transform of the energy
contained in non-axisymmetric streamwise-dependent modes
($\varepsilon_{\rm 3D}$) has been plotted in
figure~\ref{fig:TSFTUTPOk1.63Re2209.67}(\textit{b}), with the time
signal shown in the inset frame. The resolution of the spectrum is not
as good as for stable waves due to the difficulty in stabilising an
unstable wave for sufficiently long times. Nevertheless, the wave can
be claimed as sufficiently well converged and it exhibits a clear peak
of angular frequency at $\omega_0=0.0202 \; (4U/D)$, corresponding to
a period $T_0=2\upi/\omega_0=311.2 D/(4 U)$. The $mtw$ at the same
parameter values, has a slightly faster frequency $\omega_0=0.0209 \;
(4U/D)$, amounting to a shorter period $T_0=2\upi/\omega_0=299.9 D/(4
U)$.

\begin{figure}
  \begin{center}
    \begin{tabular}{cccccc}
      \raisebox{0.16\linewidth}{($1$)}\hspace{-0.6cm} &
      \includegraphics[height=0.15\linewidth,clip]{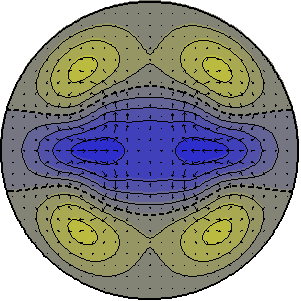} &
      \includegraphics[height=0.15\linewidth,clip]{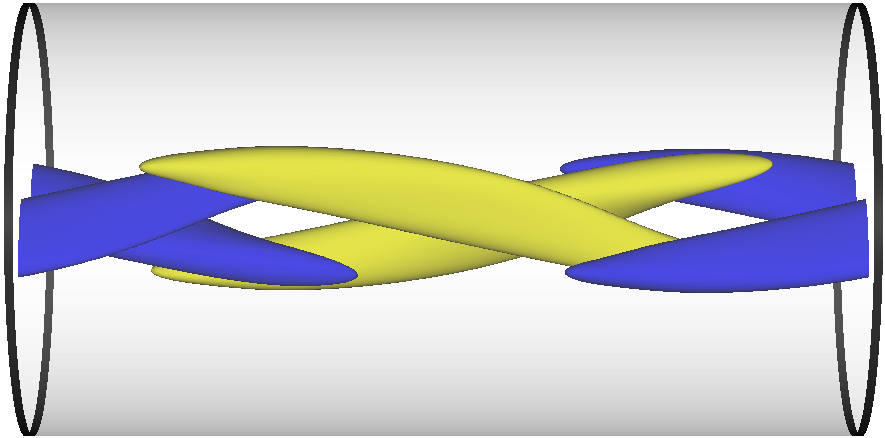} &
      \raisebox{0.16\linewidth}{($2$)}\hspace{-0.6cm} &
      \includegraphics[height=0.15\linewidth,clip]{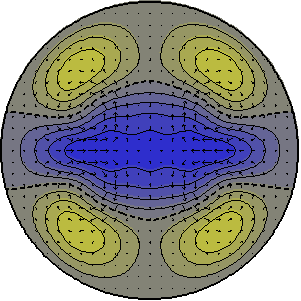} &
      \includegraphics[height=0.15\linewidth,clip]{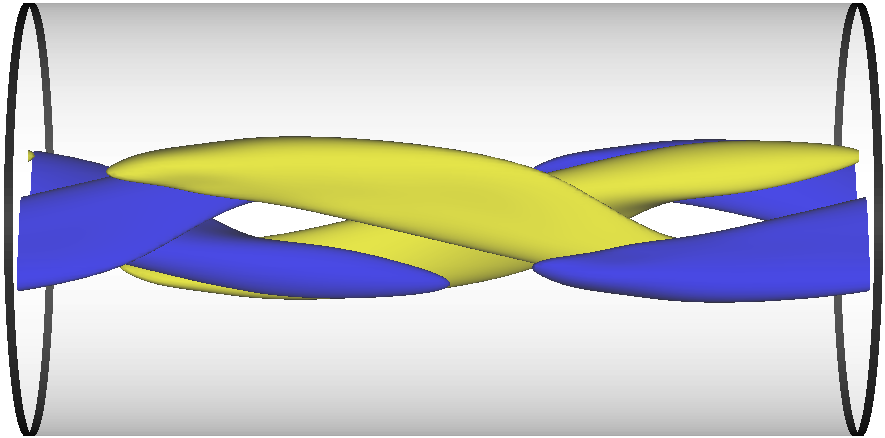}\\
      \raisebox{0.16\linewidth}{($3$)}\hspace{-0.6cm} &
      \includegraphics[height=0.15\linewidth,clip]{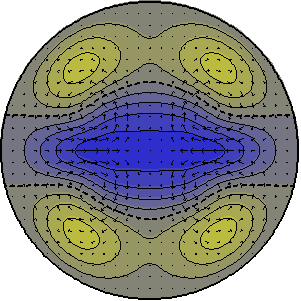} &
      \includegraphics[height=0.15\linewidth,clip]{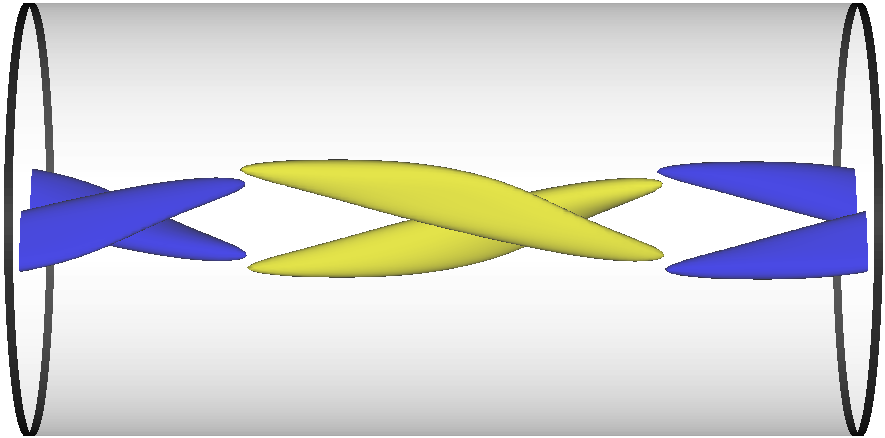} &
      \raisebox{0.16\linewidth}{($4$)}\hspace{-0.6cm} &
      \includegraphics[height=0.15\linewidth,clip]{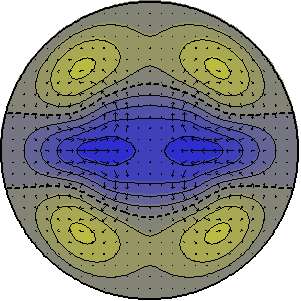} &
      \includegraphics[height=0.15\linewidth,clip]{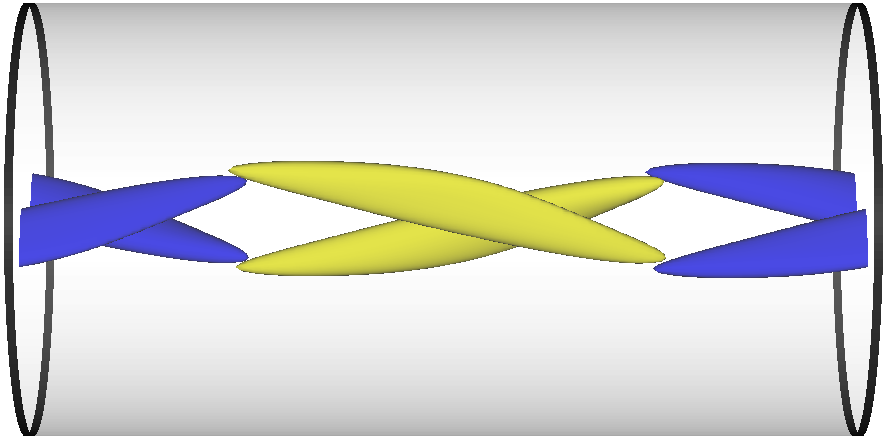}\\
    \end{tabular}
  \end{center}
  \caption{Unstable modulated travelling wave at
    $(\kappa,\Rey)=(1.63,2209.67)$. Left: $z$-averaged cross-sectional
    axial velocity contours spaced at intervals of $\Delta \langle
    u_z\rangle_z = 0.1 U$. In-plane velocity vectors are also
    displayed. Right: axial vorticity isosurfaces at $\omega_z=\pm 1
    U/D$. Fluid flows rightwards. Blue (dark gray) for negative,
    yellow (light) for positive. ($1$) $t=8$, ($2$) $t=58$, ($3$)
    $t=185$ and ($4$) $t=223$ $D/(4U)$. To avoid drift due to
    streamwise advection, snapshots are taken in a comoving frame
    moving with the instantaneous advection speed from
    figure~\ref{fig:TSFTUTPOk1.63Re2209.67}(\textit{a}). The snapshots
    have been indicated with empty circles in
    figures~\ref{fig:PhMapUTPOk1.63Re2209.67} and
    \ref{fig:TSFTUTPOk1.63Re2209.67}(\textit{a}).}
  \label{fig:SSUTPOk1.63Re2209.67}
\end{figure}

To better support the relevance of the wave in the dynamics of the
problem, a few snapshots (indicated in
figures~\ref{fig:PhMapUTPOk1.63Re2209.67} and
\ref{fig:TSFTUTPOk1.63Re2209.67}\textit{a} with circles) of the flow
field along a cycle have been represented in
figure~\ref{fig:SSUTPOk1.63Re2209.67} (see online movie).
It is clear from the snapshots that the modulation is similar but of
slightly larger amplitude than that for $mtw$ at $\Rey=2335$
(figure~\ref{fig:SSTPOk1.63Re2335}). Significantly relevant is also
the resemblance born by snapshot ($1$) to that of
figure~\ref{fig:SSChaosk1.63Re2215}($3$) corresponding to an approach
to the shift-reflect space of the chaotic wave $cw$ at
$\Rey=2215$. This fact, added to the similarities of the phase map
trajectories (see the winding of the trajectory near the shift-reflect
subspace in figure~\ref{fig:PhMap3DChaosk1.63Re2215}) clearly
underlines the influence of this wave on the dynamics even at
$\Rey$-values beyond the global bifurcation.

\begin{figure}
  \begin{center}
    \begin{tabular}{cccccccc}
      \raisebox{0.09\linewidth}{(\textit{a})}\hspace{-0.6cm} &
      \includegraphics[width=0.22\linewidth,clip]{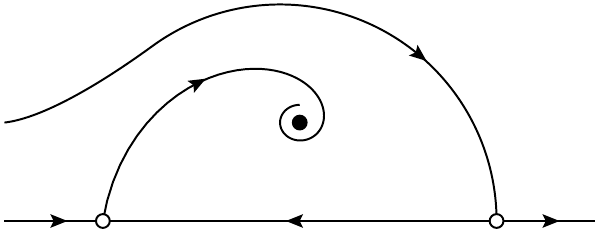} &
      \raisebox{0.09\linewidth}{(\textit{b})}\hspace{-0.6cm} &
      \includegraphics[width=0.22\linewidth,clip]{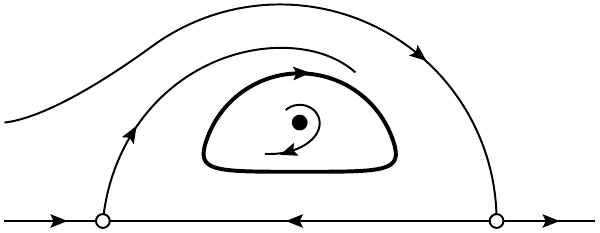}
      \raisebox{0.09\linewidth}{(\textit{c})}\hspace{-0.6cm} &
      \includegraphics[width=0.22\linewidth,clip]{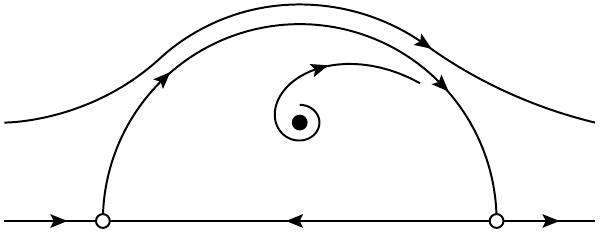}
      \raisebox{0.09\linewidth}{(\textit{d})}\hspace{-0.6cm} &
      \includegraphics[width=0.22\linewidth,clip]{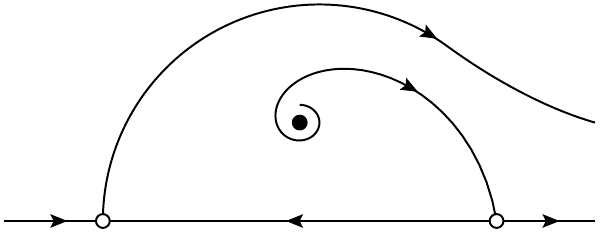}\\
      & &
      \multicolumn{4}{c}{$\overbrace{\textrm{\makebox[0.66\linewidth][c]{}}}$}\\
      & &
      \raisebox{0.08\linewidth}{(\textit{e})}\hspace{-0.6cm} &
      \includegraphics[width=0.22\linewidth,clip]{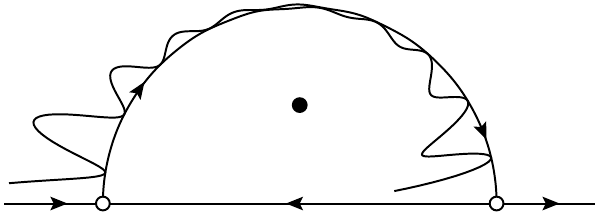}
      \raisebox{0.08\linewidth}{(\textit{f})}\hspace{-0.6cm} &
      \includegraphics[width=0.22\linewidth,clip]{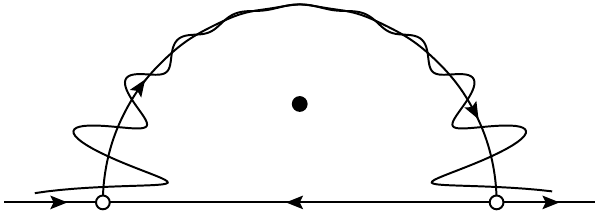}
      \raisebox{0.08\linewidth}{(\textit{g})}\hspace{-0.6cm} &
      \includegraphics[width=0.22\linewidth,clip]{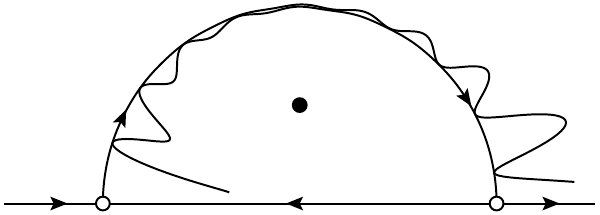}
    \end{tabular}
  \end{center}
  \caption{Schematic phase map representation on a Poincar\'e section
    of the formation and destruction of the chaotic set. (\textit{a}),
    (\textit{b}), (\textit{c}) and (\textit{d}) represent an
    approximating interpolated flow. (\textit{e}), (\textit{f}) and
    (\textit{g}) sketch the heteroclinic tangencies expected instead
    of (\textit{c}). Phase maps are symmetric with respect to the
    horizontal axis, which represents the shift-reflect subspace, but
    only the upper half is shown.}
  \label{fig:PhMapSketch}
\end{figure}

The bifurcation scenario introduced in figure~\ref{fig:BifDia} is now
schematically developed in
figure~\ref{fig:PhMapSketch}(\textit{a}-\textit{d}).
A stable modulated spiralling wave (filled circle) and two saddle
modulated travelling waves (open circles on the horizontal axis,
which represents the shift-reflect subspace) exist for sufficiently
low $\Rey$ (figure~\ref{fig:PhMapSketch}\textit{a}). Strictly
speaking, phase maps are symmetric with respect to the horizontal axis
and a symmetry-related stable modulated spiralling wave should have
been drawn. For simplicity, though, only the upper half is actually
represented. In figure~\ref{fig:PhMapSketch}(\textit{b}), the
modulated spiralling wave has undergone a Neimark-Sacker bifurcation
and a stable doubly-modulated wave has been created (thick solid
loop). This wave grows and collides with both modulated travelling
waves simultaneously in figure~\ref{fig:PhMapSketch}(\textit{c}),
generating an heteroclinic cycle that leaves this phase-space region
open in figure~\ref{fig:PhMapSketch}(\textit{d}), as observed for
$\Rey\gtrsim 2225$.

The phase maps sketched correspond to approximating the Poincar\'e
application by an interpolating flow. For the real map, heteroclinic
cycles ($hc$ in figure~\ref{fig:BifDia}) will not generically
arise. Instead, heteroclinic tangles and tangencies will typically
occur, broadening the bifurcation point (or curve) into a wedge where
complex dynamics and chaos are common feature. This is precisely what
was observed in \S\ref{sec:timedep}. The appearance of chaotic waves,
and the transition point here discussed, would then be related to the
first heteroclinic tangency of the stable and unstable modulated waves
manifolds (figure~\ref{fig:PhMapSketch}\textit{e}). Chaotic waves
exist for a finite range of the parameter
(figure~\ref{fig:PhMapSketch}\textit{f}) but end up disappearing and
causing the phase space region they inhabited to open up
(figure~\ref{fig:PhMapSketch}\textit{g}), so that the flow can leave
towards other regions. In the present case towards turbulent motion,
as described above.

In the $\Rey$-range where chaotic waves exist, the manifolds
entanglement and interweaving is complicated by the presence of the
travelling- and spiralling-wave solutions discussed in
\S\ref{sec:releq}. It is not possible with the tools at hand to gain a
complete understanding of what the role of relative equilibria and
their manifolds is in organising the dynamics around the chaotic
set. An example of the complex interminglement of manifolds is visible
in figure~\ref{fig:PhMapTrans}(\textit{b}) for $\Rey=2209.6726$, where
the reattachment to the remnants of the torus is sometimes
accomplished along the unstable manifold of the travelling wave $tw_2$
rather than along that of the modulated travelling wave $mtw$ as is
clear from the alignment of black plus signs.

All things considered, the torus breakdown scenario involving a
homoclinic or, rather, heteroclinic-cyclic connection, seems to be the
most plausible. The logarithmic fit of the period is supportive of
this surmise, except that the fitting parameter $\lambda$ needs to be
reinterpreted and it is not clear as to which of either modulated
waves it is related. The unstable multiplier associated to $mtw$ has
been measured and its value $\mu_{mtw}=1.1265$ corresponds to an
eigenvalue $\lambda=3.972 \times 10^{-4} (4U/D)$, which is one order
of magnitude smaller than the one resulting from the logarithmic
fit. It could well be that the unstable manifold of the saddle cycle
takes the lead as the unstable manifold of $mtw$ loses strength in the
presence of twisting and tangency. An indicator that this could be
happening is the deformation of the bottom left corner of the cycle in
figure~\ref{fig:PhMapTrans}(\textit{b}), which seems to be adapting to
the foldings of the tangle. Unfortunately, the unstable multiplier of
$umtw$ cannot by itself explain the divergence of the period,
either. Its value $\mu_{umtw}=3.9549$, tantamounts to an eigenvalue
$\lambda=4.419 \times 10^{-3} (4U/D)$, which is larger, by a factor of
two, than the parameter $\lambda$ ajusted by the fit.







\section{Conclusion}\label{sec:conclu}

We have studied the fate of upper-branch travelling waves of a
shift-reflect-symmetric family that belongs in the $2$-fold
azimuthally-periodic subspace of pipe Poiseuille flow. The special
stability properties of these waves, which are stable to perturbations
within their azimuthal subspace in some regions of parameter space,
assist bare time-evolution into converging time-dependent solutions
that result from ulterior destabilisation of the waves.

Despite the contorted continuation surfaces of these waves when
analysed as a function of both axial wavenumber ($\kappa$) and
Reynolds number ($\Rey$), a one-dimensional path at $\kappa=1.63$ and
varying $\Rey$ suffices to unveil a multitude of increasingly complex
solutions that are reported for the first time in pipe
flow. Successive transitions to spiralling waves, modulated travelling
waves, modulated spiralling waves and doubly-modulated travelling
waves have been reported along the way, and the arising solutions
described to some extent, in order to exemplify the extreme richness
and complexity of the problem, even within a given subspace. The
bifurcation cascade has been shown to culminate in a torus breakdown,
presumably following a cyclic heteroclinic connection of the stable
and unstable manifolds of modulated travelling waves. The usual
manifolds entanglement and heteroclinic tangencies that occur in
sufficiently high-dimensional (three or more) dynamical systems, gives
the simplest explanation to the observed phenomenology, \ie the
formation of an attracting chaotic set upon torus-breakdown and its
subsequent dismantlement at higher $\Rey$.

\begin{figure}
  \begin{center}
    \includegraphics[width=0.7\linewidth,clip]{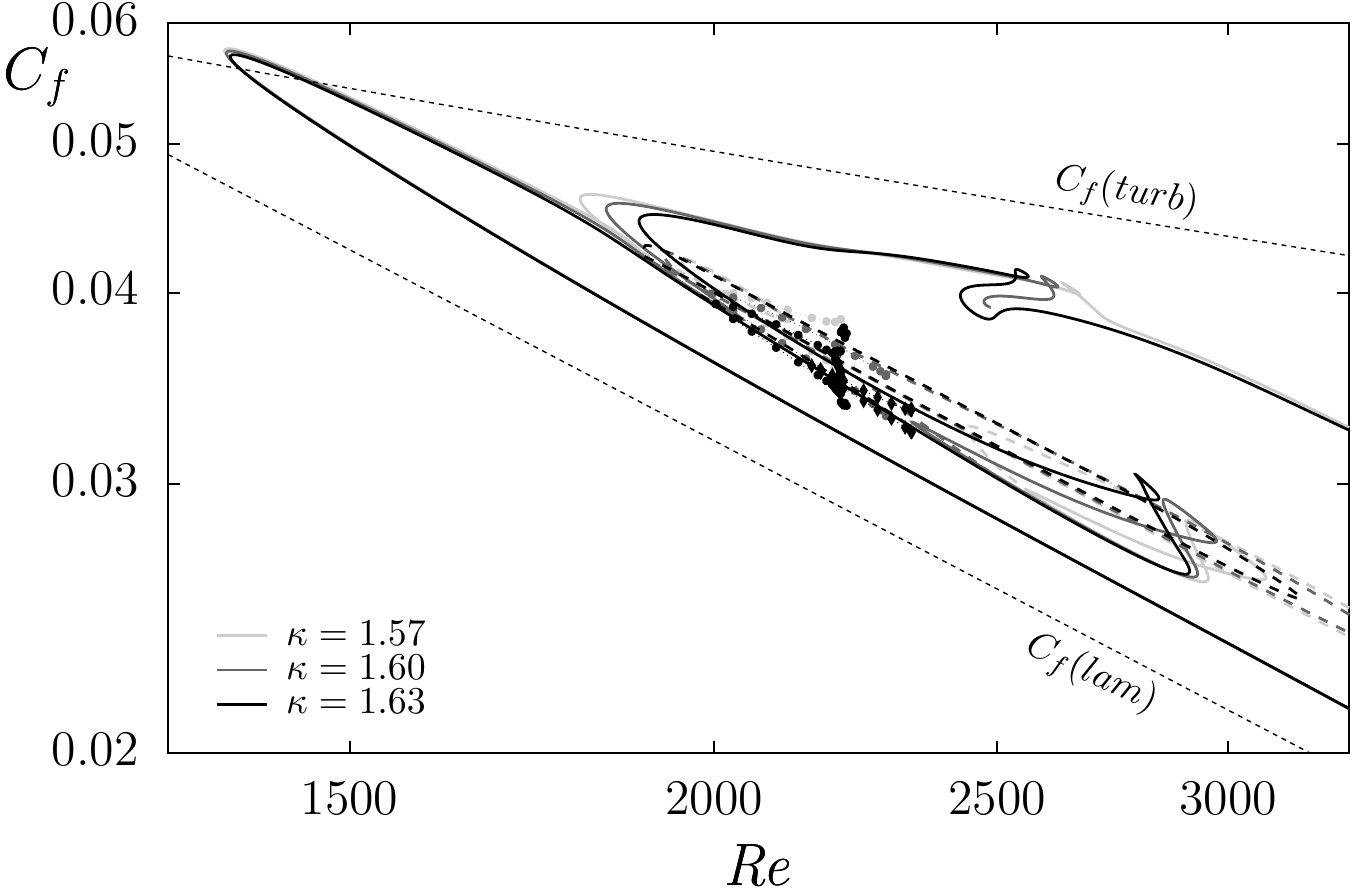}
  \end{center}
  \caption{Friction factor ($C_f$) as a function of $\Rey$ and
    $\kappa$ (shading as explained in the legend). Solid lines
    represent shift-reflect travelling waves, dashed for spiralling
    waves. Extrema of time-dependent travelling/spiralling solutions
    are marked as filled diamonds/circles. Turbulent
    \cite[$C_f(turb)$, from the experimentally obtained log-law
      $1/\sqrt{C_f}=2\log_{10}(\Rey \sqrt{C_f})-0.8$; see][page
      551]{SchGer_B_68} and laminar ($C_f(lam)=64/\Rey$) friction
    factor are indicated with dotted lines.}
  \label{fig:nuvsRe}
\end{figure}

The rupture of the chaotic set does no longer support the existence of
mildly chaotic waves, and the flow leaves to other regions of phase
space, namely $2$-fold azimuthally-symmetric turbulence. Turbulence
within the subspace is, as for full-space pipe, sustained transiently
by a chaotic saddle that already exists at lower $\Rey$ and occupies a
region of phase space far away from where all solutions here reported
lie (except possibly $tw_4$, $sw_2$ and $sw_3$). In this sense, it can
be safely asserted that the chaotic set here described plays no
significant role in the development of the phase space structures that
integrate the turbulent saddle. This is clearly supported by
figure~\ref{fig:nuvsRe}, where friction factor ($\nu$) is plotted as a
function of $\Rey$.
Laminar and turbulent typical friction factor values are indicated
with dotted lines to guide the eye. It is evident that, while
lower-branch waves approach laminar values and upper-branch waves are
close to turbulent friction factors, all other solutions reported in
the present study are confined in a narrow stripe halfway between
laminarity and turbulence.

It is nevertheless plausible that similar bifurcation cascades might
take place in other subspaces for which available simulation tools are
currently incapable of overcoming the effects of higher-dimensional
unstable manifolds. Some of the chaotic sets thus generated could well
go into the formation of a turbulent saddle, although global
bifurcations involving several subspaces simultaneously may well
complicate the picture even further.

Extending the current analysis to varying $\kappa$ could help better
understand the global bifurcation leading to chaos. In particular,
continuing time-dependent solutions to larger $\kappa$, preferably
along unstable branches where possible, would help clarify the
connections among modulated travelling and spiralling waves and,
possibly, completely unfold the fold-pitchfork bifurcation of maps
here advanced to account for the appearance of chaotic waves.

Also the direct study of $2$-fold azimuthally-symmetric turbulence,
which in itself entrains great complexity, may yield results that can
help get an intuition as to what may be going on in fully-fledged
turbulence. The $2$-fold azimuthally-symmetric subspace has already
proven to be a particularly accessible terrain in this respect.

%

\begin{acknowledgments}
This work has been supported by the Spanish Ministry of Science and
Technology, under grants FIS2007-61585, by the Catalan
Government under grant SGR-00024, and by the Deutsche
Forschungsgemeinschaft.

\end{acknowledgments}


\bibliographystyle{jfm}
\bibliography{JFM10}

\end{document}